\documentclass[%
 reprint,
 amsmath,amssymb,
 aps,
]{revtex4-2}

\usepackage{graphicx}% Include figure files
\usepackage{dcolumn}% Align table columns on decimal point
\usepackage{bm}% bold math
\usepackage{hyperref}% add hypertext capabilities

\hypersetup{
  colorlinks   = true, 
  urlcolor     = blue, 
  linkcolor    = blue, 
  citecolor   = blue 
}
\usepackage[normalem]{ulem}
\usepackage{xcolor}
\usepackage{mathdots}
\usepackage{comment}
\usepackage{verbatim}
\usepackage{bbm}
\usepackage{amsthm}

\newcommand{\abs}[1]{\lvert #1 \rvert}
\newcommand{\ket}[1]{\lvert #1 \rangle}
\newcommand{\bra}[1]{\langle #1 \rvert}
\newcommand{\rket}[1]{\lvert #1 )}
\newcommand{\rbra}[1]{( #1 \rvert}

\newcommand{\rbraket}[2]{( #1 \vert #2 )}

\newcommand{\rmd}{\mathrm{d}}
\newcommand{\tr}{\operatorname{tr}}
\newcommand{\LL}[1]{$\bar{\mathcal{L}}_{#1}$}
\newcommand{\R}{\mathcal{R}}

\newcommand{\Tr}{\operatorname{Tr}}

\usepackage{textcomp}

\usepackage[smallerops]{newtxmath}

\begin{document}

\preprint{APS/123-QED}

\title{The entanglement membrane in exactly solvable lattice models}

\author{Michael A. Rampp, Suhail A. Rather, and Pieter W. Claeys}
\affiliation{Max Planck Institute for the Physics of Complex Systems, 01187 Dresden, Germany}

\date{\today}% It is always \today, today,

\begin{abstract}
Entanglement membrane theory is an effective coarse-grained description of entanglement dynamics and operator growth in chaotic quantum many-body systems. The fundamental quantity characterizing the membrane is the entanglement line tension. 
However, determining the entanglement line tension for microscopic models is in general exponentially difficult. 
We compute the entanglement line tension in a recently introduced class of exactly solvable yet chaotic unitary circuits, so-called generalized dual-unitary circuits, obtaining a non-trivial form that gives rise to a hierarchy of velocity scales with $v_E<v_B$. 
For the lowest level of the hierarchy, \LL{2} circuits, the entanglement line tension can be computed entirely, while for the higher levels the solvability is reduced to certain regions in spacetime. This partial solvability enables us to place bounds on the entanglement velocity.
We find that \LL{2} circuits saturate certain bounds on entanglement growth that are also saturated in holographic models. Furthermore, we relate the entanglement line tension to temporal entanglement and correlation functions. 
We also develop new methods of constructing generalized dual-unitary gates, including constructions based on complex Hadamard matrices that exhibit additional solvability properties and constructions that display behavior unique to local dimension greater than or equal to three.
Our results shed light on entanglement membrane theory in microscopic Floquet lattice models and enable us to perform non-trivial checks on the validity of its predictions by comparison to exact and numerical calculations. Moreover, they demonstrate that generalized dual-unitary circuits display a more generic form of information dynamics than dual-unitary circuits.
\end{abstract}

\maketitle

\section{Introduction}

Effective descriptions form the backbone of our understanding of modern condensed matter physics~\cite{Anderson1984}. Rather than considering a full microscopic description of a many-body system, which is both intractable and uninstructive, it is generally possible to identify the most relevant degrees of freedom through an effective `coarse-graining'. In this way the important features of the model, determining its macroscopic behavior, can be determined from a tractable theory for the emergent effective degrees of freedom. In equilibrium statistical physics these features are typically the symmetry and topology of the Hamiltonian. 

Recent years have seen a substantial increase in our understanding of the universal aspects of non-equilibrium quantum dynamics~\cite{D’Alessio2016,Fisher2023}. Somewhat surprisingly, it has turned out that even chaotic dynamics, i.e., non-integrable dynamics lacking any global symmetries, displays universal macroscopic behaviors that can be captured in a hydrodynamic description. Prime examples are the emergence of a hydrodynamics of information transport with information acting as the conserved quantity~\cite{Keyserlingk2018,Nahum2018}, or the appearance of Kardar-Parisi-Zhang scaling in the fluctuations of entanglement growth in noisy systems~\cite{Nahum2017}.

Entanglement membrane theory (EMT) unifies the examples mentioned above and provides an effective coarse-grained description of the macroscopic aspects of chaotic quantum many-body dynamics~\cite{Nahum2017,Jonay2018,Zhou2019,Zhou2020}. It is based on the observation that in such non-integrable dynamics entanglement is \emph{produced} locally, as opposed to being transported by stable quasiparticles -- as is the case in integrable dynamics. The effective theory for the ``entanglement membrane'' describing the dynamics is determined by the entanglement line tension (ELT). This line tension $\mathcal{E}(v)$ constitutes a local cost function for an entanglement membrane in spacetime and depends on a local velocity $v$. The leading-order behavior of entanglement dynamics and operator growth on sufficiently large length scales is then captured by the membrane configuration minimizing the total cost. In this way the quantum many-body problem can be reduced to standard statistical mechanics. This approach has been generalized to settings involving measurements, couplings to the environment, and Clifford circuits~\cite{Li2021,Li2023,Lovas2023,Sierant2023}. However, determining the ELT from a microscopic model is generally hard. It has so far only been accomplished for random circuits in the large-local-Hilbert-space limit~\cite{Nahum2017,Zhou2019,Zhou2020}, for certain holographic models~\cite{Mezei2018,Mezei2020}, and for dual-unitary circuits~\cite{Zhou2020} (the only case of non-random Floquet spin models). While in the latter case the entanglement dynamics remain tractable, the ELT becomes trivial. Here we provide exact results for the ELT in a class of hierarchical dual-unitary circuits with nontrivial results, again avoiding the need for randomness or averaging.

\subsection{Unitary circuits}

Unitary circuits are discrete-space, discrete-time models for local unitary dynamics motivated by classical cellular automata~\cite{Fisher2023}. Such circuits have gained much attention in recent years both theoretically and experimentally, since they present minimal models for many-body dynamics that can be naturally realized in current quantum computing setups~\cite{Georgescu2014,Preskill2018}.

The basic building blocks of unitary circuits are two-site unitary gates $U$, whose matrix elements $U_{ab,cd}$ are graphically expressed as 
\begin{align}
\vcenter{\hbox{\includegraphics[width=0.8\columnwidth]{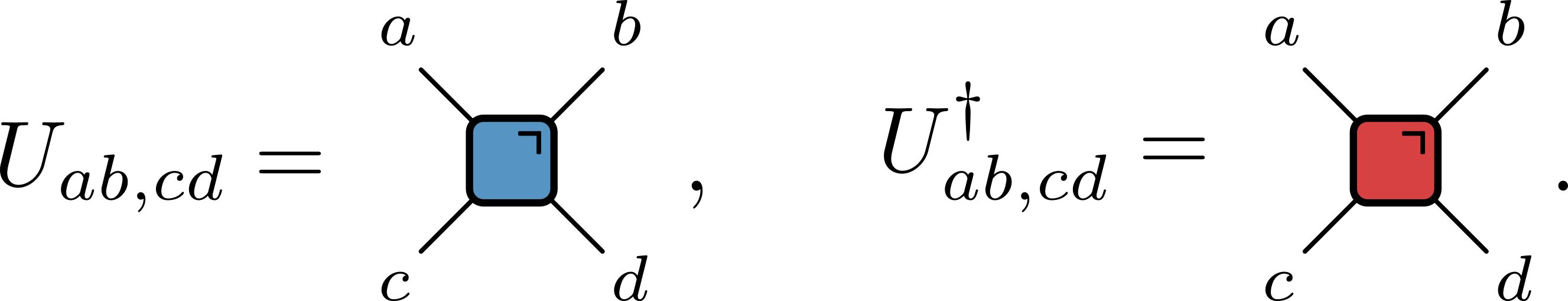}}} \label{eq:def_U}
\end{align}
If indices are suppressed, closed lines correspond to tensor contractions and open legs correspond to uncontracted indices (see also Ref.~\cite{Fisher2023}). Moving from a two-site setting to a full one-dimensional lattice, a many-body evolution operator can be constructed by arranging the two-site unitary gates in a brickwork geometry as
\begin{align}
\vcenter{\hbox{\includegraphics[width=0.7\columnwidth]{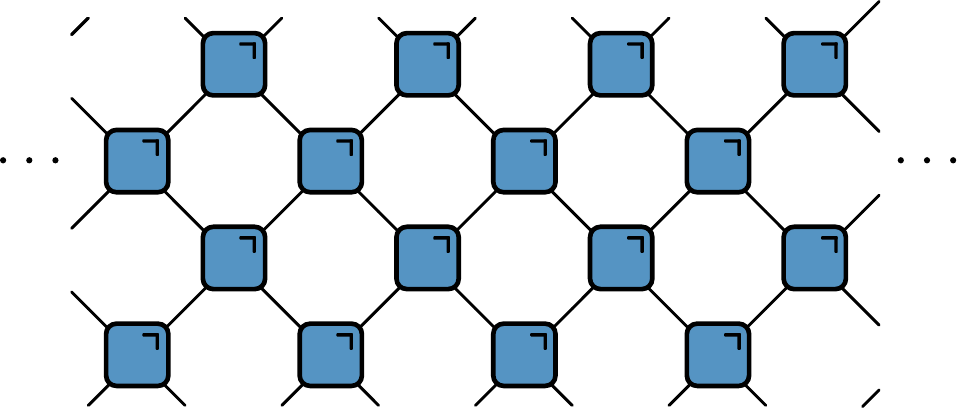}}} \label{eq:def_brickwork}\,.
\end{align}
Here discrete time runs vertically and the local degrees of freedom, e.g. qubits, are arranged horizontally, with unitary gates acting alternately on the even and odd bonds between them.

When considering entanglement properties it is often convenient to work in a ``folded'' space of replicated gates. For any number of replicas $\alpha$ the folded gate is graphically expressed as
\begin{align}
\vcenter{\hbox{\includegraphics[width=0.1\columnwidth]{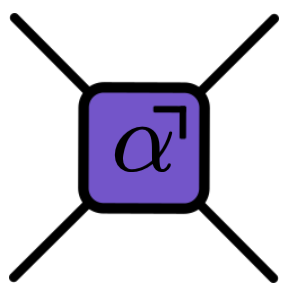}}}\,= (U \otimes U^*)^{\otimes\alpha}, \quad \alpha=1,2,3,\ldots \label{fig:folded_gate}
\end{align}
We will drop the label when $\alpha=2$.
We will make frequent use of the following vectors in the folded space
\begin{align}
\vcenter{\hbox{\includegraphics[height=0.025\textheight]{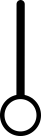}}} \,= \frac{1}{q^\frac{\alpha}{2}}\, \overbrace{\vcenter{\hbox{\includegraphics[height=0.025\textheight]{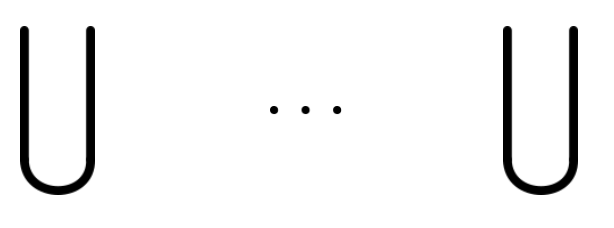}}}}^{2\alpha}\,, \qquad \vcenter{\hbox{\includegraphics[height=0.025\textheight]{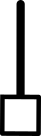}}}\, = \frac{1}{q^\frac{\alpha}{2}}\,\overbrace{\vcenter{\hbox{\includegraphics[height=0.025\textheight]{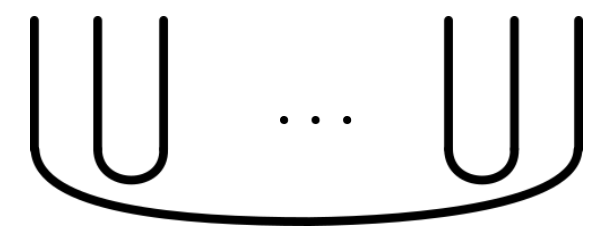}}}}^{2\alpha}\,. \,\,
\end{align}
Here we denote the local Hilbert space dimension by $q$, appearing as a normalization factor, and these two vectors correspond to permutations in replica space. The ``circle'', $\vcenter{\hbox{\includegraphics[height=0.025\textheight,angle=90,origin=c]{figs/circle.png}}}$, corresponds to the identity permutation and the ``square'', $\vcenter{\hbox{\includegraphics[height=0.025\textheight,angle=90,origin=c]{figs/square.png}}}$, corresponds to the cyclic permutation.

Unitarity results in a set of graphical identities
\begin{align}
    \vcenter{\hbox{\includegraphics[width=0.12\columnwidth]{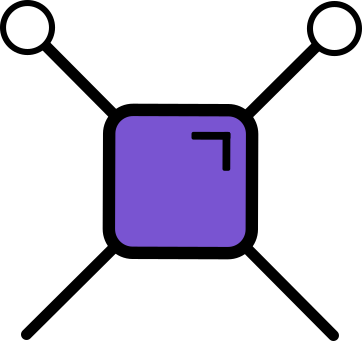}}}\, = \vcenter{\hbox{\includegraphics[width=0.008\textheight,angle=135]{figs/circle.png}}}\,\vcenter{\hbox{\includegraphics[width=0.008\textheight,angle=225]{figs/circle.png}}}, \quad
    \vcenter{\hbox{\includegraphics[width=0.12\columnwidth]{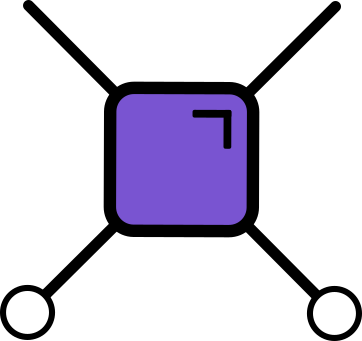}}}\, = \vcenter{\hbox{\includegraphics[width=0.008\textheight,angle=45]{figs/circle.png}}}\,\vcenter{\hbox{\includegraphics[width=0.008\textheight,angle=-45]{figs/circle.png}}}. \label{eq:unitarity}
\end{align}
The same equations hold when the cyclic permutation operator $\vcenter{\hbox{\includegraphics[height=0.025\textheight,angle=90,origin=c]{figs/square.png}}}$ is used instead of the identity permutation operator  $\vcenter{\hbox{\includegraphics[height=0.025\textheight,angle=90,origin=c]{figs/circle.png}}}$. The same applies to the equations discussed in the following section.

\subsection{Hierarchical generalization of dual unitarity}

Dual-unitary gates~\cite{Akila2016,Bertini2018,Bertini2019,Gopalakrishnan2019} satisfy an additional set of algebraic conditions that can be graphically represented as
\begin{align}
    \vcenter{\hbox{\includegraphics[width=0.12\columnwidth]{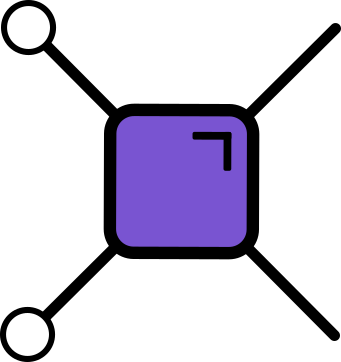}}}\, = \begin{array}{c}
\vcenter{\hbox{\includegraphics[width=0.008\textheight,angle=-45]{figs/circle.png}}}\\
\vcenter{\hbox{\includegraphics[width=0.008\textheight,angle=-135]{figs/circle.png}}}
\end{array}, \quad
\vcenter{\hbox{\includegraphics[width=0.12\columnwidth]{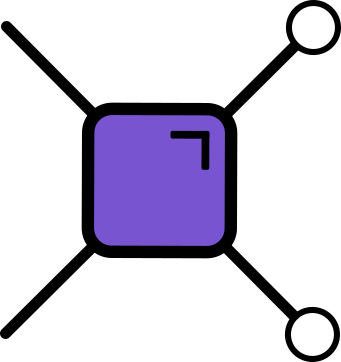}}}\, = \begin{array}{c}
\vcenter{\hbox{\includegraphics[width=0.008\textheight,angle=45]{figs/circle.png}}}\\
\vcenter{\hbox{\includegraphics[width=0.008\textheight,angle=135]{figs/circle.png}}}
\end{array}. \label{eq:dual_unitarity}
\end{align}
These identities result in a dynamics that is unitary not just along the discrete time direction but also along the spatial direction. Remarkably, dual-unitarity allow for the exact calculation of a wide range of dynamical properties including correlation functions, operator dynamics, and entanglement growth, even though the resulting circuits are generically chaotic~\cite{Bertini2019,Piroli2020,Fritzsch2021,Aravinda2021,Suzuki2022}. While most features of dual-unitary dynamics are in a sense ``generic'', the operator dynamics is known to be pathological~\cite{Bertini2020,Claeys2020}.

The hierarchical generalization of dual unitarity is a recent attempt to find models which retain some of the solvability of dual-unitary circuits while avoiding the pathological features thereof~\cite{Yu2024}. It is also an attempt to unify models which are solvable by different means. The first level of the hierarchy, $\mathcal{L}_1$, is the set of dual-unitary gates. The second level of the hierarchy, $\mathcal{L}_2$, is defined by the algebraic conditions
\begin{align}
    \vcenter{\hbox{\includegraphics[height = .16\columnwidth]{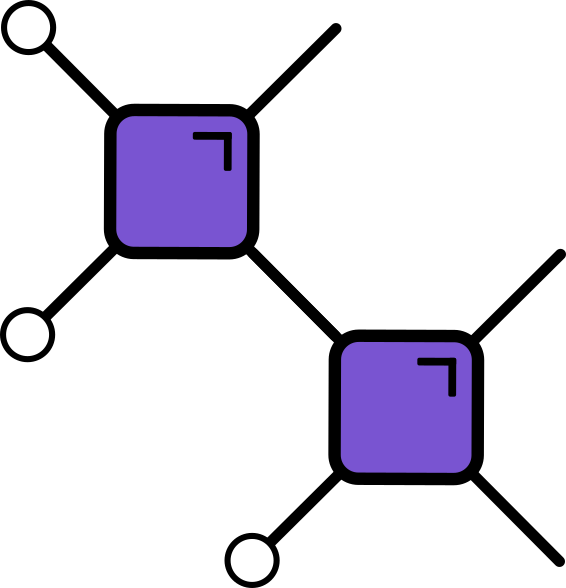}}} = \vcenter{\hbox{\includegraphics[height = .16\columnwidth]{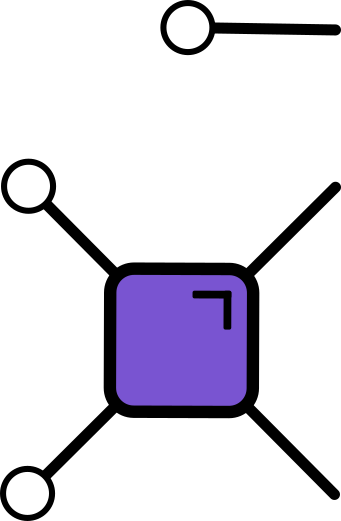}}},\,\quad \vcenter{\hbox{\includegraphics[height = .16\columnwidth]{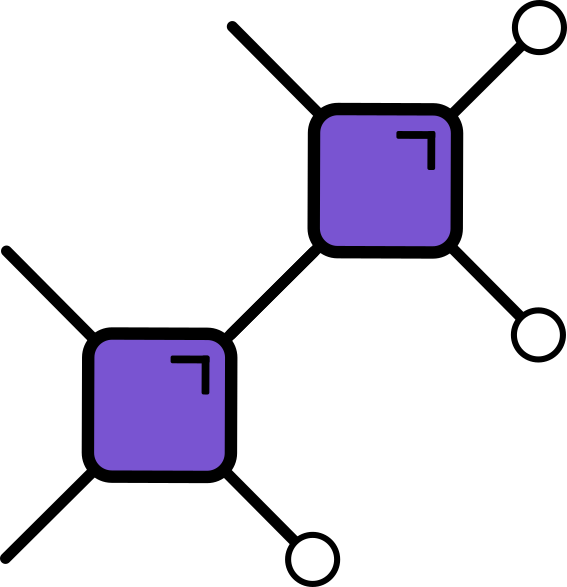}}} = \vcenter{\hbox{\includegraphics[height = .16\columnwidth]{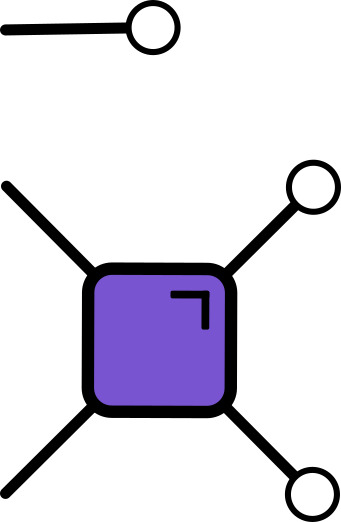}}}\,.
    \label{eq:L2DIag}
\end{align}
It encompasses the set $\mathcal{L}_1$, the set of product gates, and a class of gates locally equivalent to the CNOT gate (in particular also the CNOT gate itself). 
For qubits the above list is exhaustive, while there are also other examples for larger local Hilbert space dimension. Circuits composed out of CNOT gates have also been studied in the context of the Floquet quantum East model~\cite{Bertini2024Exact}, where the above properties were used to obtain exact predictions for the thermalization and entanglement dynamics~\cite{Bertini2024Exact}, and take a special place within this hierarchy since these gates are not just unitary but also T-dual: they satisfy
\begin{align}\label{eq:Tdual}
    \vcenter{\hbox{\includegraphics[width=0.12\columnwidth]{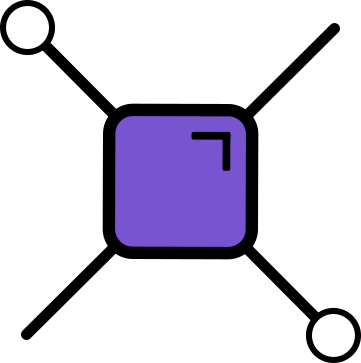}}}\, = \begin{array}{c}
\vcenter{\hbox{\includegraphics[width=0.008\textheight,angle=135]{figs/circle.png}}}\\
\vcenter{\hbox{\includegraphics[width=0.008\textheight,angle=-45]{figs/circle.png}}}
\end{array}\,, \quad
\vcenter{\hbox{\includegraphics[width=0.12\columnwidth]{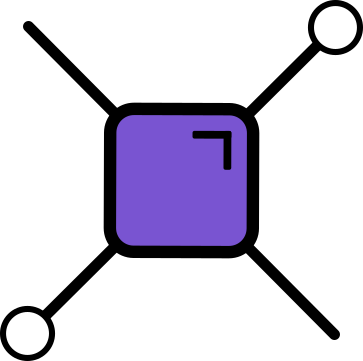}}}\, = \begin{array}{c}
\vcenter{\hbox{\includegraphics[width=0.008\textheight,angle=-135]{figs/circle.png}}}\\
\vcenter{\hbox{\includegraphics[width=0.008\textheight,angle=45]{figs/circle.png}}}
\end{array}\,.
\end{align}
This property is also satisfied by the set of product gates, i.e. two-site gates that are the direct product of two one-site unitary gates.
It is an open question -- which we answer in the positive -- if gates that are neither dual-unitary nor T-dual exist in $\mathcal{L}_2$. 
We denote by \LL{2} the set of second level gates excluding dual-unitary gates. In contrast to dual-unitary circuits, whose correlations are supported exclusively on the light-cone edge, i.e. $x=vt$ with $v=1$, the correlations in \LL{2} circuits are also supported along the $v=0$ ray, i.e. at $x=0$. Everywhere else they vanish identically. However, for gates that are also T-dual all correlations of one-site operators along the $v=1$ ray aditionally vanish~\cite{Aravinda2021}.

The $k$-th level of the hierarchy is defined by including $k$ gates in the algebraic condition. E.g., for $k=3$ the condition reads
\begin{align}
    \vcenter{\hbox{\includegraphics[height = .2\columnwidth]{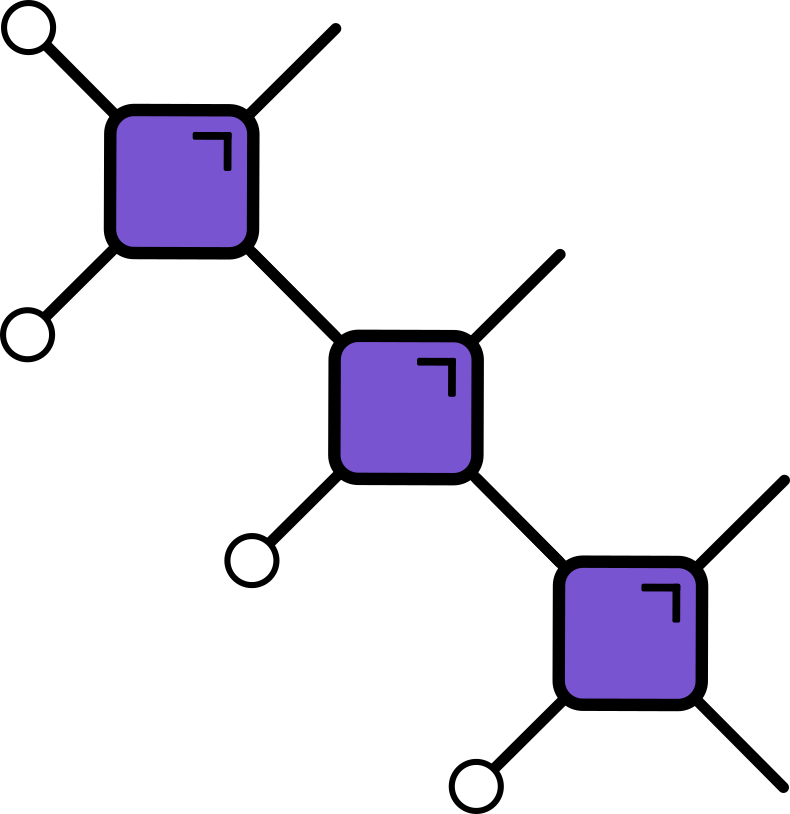}}} = \vcenter{\hbox{\includegraphics[height = .2\columnwidth]{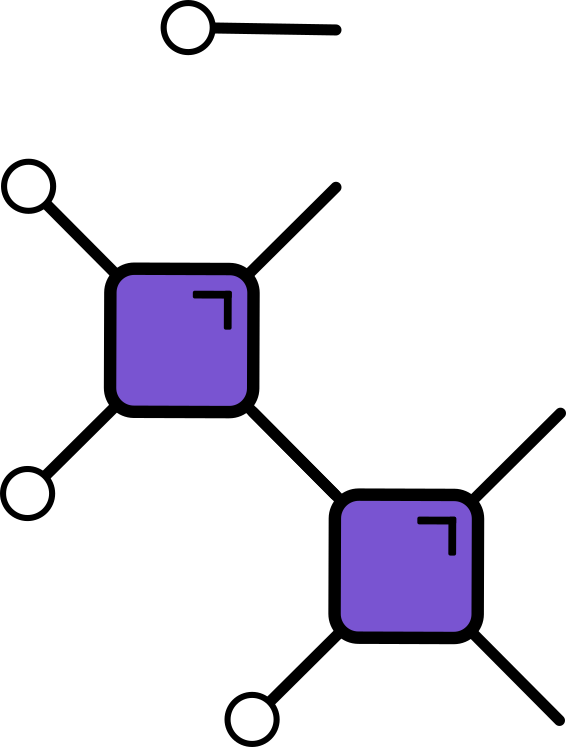}}}\,,\,\,\, \vcenter{\hbox{\includegraphics[height = .2\columnwidth]{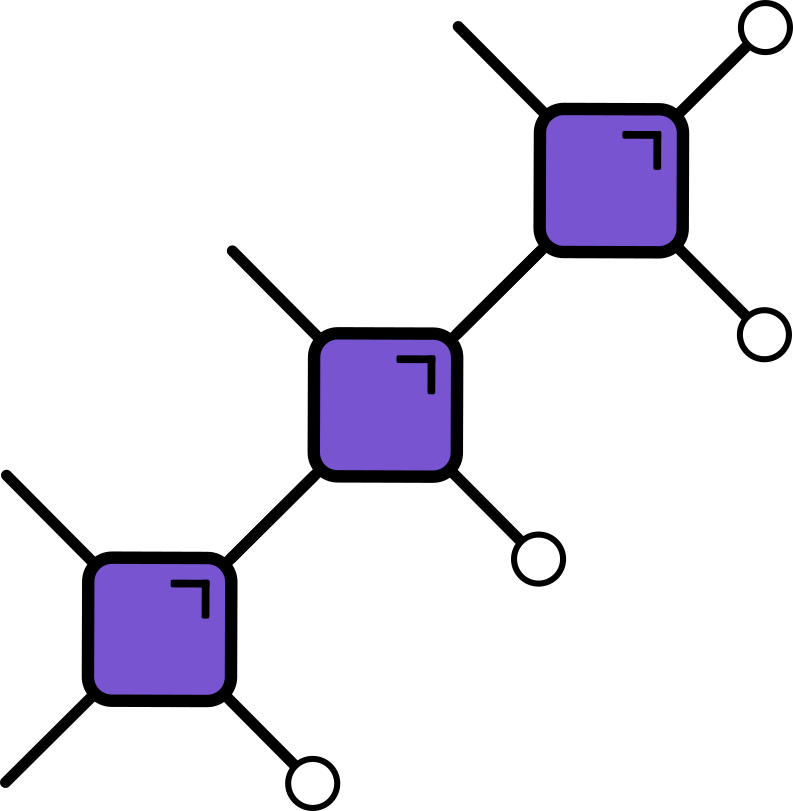}}} = \vcenter{\hbox{\includegraphics[height = .2\columnwidth]{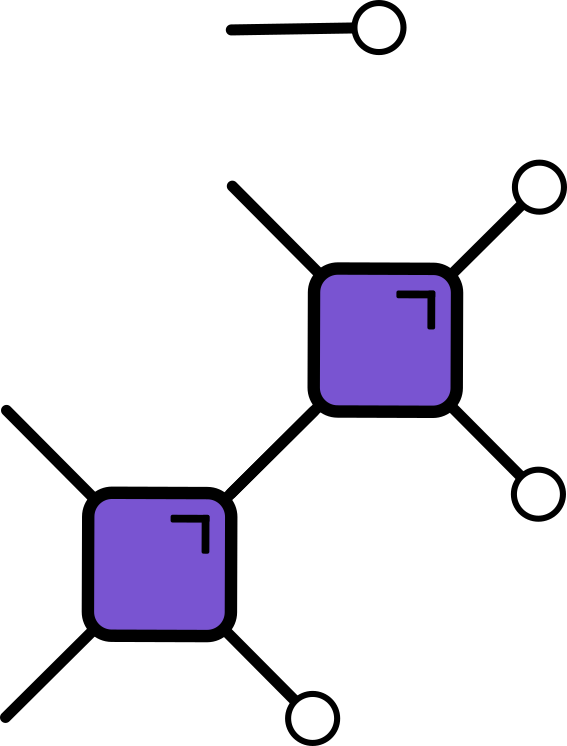}}}.
\end{align}
So far, due to the analytical and numerical challenges this equation poses, only few examples of \LL{3} gates exist, while nothing is known about $k\geq4$. Little is known about the dynamics for $k\geq3$.

\subsection{Entanglement membrane theory}

Entanglement membrane theory (EMT) is an effective coarse-grained description of entanglement growth in chaotic many-body systems. Following a quench from an initial product state, the entanglement of a region $A$ is expressed as the free energy of a minimal membrane pinned to the edges of $A$. The free energy of this membrane can be obtained by integrating a local line tension $\mathcal{E}(v)$ that depends on the local slope $v$ of the membrane as
\begin{equation}
    S_{A}(t) = \min_{\{v(t')\}} \left( \int_0^t \mathrm{d}t' \mathcal{E}(v(t'))\right).
\end{equation}
The function $\mathcal{E}(v)$ is called the entanglement line tension (ELT). This approach can be generalized to general kinds of (operator) entanglement entropies and entangled initial states.

Internal consistency of the membrane description requires that the ELT is a convex function~\cite{Jonay2018}. 
Convexity of the ELT has been proven in holographic field theories~\cite{Mezei2018}, but currently no derivation of the convexity from a physical principle exists.
EMT can be proven to apply in random circuits in the limit of a large local Hilbert space dimension ~\cite{Nahum2017,Zhou2019} and in holographic field theories~\cite{Mezei2018}. It is however not obvious that such an approach should apply to general chaotic dynamics. A physical argument for the validity of EMT can be made following Ref.~\cite{Jonay2018}. This argument rests on the assumption that chaotic systems produce entanglement locally when they are not locally equilibrated to the maximum entropy state, and that this production depends only on the gradient of the entanglement entropy. Introducing a local entanglement production rate $\Gamma(s)\geq0$ depending on the local entanglement gradient $s$, the leading behavior of the local entanglement can be expressed as
\begin{equation}
    \frac{\partial S}{\partial t} = s_{\mathrm{eq}} \Gamma\left( \frac{\partial S}{\partial x} \right),
\end{equation}
where $s_{\mathrm{eq}}$ is the equilibrium entropy density. This is a differential equation determining the leading order behavior of the entanglement profile in the scaling limit of long times and large system sizes. Such a picture is dual, by a Legendre transformation, to the picture of a minimal curve. The ELT is then given by the Legendre transformation of $\Gamma(s)$:
\begin{equation}
    \mathcal{E}(v) = \max_s \left( \Gamma(s) + \frac{vs}{s_{\mathrm{eq}}}\right).
\end{equation}

In the EMT picture, all information about the underlying microscopic physics that is relevant to the macroscopic entanglement dynamics is contained in the ELT. There are multiple ways of defining the ELT that are all expected to give the same result. For the purposes of this paper it is most convenient to define the ELT \emph{via} the operator entanglement of the time evolution operator~\cite{Jonay2018}. The intuition behind this definition is that, for a quench from a generic product state, the entanglement should be generated exclusively by the time-evolution operator. For concreteness, we consider a partition of the input and output degrees of freedom such that a membrane of slope $v$ is forced to run across the system:
\begin{align}
    U(t) = \,\vcenter{\hbox{\includegraphics[height = .48\columnwidth]{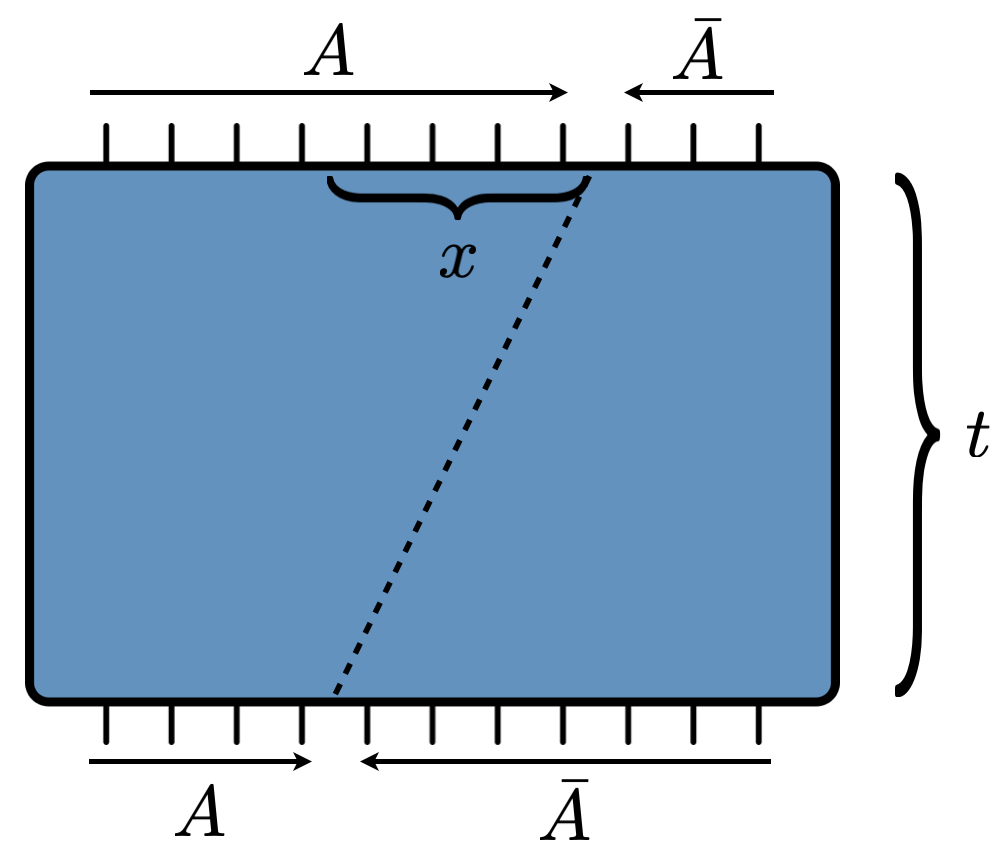}}}.
\end{align}
In order to extract the entanglement, we formally perform an operator-to-state mapping
\begin{widetext}
\begin{equation}
    U(t) = \sum_{\substack{i_1,\dots,i_N\\ j_1,\dots,j_N}} U(t)^{i_1,\dots,i_N}_{j_1,\dots,j_N} \ket{i_1,\dots,i_N}\bra{j_1,\dots,j_N}  \mapsto \ket{U(t)} \equiv \frac{1}{q^{\frac{N}{2}}}\sum_{\substack{i_1,\dots,i_N\\ j_1,\dots,j_N}} U(t)^{i_1,\dots,i_N}_{j_1,\dots,j_N} \ket{i_1,\dots,i_N}\otimes \ket{j_1,\dots,j_N},
\end{equation}
\end{widetext}
where we have fixed the computational basis and consider a chain of fixed length $N$. We take the subsystem $A$ to consist of qudits $\{i_1,\dots,i_{\lfloor(N+x)/2\rfloor}\}$ and $\{j_1,\dots,j_{\lfloor(N-x)/2\rfloor}\}$. The R\'{e}nyi-$\alpha$ operator entanglement is then defined as the R\'{e}nyi-$\alpha$ entanglement entropy of the state $\ket{U(t)}$
\begin{equation}\label{eq:Renyi_a}
    S_{\alpha} (x,t) \equiv \frac{1}{1-\alpha} \log \tr\left[ \left(\operatorname{tr}_A \ket{U(t)}\bra{U(t)}\right)^\alpha \right].
\end{equation}
Here the thermodynamic limit $N\rightarrow\infty$ should always be taken before the scaling limit $x,t\rightarrow\infty$.
We then expect that asymptotically for $x,t\rightarrow\infty$ and $x/t=v=\mathrm{constant}$ the R\'{e}nyi-$\alpha$ operator entanglement is given by
\begin{equation}
    S_{\alpha} (x,t) \approx s_{\mathrm{eq}}\, \mathcal{E}_{\alpha}(v)t. \label{eq:ELT_def}
\end{equation}
We thus define the R\'{e}nyi-$\alpha$ ELT to be the leading contribution in $t$ of the R\'{e}nyi-$\alpha$ operator entanglement. For $\alpha=1$ the ELT introduced above reduces to the ELT for the von Neumann entanglement entropy, $\mathcal{E}(v)=\mathcal{E}_{1}(v)$. This definition is typically not the most convenient one for numerical purposes, as the boundaries of the operator entanglement are such that the membrane is not exactly pinned to the edge of $A$ on the lower boundary, but only suffers a free energy penalty when deviating from it, such that there may be significant finite size corrections~\cite{Zhou2020}. Still, the result in the scaling limit remains unchanged. We will show that Eq.~\eqref{eq:ELT_def} can be exactly evaluated for any $x,t$ in the models we are considering in this work, thus enabling the exact extraction of the ELT.

\begin{figure}[t]
    \centering
    \includegraphics[width = 0.3\textwidth]{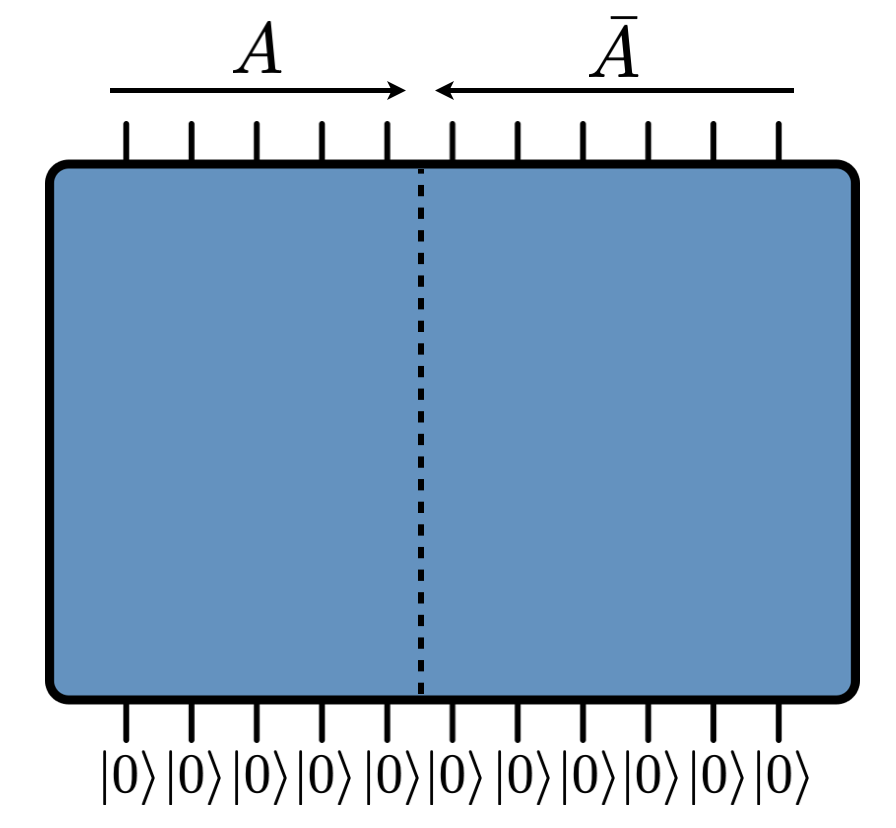}
    \caption{Illustration of the relation $v_E=\mathcal{E}(0)$. For a quench from an initial unentangled state, the membrane is only pinned on the top boundary and can end at any point on the lower boundary. For translationally invariant unitary dynamics, the minimal configuration corresponds to a vertical membrane.}
    \label{fig:vertical}
    %\vspace{-\baselineskip}
\end{figure}

For our purposes, we will use that in unitary circuits the R\'{e}nyi operator entanglement~\eqref{eq:Renyi_a} reduces to a tensor network diagram
\begin{equation}
    S_{\alpha} (x,t) = -\frac{1}{\alpha-1}\log Z_\alpha(m,n)\,,
\end{equation}
where $Z_\alpha$ is given by
\begin{align}
    Z_\alpha(m,n) = \,\vcenter{\hbox{\includegraphics[height = .48\columnwidth]{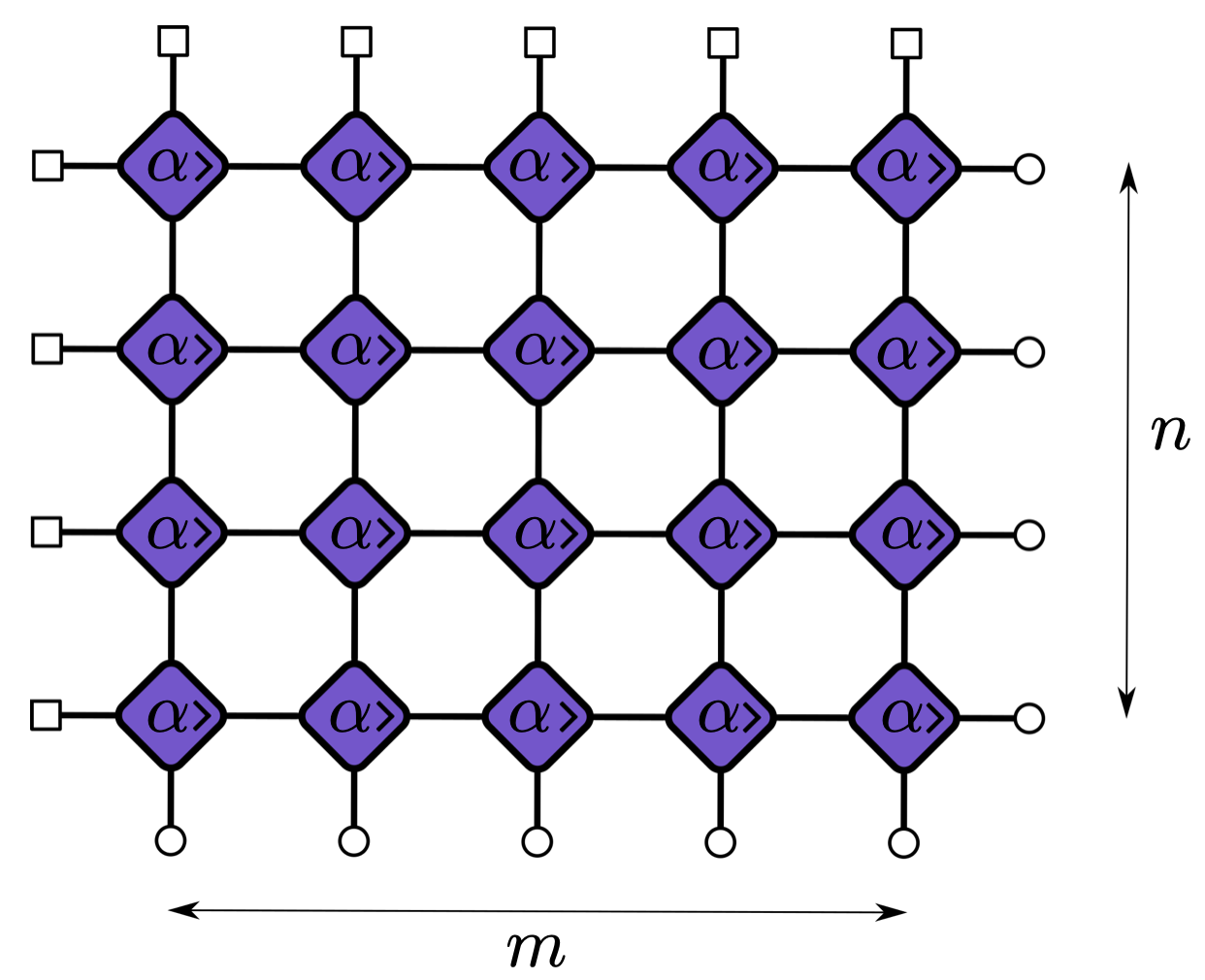}}}. \label{eq:z_alpha}
\end{align}
The size of this tensor network is set by the coordinates of the entanglement cut as
\begin{equation}
    n = \frac{t-x -(x \,\mathrm{mod}\, 2)}{2}, \quad m = \frac{t+x -(x\, \mathrm{mod}\, 2)}{2}.
\end{equation}
In the scaling limit $x,t \to \infty$ with constant velocity $v=x/t$, acting as the slope of the cut, the ELT follows from the leading-order behavior of  $\log Z_\alpha$ as
\begin{align}
    -\frac{1}{\alpha-1}\log Z_\alpha(m,n) \approx \, s_{\mathrm{eq}}\, \mathcal{E}_\alpha(v) t\,,
\end{align}
where $s_{\mathrm{eq}}=\log(q)$ is the equilibrium entanglement entropy density. We note that while the ELT is strictly speaking only meaningful in ergodic models, the operator entanglement of the time-evolution operator is a physical quantity for any model. Generally, it provides state-independent information on entanglement growth. The calculations presented in Sec.~\ref{sec:ELT} are hence valid independent of the degree of ergodicity of the underlying circuit.

Several important physical quantities can be directly extracted from the ELT, in particular the entanglement velocity $v_E$ and the butterfly velocity $v_B$. The entanglement velocity is defined as the asymptotic growth rate of the half-chain entanglement entropy from a generic initial state
\begin{equation}
    S(t) = s_{\mathrm{eq}} v_E t.
\end{equation}
For a quench from an infinite translationally invariant state with low entanglement, the membrane configuration minimizing the cost function in the computation of the half-chain entropy is that of a vertical membrane, as illustrated in Fig.~\ref{fig:vertical}. The entanglement velocity is hence identified with $v_E=\mathcal{E}(0)$. Note that the entanglement velocity generally depends on the R\'{e}nyi index $\alpha$.

The butterfly velocity $v_B$ defines an effective causal light cone in a many-body system. Outside of this causal light cone the action of the time-evolution operator does not generate any (operator) entanglement. Thus, $v_B$ is determined by the equation $v_B=\mathcal{E}(v_B)$, corresponding to the velocity above which the operator entanglement coincides with that of the identity. Consequently the ELT satisfies $\mathcal{E}(v)=v$ for $v\geq v_B$. It has been argued that this point coincides for all values of the R\'{e}nyi index $\alpha$, leading to the identification of a single causal speed for all R\'{e}nyi entropies~\cite{Zhou2019}. 
An alternative point of view on the causal light cone is given by the Heisenberg evolution of local operators. Beyond the operator front $\abs{x}=v_B t$, the Heisenberg evolved operator acts trivially. This is typically diagnosed using the out-of-time-ordered correlator (OTOC)~\cite{Larkin1969,Shenker2014}, which is accessible experimentally in current quantum simulation setups~\cite{Garttner2017,Vermersch2019,Mi2021}, and where the late-time decay of the OTOC can also be computed from EMT~\cite{Zhou2020}. The curvature at $v=v_B$, $\mathcal{E}''(v_B)$, determines the broadening of the operator front, as it controls the contribution of subleading fluctuation configurations of the membrane.

\subsection{Outline}

This article is organized as follows. In Sec.~\ref{sec:ELT} we compute the ELT in generalized dual-unitary circuits and discuss the physical implications of the result. Focusing first on \LL{2} circuits we extract the entanglement and butterfly velocity and discuss the relation of the particular simple form of the ELT to general bounds on entanglement growth. We show that the entanglement velocity is related to information spreading. Moreover, we discuss the relation of the ELT to influence matrices, thus providing a link to correlation functions. In this way the solvability of the \LL{2} circuits is related to the area-law (temporal) entanglement of the influence matrix. Finally, we generalize our results to \LL{k} circuits with $k\geq3$ and show that solvability of both correlation functions and ELT is restricted to rays in spacetime with velocities $v_*\leq \abs{v}\leq 1$. In Sec.~\ref{sec:op_growth} we verify the predictions of EMT with results extracted from probes of operator dynamics, focusing on the out-of-time-order correlator and the tripartite information. Furthermore, in Sec.~\ref{sec:ent_growth} we present numerical results on entanglement growth consistent with the predictions of EMT. In Sec.~\ref{sec:Hadamard} we discuss constructions of hierarchical unitary gates based on complex Hadamard matrices, motivated by similar constructions for dual-unitary gates, and show how these constructions can allow for further simplifications in the calculation of the ELT. Finally, in Sec.~\ref{sec:construction} we develop alternative methods to construct new \LL{2} gates based on constructions of gates with flat Schmidt spectrum. 
We conclude in Sec.~\ref{sec:conc}.

\section{Entanglement line tension}
\label{sec:ELT}

\begin{figure}[t]
    \centering
    \includegraphics[width = 0.45\textwidth]{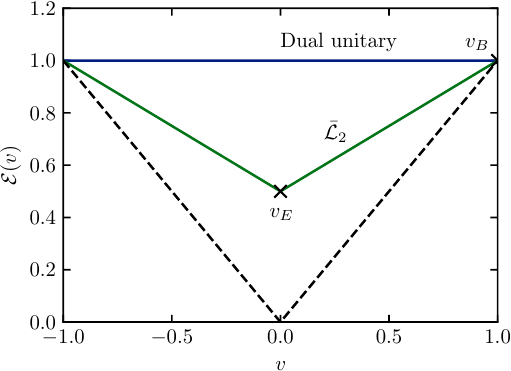}
    \caption{In \LL{2} circuits the ELT assumes a linear form. In constrast to dual-unitary circuits the slope is nonzero and \LL{2} circuits are thus characterized by an entanglement velocity $v_E<1$. Both \LL{2} and dual-unitary circuits have a maximal butterfly velocity $v_B=1$.}
    \label{fig:elt_sketch}
    %\vspace{-\baselineskip}
\end{figure}

In this section we show that the ELT can be computed exactly for all velocities in \LL{2} circuits and for a range of velocities in higher levels of the hierarchy. Two velocities that can be directly extracted from the ELT are the entanglement velocity $v_E$, setting the entanglement growth, and the butterfly velocity $v_B$, setting the operator growth. The existence of a non-flat ELT implies a hierarchy $v_E<v_B$ between these two central velocities characterizing quantum many-body dynamics. Dual-unitary circuits exhibit a flat ELT and have both maximal entanglement velocity~\cite{Piroli2020,Zhou2022} and a maximal butterfly velocity~\cite{Claeys2020}, $v_E = v_B = 1$. Conversely, in \LL{k} circuits the butterfly velocity is generically maximal, $v_B=1$, while the entanglement velocity is smaller than one, $v_E < 1$. We show that in \LL{2} circuits the ELT takes a particularly simple linear form as illustrated in Fig.~\ref{fig:elt_sketch}. Due to this property \LL{2} circuits saturate certain general bounds on entanglement growth that are also known to be saturated in some holographic theories.

\subsection{Entanglement line tension in \LL{2} circuits}

\subsubsection{Determination of the line tension}

In the following we focus on the R\'{e}nyi-2 ELT ($\alpha=2)$. 
The extension to integer R\'{e}nyi index $\alpha>2$ is straightforward in such a way that all presented results are independent of the R\'{e}nyi (replica) index $\alpha$.

To reduce Eq.~\eqref{eq:z_alpha} graphically we execute the following steps.
Without loss of generality we first take $m\geq n$ corresponding to $v\geq0$. The diagram~\eqref{eq:z_alpha} can be reduced as
\begin{align}
 Z_2(m,n)
 =& \,\vcenter{\hbox{\includegraphics[height = .4\columnwidth]{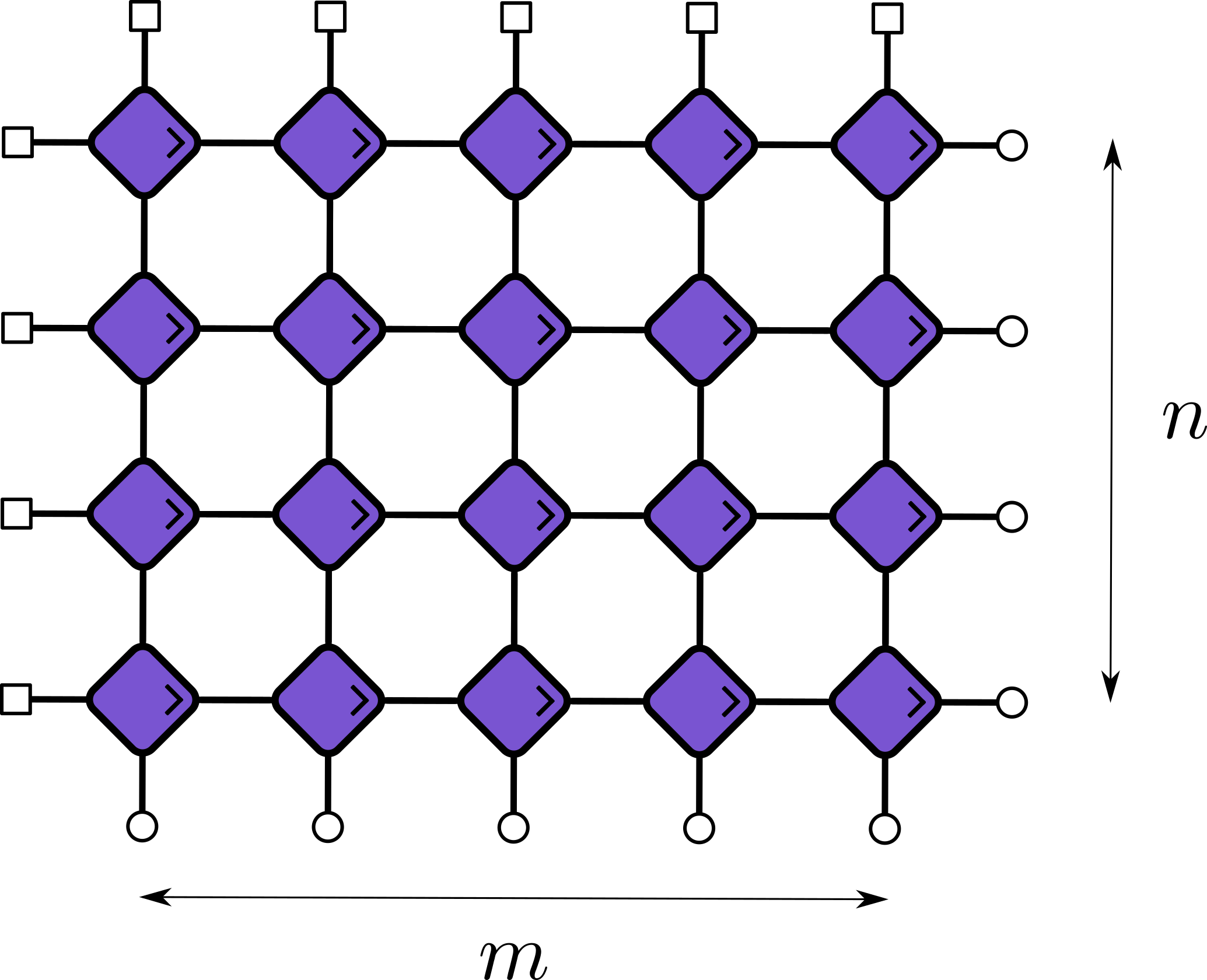}}} \nonumber\\
=& \left(\frac{1}{q}\right)^{m-n}\,\vcenter{\hbox{\includegraphics[height = .42\columnwidth]{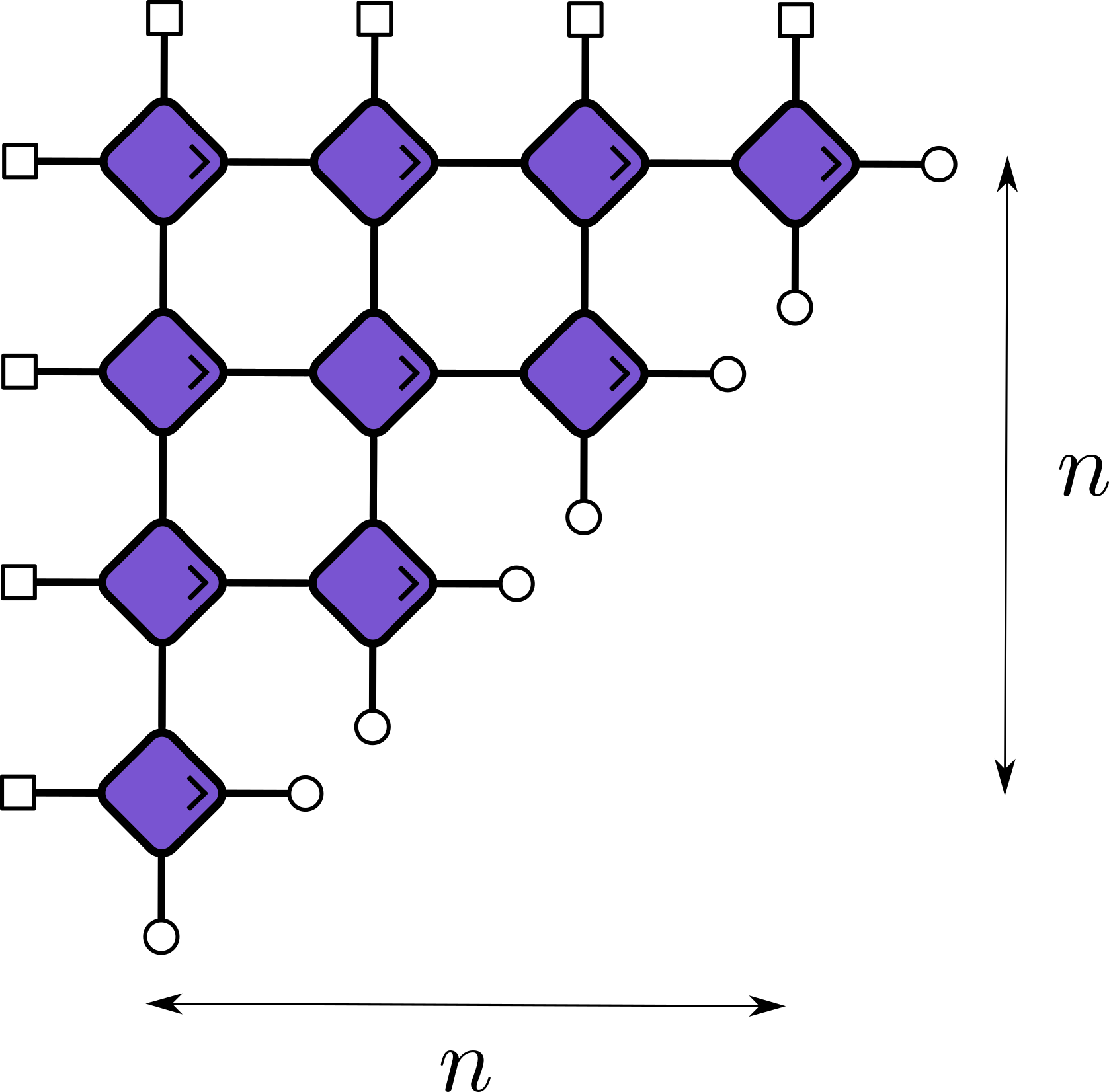}}} 
\end{align}
using the hierarchical condition~\eqref{eq:L2DIag} for the identity permutation starting from the bottom right, leading to a ``staircase'' structure. The prefactor appears because of the overlap $\vcenter{\hbox{\includegraphics[height=0.01\textheight,origin=c]{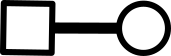}}} = q^{1-\alpha} = q^{-1}$ for $\alpha=2$. This diagram can no longer be simplified using the algebraic conditions for the identity permutation, but the hierarchical condition~\eqref{eq:L2DIag} for the cyclic permutation can now be applied to simplify the diagram starting from the top left. The final diagram factorizes and evaluates to
\begin{align}
Z_2(m,n)=& \left(\frac{1}{q}\right)^{m-n} \left( \,\vcenter{\hbox{\includegraphics[width = .1\columnwidth]{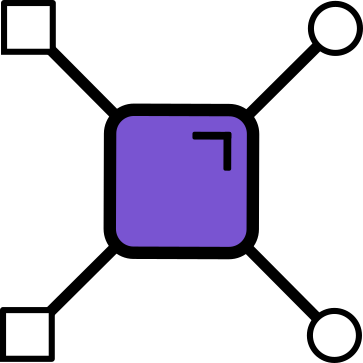}}}\,\right)^n\,.
\end{align}
For each row we get a factor of 
\begin{align}
    B_1/q^2\equiv\vcenter{\hbox{\includegraphics[width = .1\columnwidth]{figs/bubble.png}}}\,,
\end{align}
and for each column that exceeds the number of rows a factor of $1/q$, resulting in
\begin{equation}
    Z_2(m,n) = \frac{B_1^{\min(m,n)}}{q^{m+n}} \label{eq:l2_diagram_te_opent}\,.
\end{equation}
The quantity $B_1$ is directly related to the operator entanglement of the two-site gate. It can be expressed through the Schmidt values $\lambda_i$ of $U$, where
\begin{align}\label{eq:Schmidt}
U=\sum_{i=1}^{q^2}\lambda_i X_i\otimes Y_i,
\end{align}
with $\tr(X_i^{\dagger}X_j)=\tr(Y_i^{\dagger}Y_j) = \delta_{ij}$, as 
\begin{equation}\label{eq:B1_Schmidt}
    B_1 = \frac{1}{q^2}\sum_{i=1}^{q^2}\lambda_i^4.
\end{equation}
It has been shown in Ref.~\cite{Rather2020} that a gate is dual unitary if and only if it has maximal operator entanglement $E(U) = 1-B_1$, and this operator entanglement was also shown to be the relevant quantity when characterizing the operator growth of perturbed dual-unitary circuits through calculations of out-of-time-order correlation functions~\cite{Rampp2023a}.

Extraction of the leading behavior of Eq.~\eqref{eq:l2_diagram_te_opent} in $t$ yields the entanglement line tension as
\begin{equation}
    \mathcal{E}_{2}(v) = 1- \left(1-\abs{v}\right)\frac{\log B_1}{\log q^2}.
\end{equation}
We find that the ELT has a nonzero slope [Fig.~\ref{fig:elt_sketch}] and is determined by a single microscopic parameter, $B_1$, quantifying the operator entanglement. The entanglement velocity $v_E$ directly follows as
\begin{equation}\label{eq:vE_fromB1}
    v_E^{(2)} = \mathcal{E}_{2}(0) = 1 - \frac{\log B_1}{\log q^2}. 
\end{equation}
The line tension can be expressed in terms of the entanglement velocity as
\begin{equation}
    \mathcal{E}_{2}(v) = v_E^{(2)} + (1- v_E^{(2)})\abs{v}.
\end{equation}
The same functional form holds for higher replica index $\alpha>2$ upon replacing $v_E^{(2)}\rightarrow v_E^{(\alpha)}$, where 
\begin{equation}
    v_E^{(\alpha)} = 1 -  \frac{1}{\alpha-1}\frac{ \log \left(\sum_j \lambda_j^{2\alpha}\right)}{\log(q^2)},
\end{equation}
in terms of the Schmidt values of the gate. As we will discuss later, all nonzero Schmidt values are identical, i.e. the Schmidt spectrum is flat. As such we have that $v_E^{(\alpha)}=v_E^{(2)}\equiv v_E$ and consequently $\mathcal{E}_\alpha(v)=\mathcal{E}_2(v)\equiv\mathcal{E}(v)$, enabling us to write
\begin{equation}
    \mathcal{E}(v) = v_E + (1- v_E)\abs{v}.
\end{equation}

The entanglement velocity generally satisfies $v_E < 1$ and for the known family of two-qubit gates we find $v_E=1/2$. For the CNOT gate we can read off $B_1$ since
\begin{align}
    \textrm{CNOT} = \sqrt{2}\left(\frac{\ket{0}\bra{0} \otimes \mathbbm{1}}{\sqrt{2}}+ \frac{\ket{1}\bra{1} \otimes \sigma^x}{\sqrt{2}}\right),
\end{align}
and $q=2$, such that $B_1 = 2$ and hence $v_E = 1/2$. Conversely, product gates have no operator entanglement, $B_1=q^2$, and the above arguments predicts $v_E=0$ consistent with the absence of entangling power. 
That $v_E < 1$ away from the dual-unitary point follows from the observation that $B_1=1$ implies a maximal operator entanglement and hence dual-unitarity, for which $v_E=1$ and the ELT is flat. Note that the above derivation also holds for dual-unitary models, since dual-unitarity implies the hierarchical conditions Eq.~\eqref{eq:L2DIag}.

The butterfly velocity follows as a solution to $\mathcal{E}(v_B)=v_B$, which is only satisfied for $v_B=1$, implying the maximal velocity also observed in dual-unitary circuits. Moreover, the absence of curvature at $v=v_B$, $\mathcal{E}''(v_B)=0$, implies that the operator front does not broaden.

\begin{figure}[t]
    \centering
    \includegraphics[width = 0.45\textwidth]{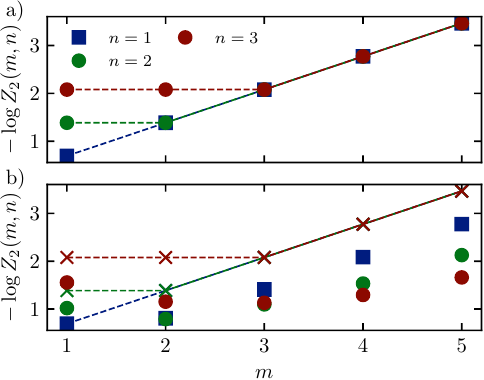}
    \caption{Comparison of the R\'{e}nyi-2 operator entanglement of the time-evolution operator (solid markers) with theoretical predictions (dashed lines and crosses). (a) \LL{2} circuits show excellent agreement with Eq.~\eqref{eq:l2_diagram_te_opent}. The gate is chosen randomly from the known set of qubit \LL{2} gates (see App.~\ref{app:parametrization}). (b) A circuit made of gates locally equivalent to those of (a) but without the \LL{2} property. Despite the gate having the same operator entanglement, the properties of the global time evolution operator differ drastically. This distinction highlights the importance of the local structure of \LL{k} gates.
    }
    \label{fig:te_opent}
    %\vspace{-\baselineskip}
\end{figure}

Comparison of Eq.~\eqref{eq:l2_diagram_te_opent} with numerical data shows excellent agreement [Fig.~\ref{fig:te_opent}(a)]. While the final expressions only depend on the operator entanglement, we emphasize that the presented results are not expected to hold for generic gates. The importance of the \LL{2} property is highlighted by considering gates that differ from \LL{2} gates by one-site unitary transformations
\begin{align}
    V = \,\vcenter{\hbox{\includegraphics[height = 0.15\columnwidth]{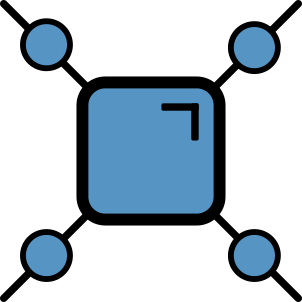}}}, \qquad \vcenter{\hbox{\includegraphics[height = 0.07\columnwidth]{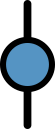}}}\in \mathbb{U}(q).
\end{align}
(The local unitary transformations can all be distinct.) Two gates differing by such transformations are called locally equivalent. Even though locally equivalent gates have identical operator entanglement they may display strongly different behavior [Fig.~\ref{fig:te_opent}(b)]. This is a reflection of the \LL{2} condition's [Eq.~\eqref{eq:L2DIag}] lack of invariance under local transformations. Contrary to this the dual-unitarity condition Eq.~\eqref{eq:dual_unitarity} is a condition on the entanglement properties of the gate alone~\cite{Rather2020}, and hence it is locally invariant. Nevertheless, local transformations may change the ergodicity properties of the resulting circuit drastically~\cite{Claeys2021}.

\begin{figure}[t]
\centering
\includegraphics[width=0.45\textwidth]{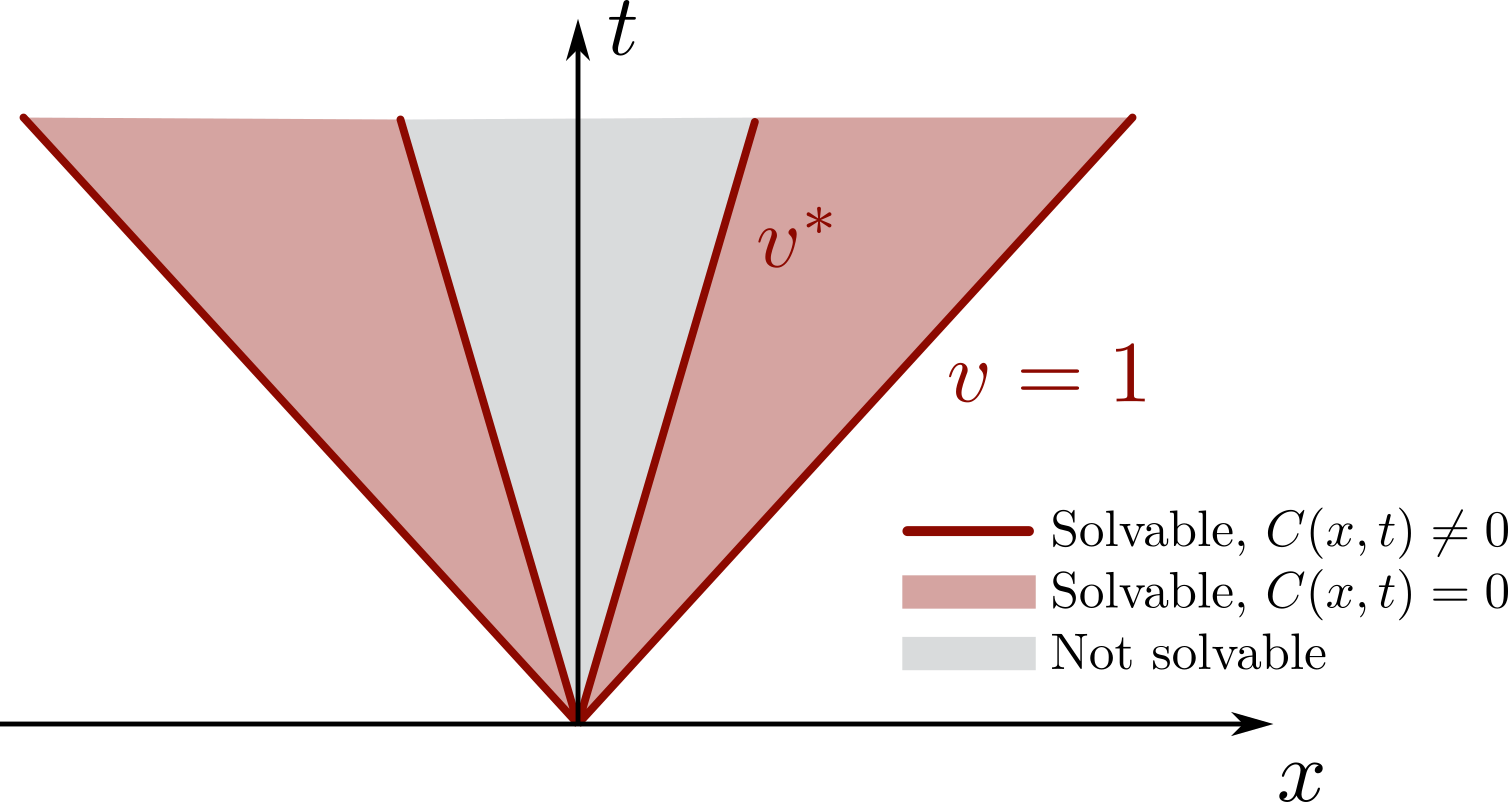}
\caption{In \LL{k\geq3} circuits the correlations are solvable above a threshold velocity $v_*=(k-2)/k$. Inside this range the correlations vanish exactly. Non-vanishing but solvable correlations occur on the boundaries of the solvable region for $\abs{v}=v_*$ and $\abs{v}=1$.}
\label{fig:lk_overview}
\end{figure}

\subsubsection{Maximal butterfly velocity and information recovery}

The ELT contains two crucial pieces of physical information. First, the butterfly velocity $v_B$ is determined by $\mathcal{E}(v_B)=v_B$. For all entangling \LL{2} circuits ($B_1<q^2$) the only solution to this equation is $v_B=1$. Therefore, in all entangling \LL{2} circuits information spreads at the maximal speed allowed by causality. This highlights that dual-unitary circuits are not the only chaotic models with maximal butterfly velocity, as previously discussed in Ref.~\cite{Claeys2020}.

The conclusion that $v_B=1$ can also be reached in a more direct manner by relating the operator entanglement to the Hayden-Preskill protocol. This protocol is concerned with the question of recovering quantum information from a small subsystem of a thermodynamically large quantum system following unitary evolution~\cite{Hayden2007}. The information is injected in subsystem $A$ and is supposed to be recovered from $D$ at a later time~\footnote{The special feature of the Hayden-Preskill protocol is that the recovery is assisted by entanglement with the initial state in the complement of $A$.}. Here we consider the setup where the subsystems $A$ and $D$ are connected by a light ray at distance $\ell$. Additionally, we assume that $A$ and $D$ are both composed of a single qudit, resulting in a setup of the form
\begin{align}
    \vcenter{\hbox{\includegraphics[height = .5\columnwidth]{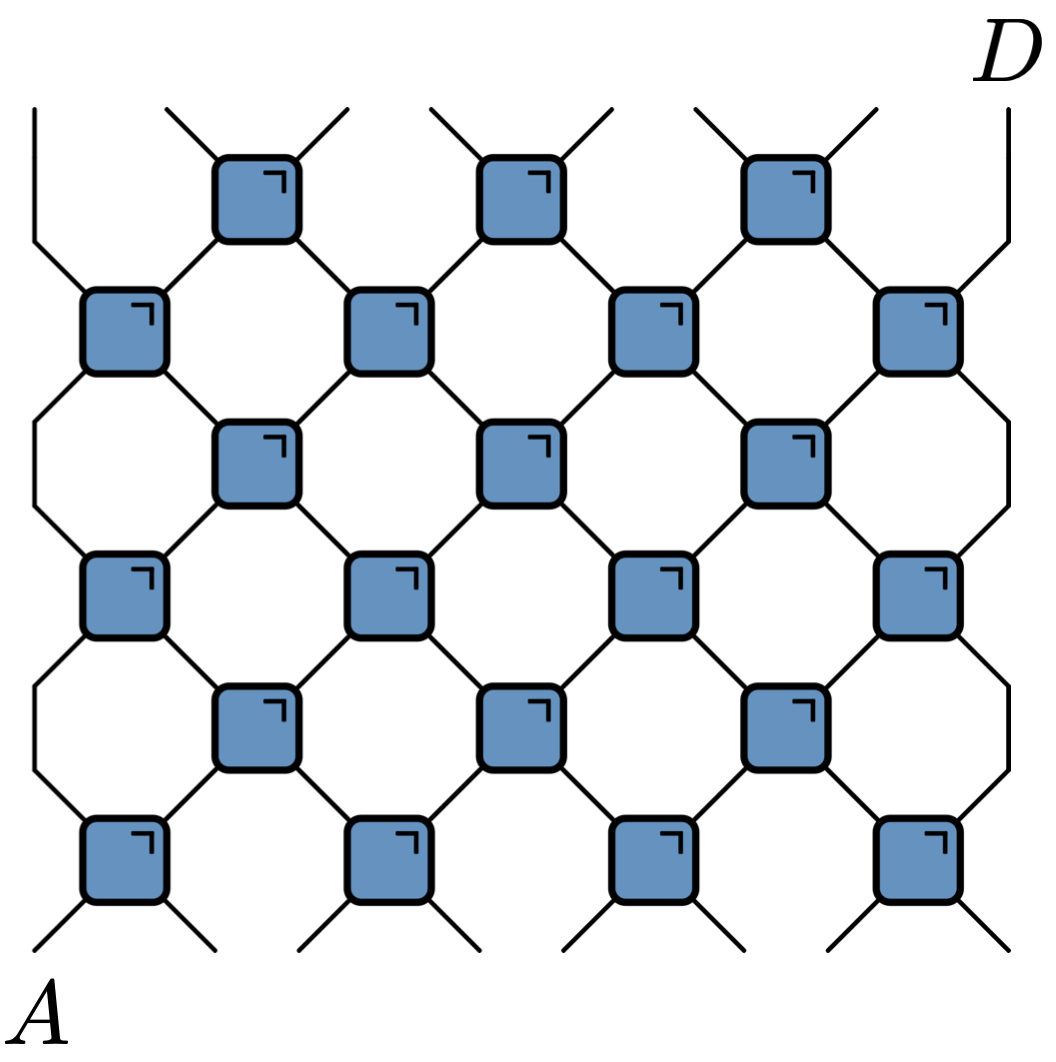}}}.
\end{align}
As recently discussed in Ref.~\cite{Rampp2023}, the fidelity of information recovery is related to the operator purity of the diagonal composition of $\ell-1$ gates as given by
\begin{align}
    B_{\ell-1} \equiv q^\ell\, \underbrace{\vcenter{\hbox{\includegraphics[width = .55\columnwidth]{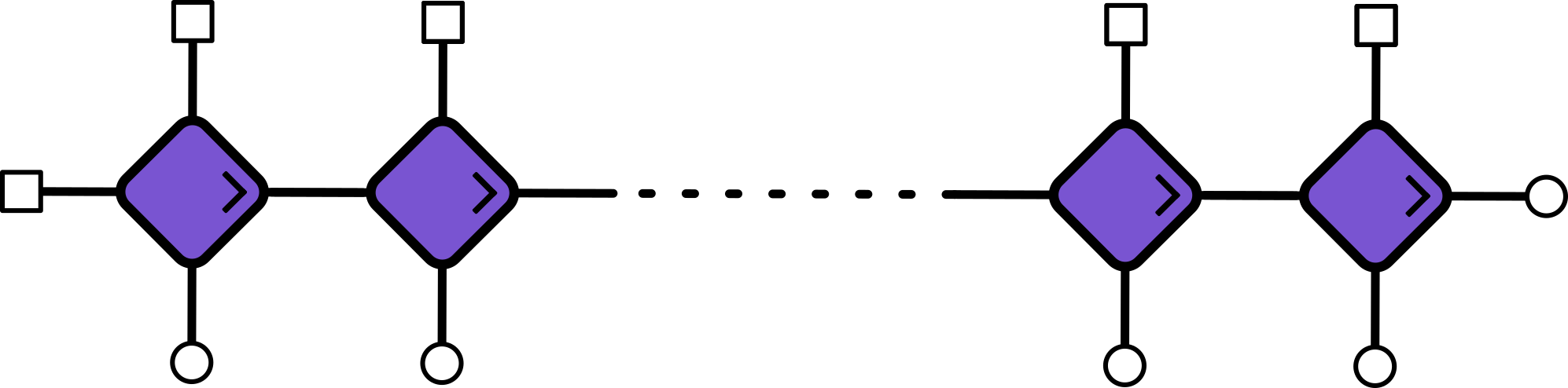}}}}_{\ell-1}. \label{eq:info_transport}
\end{align}
While for generic gates this quantity tends to the same value that non-entangling gates take as $\ell\rightarrow\infty$, indicating generic operator spreading with a non-maximal butterfly velocity, the \LL{2} property implies that the operator purity remains non-trivial for $\ell\rightarrow\infty$. Applying the \LL{2} condition to Eq.~\eqref{eq:info_transport} repeatedly enables to reduce the expression to a single gate yielding $B_{\ell-1}=B_1$. For non-entangling gates, where $B_1<q^2$, this signifies that \LL{2} circuits transport a finite fraction of information along the light-like direction, directly implying a maximal butterfly velocity. As the entanglement velocity is also determined by the operator entanglement of the gate, this point of view provides a link between the dynamics along the light cone and the local entanglement production. This thought experiment also enables a crucial distinction between dual-unitary and \LL{2} circuits: dual-unitary circuits transport the complete amount of information along the light cone, while for \LL{2} circuits only a finite fraction of information is transported along the light cone.

\begin{figure*}[t]
\centering
\includegraphics[width=0.75\textwidth]{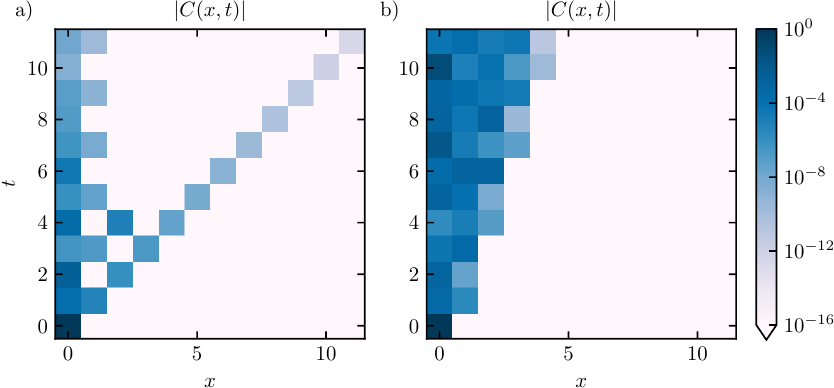}
\caption{Numerically computed correlation functions of three-site operators in generalized dual-unitary circuits. The gates are randomly selected from the given parametrizations. (a) Qubit \LL{2} gate: the correlator is non-vanishing on the $\abs{v}=0$ and $\abs{v}=1$ rays in spacetime. (b) Qubit \LL{3} gate with one-site dressing: the correlator vanishes in the range $1/3<\abs{v}<1$. The vanishing of the $\abs{v}=1$ component is a special feature of the qubit parametrization. The result is consistent with a butterfly velocity $v_B=1/3$ extracted from the OTOC (see Sec.~\ref{sec:op_growth}).}
\label{fig:correlations}
\end{figure*}

\subsubsection{Entanglement growth}

The second important piece of information is the entanglement velocity as given by $v_E=\mathcal{E}(0)$. In \LL{2} circuits the entanglement velocity is smaller than $v_B$ and it is determined only by the operator entanglement of the gate. As we will discuss in more detail in Sec.~\ref{sec:construction}, \LL{2} gates have a flat Schmidt spectrum, i.e. all nonzero Schmidt values $\lambda_i$ from Eq.~\eqref{eq:Schmidt} are equal. Together with the normalization condition $\sum_{i} \lambda_i^2 = q^2$, this result implies that the entanglement velocity can only take a discrete set of possible values for a given local Hilbert space dimension $q$. Fixing the total number of nonzero Schmidt values as the Schmidt rank $\R$, these nonzero Schmidt values necessary equal $\lambda_i = q/\sqrt{\R}, i = 1 \dots \R$, resulting in $B_1 = q^2/\R$. Eq.~\eqref{eq:vE_fromB1} then reduces to
\begin{align}
  v_E = \frac{\log \R}{\log q^2}\,. 
\end{align}
For bipartite unitary gates acting on $\mathbb{C}^q \otimes \mathbb{C}^q$ all Schmidt ranks from $1$ to $q^2$ are possible except for $q=2$ \cite{Dur2002,Hermes2018}. For qubits the only possible values are $\R = 1,2,4$ leading to possible entanglement velocities $v_E=0$ (non-entangling), $v_E=1$ (dual-unitary), and $v_E=1/2$. The latter is realized by CNOT gates, illustrating the general result that bipartite unitaries having Schmidt rank 2 and 3 can be written as controlled unitaries~\cite{Cohen2013,Chen2014}.   

The form of the ELT in \LL{2} circuits is particularly simple as it has no curvature. It is thus extremal in the sense that it takes the maximal values allowed by convexity given $v_E$ and $v_B$. Interestingly, this piecewise linear form saturates two general bounds on entanglement growth that were conjectured by Mezei and Stanford~\cite{Mezei2017a,Mezei2018}. The first bound, originally formulated in Ref.~\cite{Hartman2015}, is 
\begin{equation}
    S_A(t) \leq s_{\mathrm{eq}} \mathrm{vol}(\mathrm{tsunami}(t)),
\end{equation}
where $\mathrm{vol}(\mathrm{tsunami}(t))$ is the volume of the region that is causally connected to $A$, called the ``entanglement tsunami''~\cite{Hartman2013,Liu2014,Liu2014a,Leichenauer2015}. 
While it is \emph{a priori} not obvious if the relevant causal speed here is $v_B$ or the (generally larger) Lieb-Robinson velocity~\cite{Mezei2017a}, if membrane theory applies the correct causal speed is $v_B$, as membranes with $\abs{v}>v_B$ do not contribute to entanglement growth. This bound states that entanglement growth is limited by causality, i.e., entanglement cannot be produced between causally disconnected regions. The second bound concerns the growth rate of the entropy
\begin{equation}
    \frac{\rmd S_A(t)}{\rmd t} \leq v_E \lvert\partial A\rvert s_{\mathrm{eq}} .
\end{equation}
This bound contains the insight that entanglement is only produced at the boundary between subsystems. These bounds were found to be saturated in certain holographic theories~\cite{Mezei2017}. It is somewhat surprising that this is also the case in unrelated lattice models with finite local Hilbert space dimension. However, we note that \LL{2} circuits share a curious feature with the operator spreading picture invoked in Ref.~\cite{Mezei2017a} to explain the saturation of the bounds: as a consequence of their tri-directionality and as shown in the previous section the entanglement velocity is determined by the amount of information passed along the light cone. \LL{2} circuits provide a natural setting in which this physics is realized.

As most interesting physical consequences of the Mezei-Stanford bounds occur in spatial dimensions $d>1$, where entanglement growth depends on the geometric shape of the bipartition~\cite{Mezei2017a}, it would be interesting to find higher-dimensional generalizations of \LL{2} circuits.

\subsection{Connection to temporal entanglement and correlation functions}

What lies behind the solvability of the \LL{2} circuits? It turns out that the \emph{influence matrix} (IM)~\cite{Lerose2021,Sonner2021,Giudice2022}, an object that encodes the back action of a system on its local subsystems, is area-law entangled and can be explicitly constructed -- taking a simple multi-site product state form. The influence matrix can be generalized to arbitrary time-like surfaces with slope $0\leq v\leq 1$~\cite{Foligno2023a}. Such a generalized IM can then be used to compute dynamical two-point correlations along rays with velocity $v$ in space-time. We focus on influence matrices for correlations at infinite temperature here, which can be be directly related to the ELT. For a quench from an initial state it is not guaranteed that the IM follows the area law, unless the state satisfies a solvability condition (see Ref.~\cite{Yu2024}).

For \LL{2} circuits the IM follows an area law for all $v$. For $v=0$ we can consider, e.g., the right influence matrix, which here simplifies as
\begin{align}
    \vcenter{\hbox{\includegraphics[width = .25\columnwidth]{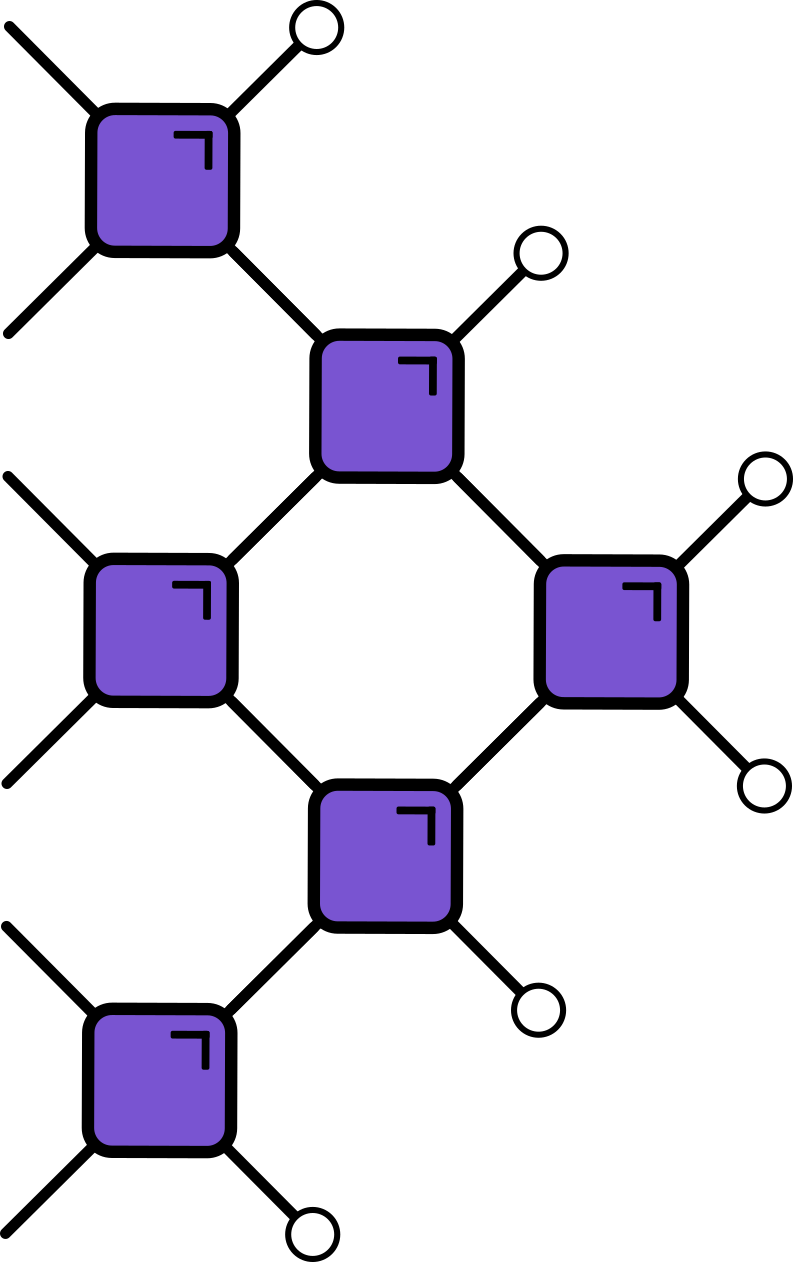}}}\,=\,\vcenter{\hbox{\includegraphics[width = .1\columnwidth]{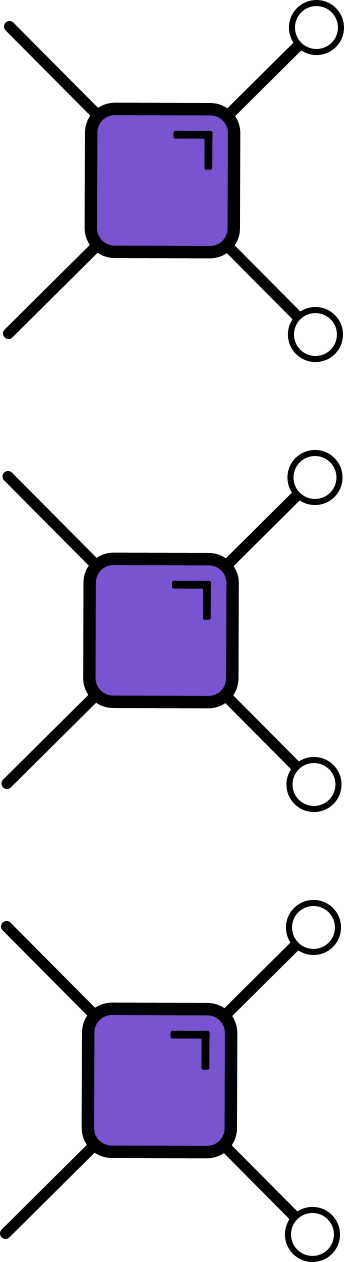}}}.
\end{align}
The left-hand side is the definition of the influence matrix, as illustrated for six discrete time steps, and we have applied the hierarchical conditions to reduce the general expression on the left to the final result on the right. Crucially, while the IM on the left generally exhibits volume-law entanglement, the expression on the right can clearly be decomposed in a product of two-site operators and has area-law entanglement. For general $v$ the definition of the IM has a nonzero slope $v$ and a similar decomposition follows using a similar algorithm as in the computation of $Z_\alpha$.

The IM can be used to extract the ELT along $v$, as we explain now. Consider $Z_\alpha$ for integer $\alpha\geq2$ and notice that it can be broken up into parts such as
\begin{align}
    \vcenter{\hbox{\includegraphics[width = .75\columnwidth]{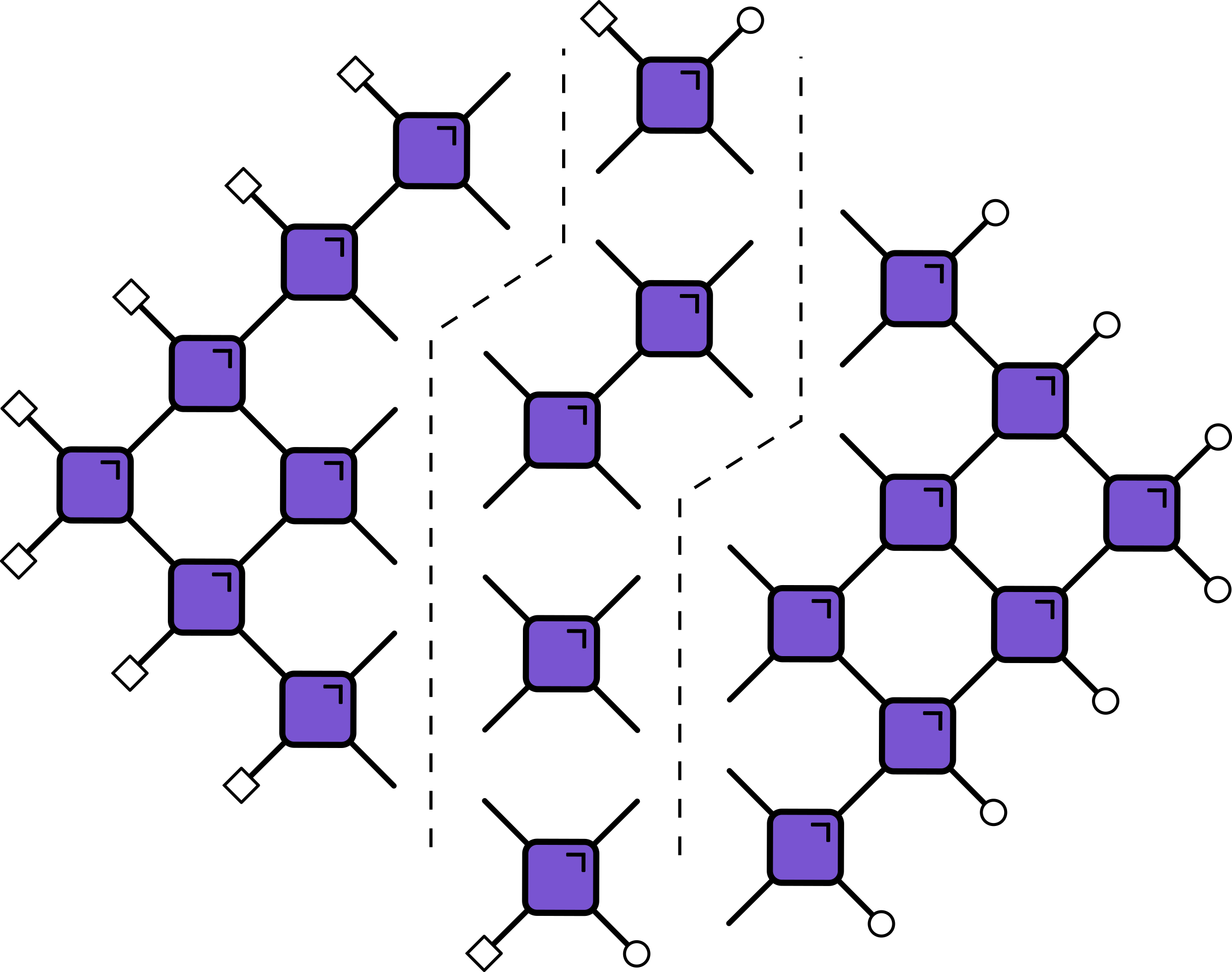}}}.
\end{align}
These parts can be identified with tensor products of influence matrices with corresponding slope $v$, as
\begin{equation}
    Z_\alpha(x,y) = \bra{L_v}^{\otimes\alpha} \mathcal{S} T_v \ket{R_v}^{\otimes\alpha},
\end{equation}
where $\mathcal{S}$ is a cyclic permutation in replica space and $T_v$ is a dual transfer matrix. $\ket{R_v}$ and $\bra{L_v}$ denote right and left IMs, respectively. This decomposition is not unique, but for a coarse grained description only the slope $v$ is relevant. Clearly, if the IM follows an area law the ELT can be computed efficiently. We see that the ELT is solvable if and only if the dynamical two-point correlator is solvable.

The above construction establishes a direct link between the ELT and dynamical correlation functions by showing that both quantities originate from the same fundamental object, the infinite temperature IM. It remains an interesting question if this link can lead to further insight on the connections between these quantities. In the case of \LL{2} circuits the IM not only follows an area law, it takes the form of a product state of, in general, dimers and monomers. This property lies behind the vanishing of correlation functions, since -- loosely speaking -- the lack of temporal entanglement between sites at the initial and final time leads the correlation to vanish. (This argument fails only when the correlation is produced by the dual-transfer matrix carrying the operators, as is the case at $v=0$ and $v=\pm1$.) On the other hand, the factorization of the IM also leads to the linear form of the ELT. 

The area-law IM also gives some insight into the operator spreading picture of correlations. In the language of Ref.~\cite{Nahum2022} it implies that there are no operators contracting to a point for $0<\abs{v}<1$ and that the operator trajectories contributing to the correlator for $v=0,\pm1$ are ``thin'' (see Ref.~\cite{Nahum2022} for details).

\begin{figure*}[t]
\centering
\includegraphics[width=0.7\textwidth]{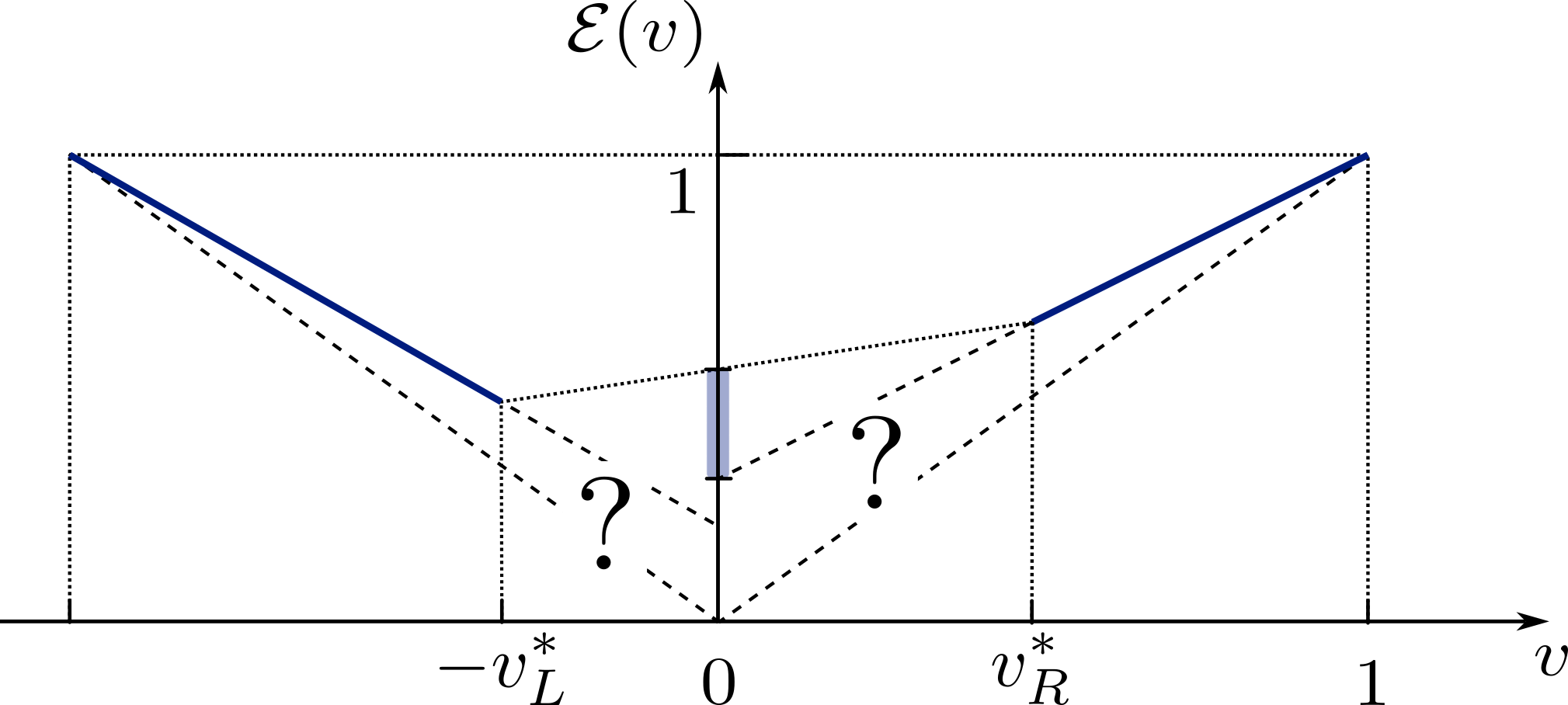}
\caption{The ELT is solvable above the (possibly distinct) threshold velocities and takes a linear form there. Bounds on the entanglement velocity $v_E$, indicated on the vertical axis, can be inferred from its partial knowledge by using the convexity of the ELT .}
\label{fig:lk_bounds}
\end{figure*}

\subsection{Higher levels of the hierarchy}

The ideas developed in the above sections can be applied, with minor modifications, to higher levels of the hierarchy. However, there is a crucial difference. While for \LL{2} circuits the ELT and correlations are solvable in the full spacetime, for higher levels the solvability is constrained to a range of velocities $v_*\leq\abs{v}\leq1$, as schematically depicted in Fig.~\ref{fig:lk_overview}. In this range, the ELT is linear and the correlations vanish exactly (apart from the edge rays $\abs{v}=v_*$ and $\abs{v}=1$), but outside of this range their evaluation is in general exponentially hard. This result can be thought of as the temporal entanglement exhibiting a transition in scaling when varying $v$. We show that generic \LL{k} circuits have maximal butterfly velocity and we provide bounds on the entanglement velocity.

Consider now an \LL{k} circuit with $k>2$. The ELT is computed using the same algorithm as outlined above. However, the diagram only factorizes if its width and height satisfy the inequality
\begin{align}
    m \geq (k-1)n \iff v\geq v_*\equiv\frac{k-2}{k}. \label{eq:Lk_ineq}
\end{align}
If this inequality is met we find
\begin{equation}
    \mathcal{E}(v) = 1- (1-\abs{v})\frac{\log B_{k-1}}{\log q^2}, \quad \abs{v}\geq v_*. \label{eq:elt_lk}
\end{equation}
We can again conclude that for all levels of the hierarchy the butterfly velocity is maximal, $v_B=1$, unless $B_{k-1}=q^2$. In distinction to the $k=2$ case, this condition allows gates to be entangling. If it is met $v_*$ defines an exact light cone, demonstrating that generalized dual unitarity constrains the causal propagation of signals for any level of the hierarchy.

The same arguments as above lead to the conclusion that the infinite-temperature IM factorizes for $v_*\leq \abs{v}\leq 1$. The correlations thus vanish identically in the range $v_*< \abs{v}< 1$, and they are solvable but non-trivial for $\abs{v}=v_*$ and $\abs{v}=1$. This behavior is confirmed by numerical results shown in Fig.~\ref{fig:correlations}.

As $\mathcal{E}(v)$ is inaccessible for $v=0$ we cannot compute the entanglement velocity $v_E$ exactly. However, convexity of the ELT implies certain bounds on $v_E$. Physically, degrees of freedom of all velocities contribute to entanglement growth. Knowledge of the contribution of degrees of freedom above the threshold velocity can be used to infer bounds on the total amount of entanglement growth. Such a bound can be thought of as a refinement of the general statement that causality constrains entanglement growth~\cite{Hartman2015}; \LL{k} circuits are endowed with stronger causal constraints than merely unitarity, in turn enabling stronger constraints on entanglement growth.

We begin by stating a lower bound on $v_E$. It follows from the observation that the convex curve with minimal $v_E$ emanating from $\mathcal{E}(v_*)$ is a straight line with slope $\mathcal{E}'(v_*)$. This yields
\begin{equation}
    v_E \geq 1 - \frac{\log B_{k-1}}{\log q^2}, \quad U\in\bar{\mathcal{L}}_k.
\end{equation}
If the circuit possesses a form of generalized dual unitarity in both light-cone directions, the stronger bound is pertinent.

To obtain an upper bound, we consider the general situation in which the circuit satisfies the \LL{k_R} condition along the right light cone and \LL{k_L} along the left light cone with $k_R,k_L$ not necessarily equal. The upper bound then follows from the ``worst guess'' for the ELT, i.e. a straight line connecting the two exactly solved intervals. Denoting the threshold velocities to the right and left as $v_{*R}=(k_R-2)/k_R$ and $v_{*L}=(k_L-2)/k_L$, the resulting upper bound reads
\begin{equation}
    v_E \leq \frac{v_{*R}\mathcal{E}(-v_{*L}) + v_{*L}\mathcal{E}(v_{*R})}{v_{*R}+v_{*L}}.
\end{equation}
There are two important special cases: (i) if the exactly solved parts of the ELT coincide in both directions, the bound acquires the simple form
\begin{equation}
    v_E \leq \mathcal{E}(v_*) = 1 - \frac{2}{k}\frac{\log B_{k-1}}{\log q^2}, \quad U\in\bar{\mathcal{L}}_k.
\end{equation}
This assumption holds in all parity symmetric unitary circuits, where $\mathcal{E}(-v)=\mathcal{E}(v)$, and thus $v=0$ is always a minimum. 
This bound also follows if the further assumption that $\mathcal{E}'(v)\geq0$ for all $v>0$ is met, i.e., if $\mathcal{E}(v)$ is monotonically increasing for $v>0$. 
It is known to be violated in quantum cellular automata with a finite topological index, where it is associated to a background entanglement current~\cite{Gong2022}. (ii) In the case where the circuit only satisfies the \LL{k} condition in a single light-cone direction -- without loss of generality we take this to be the right light cone -- the bound reads
\begin{equation}
    v_E \leq \frac{v_{*} + \mathcal{E}(v_{*})}{v_{*}+1}.
\end{equation}
These bounds are indicated graphically in Fig.~\ref{fig:lk_bounds}.

\section{Operator growth}
\label{sec:op_growth}

\begin{figure}[t]
    \centering
    \includegraphics[width = 0.45\textwidth]{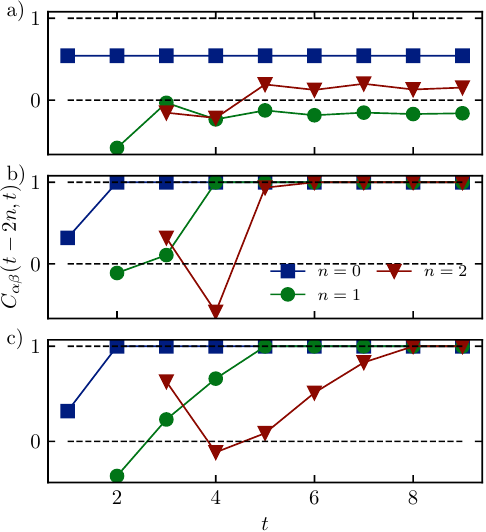}
    \caption{OTOC in qubit \LL{k} circuits for randomly selected gates of the given parametrizations. (a) In \LL{2} circuits the OTOC does not relax to the trivial value $C=1$, indicating $v_B=1$. (b) In the class of \LL{3} circuits defined by Eq.~\eqref{eq:l3_qubit} the OTOC relaxes to the trivial value exactly for $x\geq2$. This indicates $v_B=0$. (c) Dressing one leg of the gate used in (b) changes its dynamical properties drastically. Now, the OTOC relaxes exactly for $x=(t+4)/3$ indicating $v_B=1/3$. }
    \label{fig:otoc}
    %\vspace{-\baselineskip}
\end{figure}

\begin{figure*}[t]
    \centering
    \includegraphics[width = 0.8\textwidth]{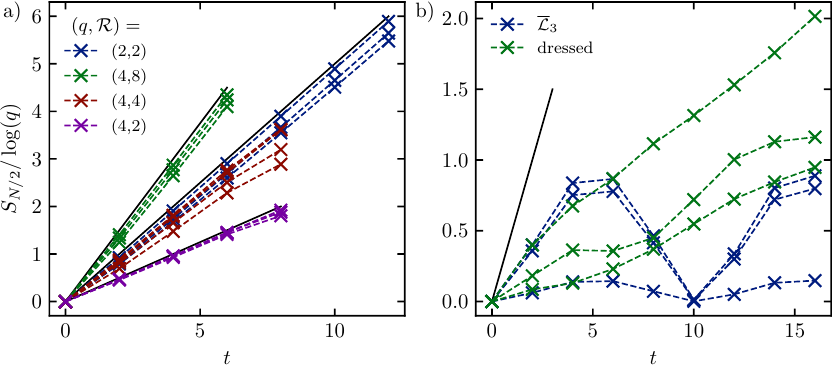}
    \caption{Growth of the half-chain entanglement entropy from random translationally invariant product states. The numerics are performed on chains of $N=24$ qubits and $N=12$ ququads, respectively. (a) The possible values of $v_E$ in \LL{2} circuits with $q=2$ and $q=4$ are illustrated. The qubit \LL{2} gate is randomly selected from the known parametrization. The ququad gates used are: Eq.~\eqref{eq:Rank2UniX} for $\mathcal{R}=2$, Eq.~\eqref{eq:schmidtrankq_gate} for $\mathcal{R}=4$, and Eq.~\eqref{eq:schmidtrank2q_gate} for $\mathcal{R}=8$. The solid lines are guides to the eye. (b) An \LL{3} gate randomly selected from the known parametrization shows saturation of the entanglement followed by oscillations indicating non-ergodicity. Upon dressing one leg with a randomly selected generic unitary entanglement growth is observed, with a rate satisfying the upper bound $v_E\leq1/2$ imposed by EMT.}
    \label{fig:entgrowth}
    %\vspace{-\baselineskip}
\end{figure*}

As was shown in section~\ref{sec:ELT}, EMT predicts that operators in \LL{k} circuits grow at maximal speed, and that due to the absence of curvature of the ELT there is no diffusive broadening of the operator front. Independently, an argument for maximal operator spreading based on certain quantum-information theoretic properties of \LL{k} gates was given. In this section we substantiate these results with more traditional calculations of the OTOC.

Given a basis of orthonormal local operators $\{\sigma_{\alpha},\alpha=0,\ldots,q^2-1 \}$ satisfying $\tr[\sigma_\alpha\sigma_\beta]=q\delta_{\alpha\beta}$, $\tr[\sigma_\alpha]=\delta_{\alpha 0}$, the OTOC is defined as
\begin{align}
    C_{\alpha\beta}(x,t) &= \langle \sigma_{\alpha}(0,t)\sigma_\beta(x,0)\sigma_{\alpha}(0,t)\sigma_\beta(x,0)\rangle \label{eq:otoc}.
\end{align}
Here $\sigma_{\alpha}(0,t) = U(t)\sigma_{\alpha}(x)U(t)^{\dagger}$ and $\sigma_{\alpha}(x)$ acts nontrivially as $\sigma_{\alpha}$ on site $x$ and as the identity everywhere else. The OTOC can again be represented as a tensor network (see Ref.~\cite{Claeys2020}):
\begin{align}
    C_{\alpha\beta}(x,t) = q^{m+n}\, \vcenter{\hbox{\includegraphics[width = .58\columnwidth]{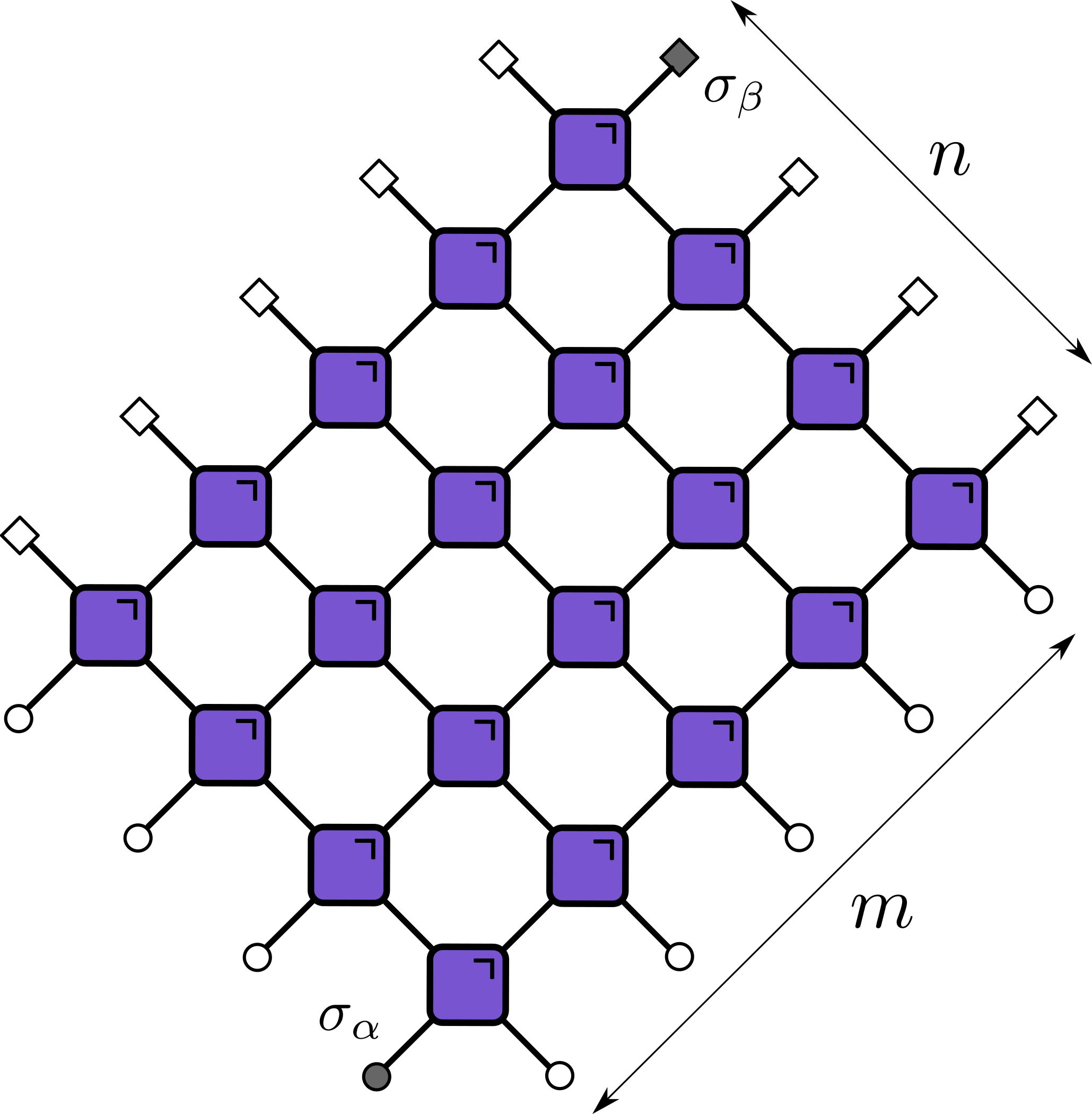}}}
\end{align}
with light-cone coordinates
\begin{subequations}
\begin{align}
    &n = \frac{t-x+2}{2},\,\, &m = \frac{t+x}{2},\qquad &t-x\in2\mathbb{Z}, \\
    &n = \frac{t-x+1}{2},\,\, &m = \frac{t+x+1}{2},\,\, &t-x\in2\mathbb{Z}+1.
\end{align}
\end{subequations}
In the limit $t+x\rightarrow\infty$ the problem of its evaluation reduces to the determination of the leading eigenspace of the light-cone transfer matrix (LCTM)~\cite{Bertini2020,Claeys2020}
\begin{align}
    T_n = \frac{1}{q}\, \underbrace{\vcenter{\hbox{\includegraphics[height=0.05\textheight]{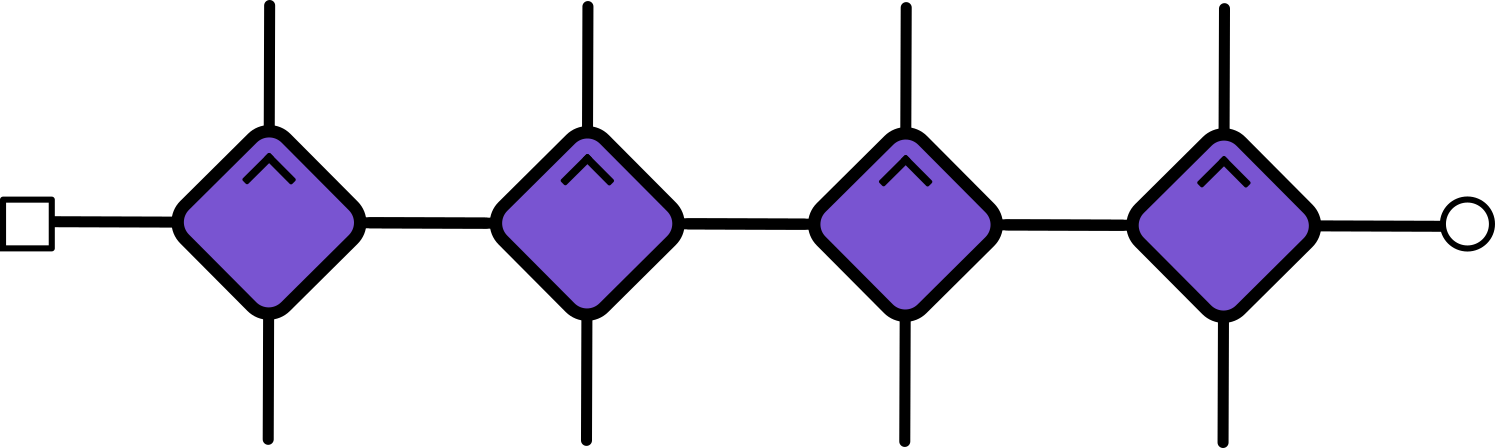}}}}_n\,.
\end{align}
In generic unitary circuits, there is a unique leading eigenvector determined purely by unitarity. The absence of further leading eigenvectors reflects a non-maximal butterfly velocity, and an analysis of the subleading eigenvectors is needed to determine the behavior of the OTOC~\cite{Rampp2023a,Huang2023}. On the other hand, if the leading eigenspace is degenerate the butterfly velocity is maximal since it leads to a nontrivial OTOC along the light ray $|x|=t$, and the asymptotic profile of the OTOC is determined by the projection of the boundaries of the tensor network on the leading eigenspace. This scenario occurs in dual-unitary circuits, where additional leading eigenvectors can be determined from dual-unitarity, resulting in a set of leading eigenvectors known as a ``maximally chaotic subspace''~\cite{Bertini2020,Claeys2020}.

In \LL{k} circuits a set of leading eigenvectors of the LCTM can be constructed analytically, generalizing the maximally chaotic subspace of dual-unitary circuits. Absent any further symmetries we expect this set to be exhaustive. There are however two key differences to DUCs: (i) the left and right leading eigenstates are not the transposed of each other, (ii) the leading eigenstates are in general highly entangled. The latter point implies that even though the leading eigenstates can be constructed, the OTOC cannot be computed exactly beyond small values of $t-x$, as the computation of the overlap with the boundary conditions is exponentially hard.

First, we show how to construct the non-trivial leading eigenvectors for the case of \LL{2} circuits. It is easy to check that the staircase vector $\rket{s_n}$ (shown here for $n=1$ and $n=2$)
\begin{align}
    \rket{s_1}=\vcenter{\hbox{\includegraphics[width = .15\columnwidth]{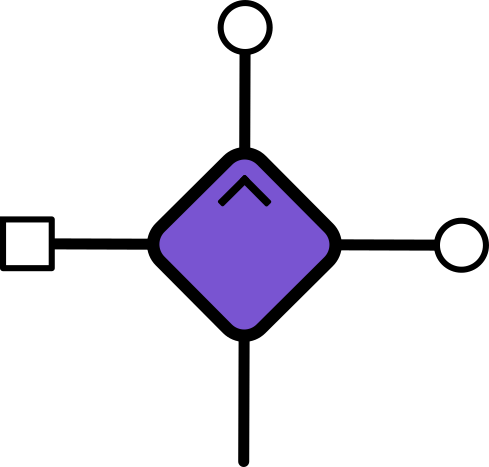}}}, \quad \rket{s_2}=\vcenter{\hbox{\includegraphics[width = .25\columnwidth]{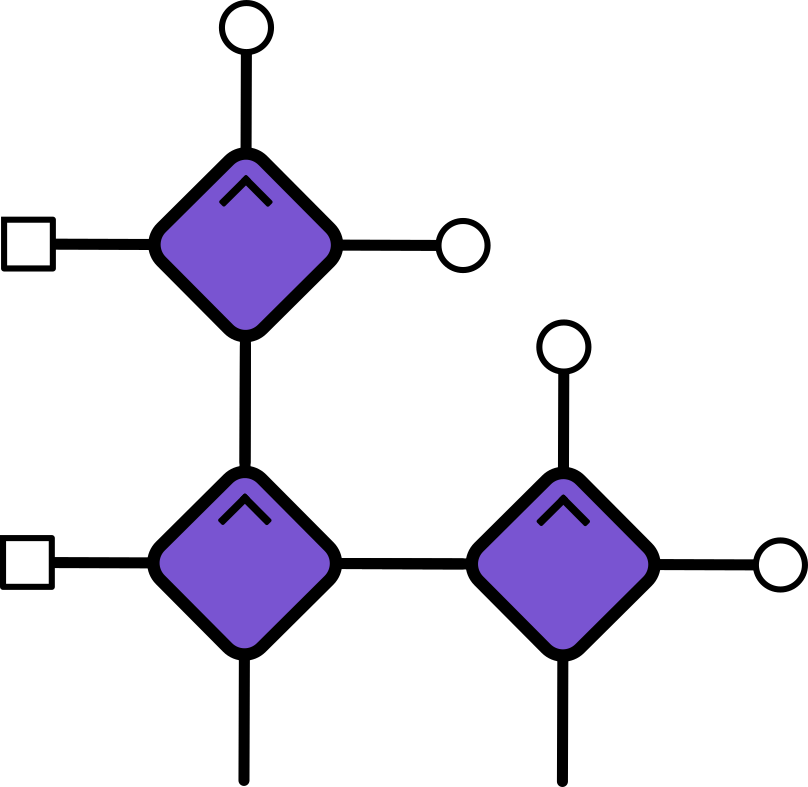}}}.
\end{align}
is a leading eigenvector of the LCTM of width $n$. The remaining leading eigenvectors can be obtained from smaller staircases by forming a tensor product with the trivial leading eigenvector, e.g.,
\begin{equation}
    \rket{\Box_{n-1}s_1} = \,\vcenter{\hbox{\includegraphics[width = .4\columnwidth]{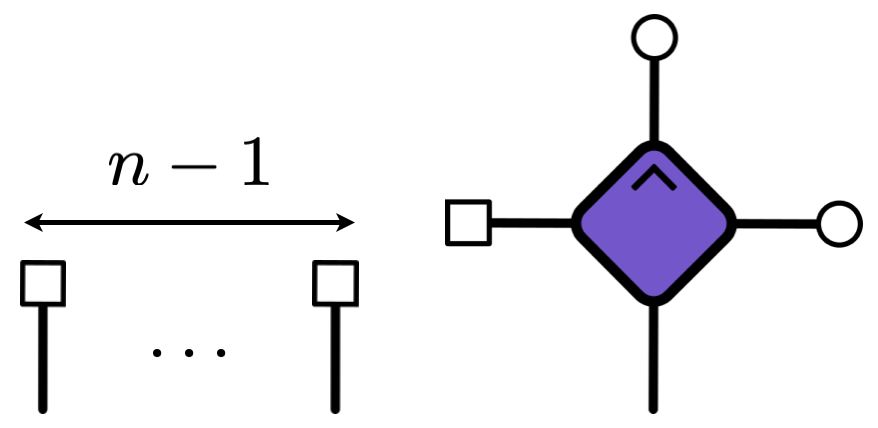}}}.
\end{equation}
That this is an eigenvector can be directly checked only using unitarity. This construction provides in total $n+1$ leading eigenvectors, as in the dual-unitary case. The left eigenvectors are constructed in an analogous manner
\begin{align}
    \rbra{\Tilde{s}_1}=\vcenter{\hbox{\includegraphics[width = .15\columnwidth]{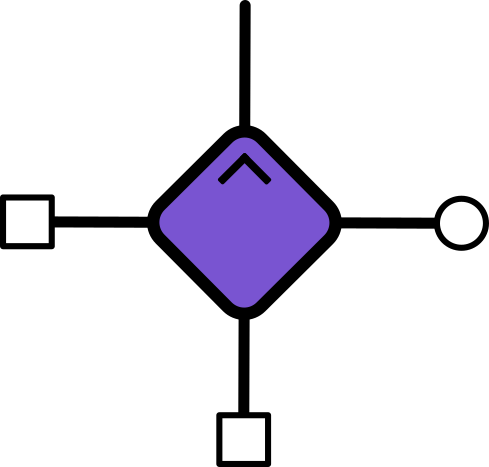}}}, \quad \rbra{\Tilde{s}_2}=\vcenter{\hbox{\includegraphics[width = .25\columnwidth]{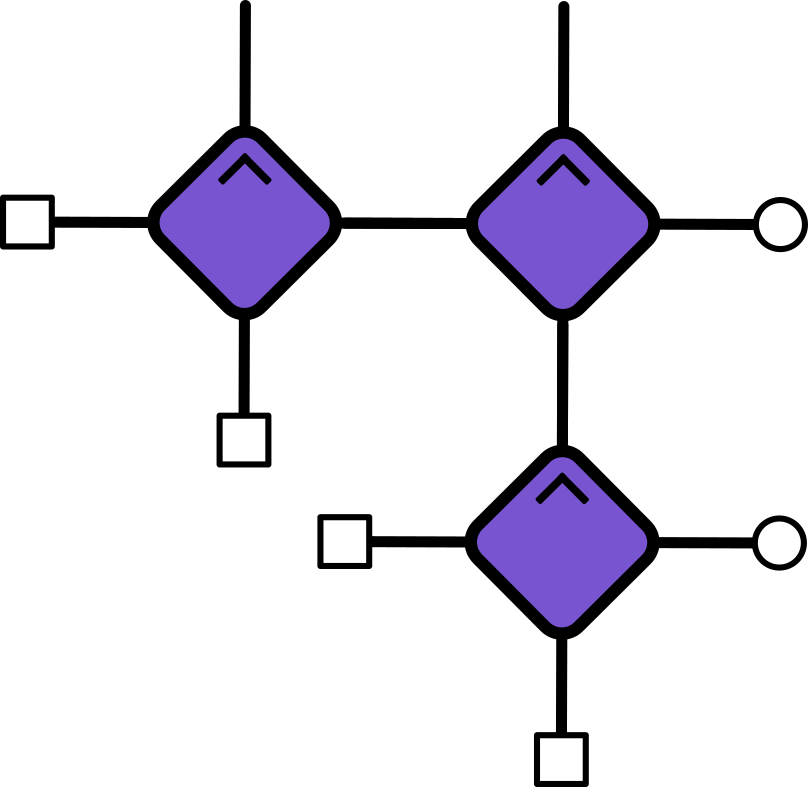}}}.
\end{align}
Note that the right and left eigenvectors are not the mutual transposed of each other.

To construct the projector on the leading eigenspace the overlap of the eigenvectors is needed. We denote $\rket{r_k} = \rket{\Box_{n-k}s_k}$ and $\rbra{\ell_k}=\rbra{\Tilde{s}_k\circ_{n-k}}$. As is shown in Appendix~\ref{app:overlaps} the overlap matrix (for a channel of width $n$) takes the general form
\begin{equation}
    \rbraket{\ell_i}{r_j} = \frac{1}{q^n}\begin{pmatrix}
        1 & 1 & \ldots & & 1 \\
        1 & & & 1 & b_1 \\
        \vdots & & 1 & b_1 & b_1^2 \\
        \vdots & \iddots & \iddots & \iddots & \vdots\\
        1 & b_1 & b_1^2 &\ldots & b_1^n    
    \end{pmatrix},
\end{equation}
where $b_1=B_1/q^2$. This matrix can be identified as a Hankel matrix. If this matrix has full rank all eigenvectors constructed in this way are linearly independent and the LCTM has (at least) $n-1$ nontrivial leading eigenvectors, indicating a maximal butterfly velocity. We see immediately that the above Hankel matrix becomes rank deficient if and only if $b_1=1$ and hence $B_1=q^2$, i.e., only for non-entangling gates. In this case the staircase vectors are not linearly independent and thus do not provide any additional leading eigenvectors beyond the trivial one.

The last missing ingredient for the calculation of the asymptotic profile of the OTOC is the overlap with the boundary conditions. 
Unfortunately, the overlaps between the staircase vectors and the boundary conditions cannot be simplified to an efficiently computable form. Generically, the staircase vectors are volume-law entangled, in contrast to dual-unitary circuits, where they reduce to product states. 

To gain information about the behavior of operator dynamics deep inside the light cone, we instead focus on the tripartite information~\cite{Hosur2016,Schnaack2019,Bertini2020a}, a quantity that is amenable to exact calculation in ergodic \LL{2} circuits. Details of the calculation are presented in App.~\ref{app:tripartite}. We find
\begin{equation}
    I^{(3)}(x,t) \approx (t-x) \log(b_1), \quad t+x \gg 1. \label{eq:ti_l2}
\end{equation}
For $x<t$ the tripartite information is negative and large, indicating scrambling. Moreover, the front defined by $x=t$ is sharp, i.e. it does not broaden diffusively.

For higher levels of the hierarchy the construction is analogous, merely the height of the steps changes,
\begin{align}
    \rket{s_1}=\vcenter{\hbox{\includegraphics[width = .15\columnwidth]{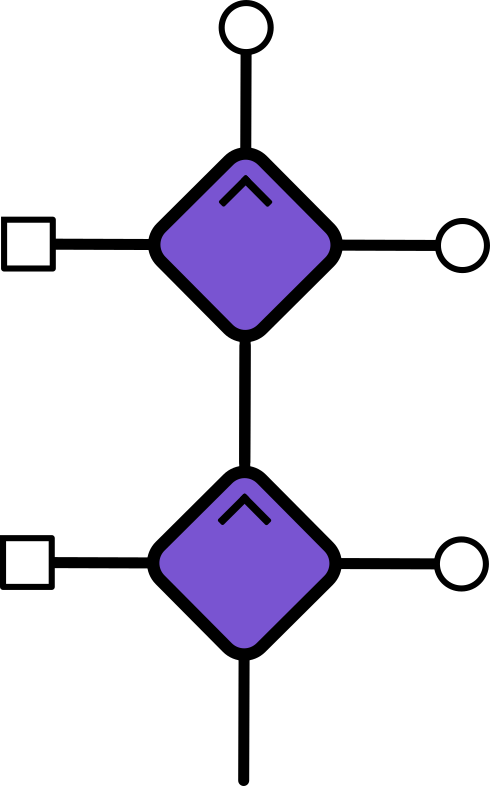}}}, \quad \rket{s_2}=\vcenter{\hbox{\includegraphics[width = .25\columnwidth]{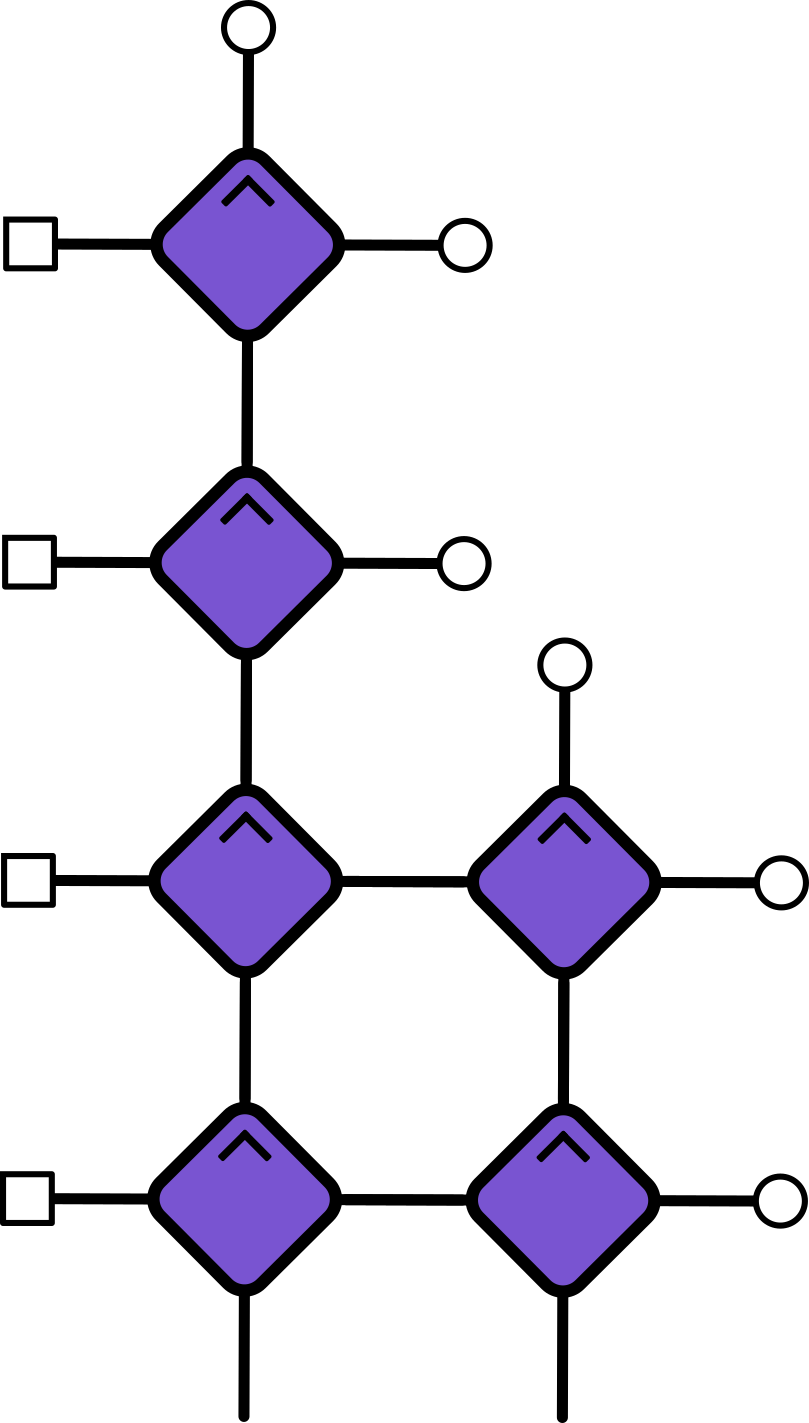}}}.
\end{align}
Analysis of the overlaps shows that the Hankel matrix structure is preserved and the condition for linear dependence becomes $B_{k-1}=q^2$ for gates from \LL{k}. This implies that for $k\geq3$ there can be \LL{k} circuits which are not product gates, but still have $v_B<1$. In this case, for ergodic circuits the ELT gives the upper bound $v_B\leq v_*$. The \LL{3} gates known so far all have the property $B_2=q^2$, thus a non-maximal butterfly velocity is expected. The parametrization of qubit \LL{3} gates is reviewed in App.~\ref{app:parametrization}. For these circuits our numerical results are consistent with a vanishing butterfly velocity, $v_B=0$, indicating strong non-ergodicity [Fig.~\ref{fig:otoc}(b)]. In fact, it can be shown that they display localized behavior by a slight modification of the argument presented in Ref.~\cite{Bertini2022}. Ergodic gates retaining some of the solvability can be constructed starting from this class of gates by dressing only one of the legs with a generic local unitary transformation. This preserves the \LL{3} property along one of the light-cone directions but breaks it along the other. As a result the solvability is preserved only for positive (negative) velocities. Along the remaining solvable direction the \LL{3} structure enforces the bound $v_B\leq1/3$. Our numerical results are consistent with $v_B=1/3$ [Fig.~\ref{fig:otoc}(c)]. Curiously, this class of gates seem to have a non-zero butterfly velocity even though all subleading eigenvalues of the LCTM vanish, contrary to expectations. This is possible if the largest Jordan block grows fast enough with the width of the LCTM. Details on this mechanism are provided in App.~\ref{app:butterfly}.

\section{Entanglement growth}
\label{sec:ent_growth}

In this section we check the prediction of the entanglement velocity provided by EMT. We present numerical results supporting the EMT picture for entanglement growth in \LL{2} and \LL{3} circuits in Fig.~\ref{fig:entgrowth}. We consider random translationally invariant product states as initial states. Analytical calculations from special solvable states supporting the EMT picture were presented in Ref.~\cite{Foligno2023}. 

In \LL{2} circuits, for a given local Hilbert space dimension only a discrete set of entanglement velocities, corresponding to allowed Schmidt ranks $\mathcal{R}$ are possible. E.g., for qubits the only non-trivial possibility is $v_E=1/2$ ($\mathcal{R}=2$). For ququads ($q=4$) the possible values are $v_E=1/4$ ($\mathcal{R}=2$), $v_E=1/2$ ($\mathcal{R}=4$), and $v_E=3/4$ ($\mathcal{R}=8$). The gates constructed in Sec.~\ref{sec:construction} provide the first examples of \LL{2} gates falling into the classes $(q,\mathcal{R})=(4,2)$ and $(q,\mathcal{R})=(4,8)$. Unfortunately they are non-ergodic, but it is likely that locally equivalent ergodic gates exist. Despite their non-ergodicity, the entanglement growth from generic states is consistent with the predictions of ELT [Fig.~\ref{fig:entgrowth}(a)]. This result reflects that the operator entanglement of the time-evolution operator encodes generic aspects of entanglement growth.

The strong non-ergodicity of the qubit \LL{3} circuits is reflected in the rapid saturation of the half-chain entanglement to an $\mathcal{O}(1)$ value. On the other hand, the one-sided dressing induces a finite entanglement velocity consistent with the bound $v_E\leq1/2$ [Fig.~\ref{fig:entgrowth}(b)].

\section{Constructions based on complex Hadamard matrices}
\label{sec:Hadamard}

The pattern of correlations in \LL{2} circuits along three directions in spacetime suggests a geometric interpretation making apparent an underlying hexagonal symmetry. Constructions exploiting this connection have recently appeared in the literature~\cite{Liu2023,Sommers2024}. In the following we present generalizations of these existing constructions, as well as a novel construction, in a unified manner based on the common framework of complex Hadamard matrices (CHMs). Complex Hadamard matrices are (rescaled) unitary matrices where all entries have equal magnitude. When arranged on the links of regular lattices, this property endows the circuits with additional unitary directions along symmetry axes of the lattice. 

In the case of dual-unitary circuits, the connection between dual-unitarity and complex Hadamard matrices goes back to initial works on space-time duality in kicked spin chains. The kicked Ising model at specific points in parameter space was originally shown to be self-dual in Ref.~\cite{Akila2016}, with the unitary Floquet operator at the self-dual point admitting a decomposition in terms of Hadamard matrices arranged on a square lattice~\cite{Gutkin2020}. It was subsequently realized that these kicked models could be reinterpreted as brickwork circuits of the form~\eqref{eq:def_brickwork}, with composite two-site unitary gates constructed out of Hadamard matrices~\cite{Gopalakrishnan2019,Bertini2019}. 
The construction of dual-unitary gates out of Hadamard matrices is not exhaustive, but rather results in dual-unitary gates with additional properties that allow for simplified derivations of results on e.g. the spectral form factor~\cite{Bertini2018}, entanglement dynamics~\cite{Bertini2019a,Gopalakrishnan2019,Piroli2020}, deep thermalization~\cite{Ho2022,Claeys2022b}, measurement-based quantum computing~\cite{Stephen2022}, and extensions of dual-unitarity to higher-dimensional lattices~\cite{mestyan_multi-directional_2022,Osipov2023}. These circuits also support additional `solvable' product states for which the dynamics of correlations and entanglement can be exactly calculated, in a way that directly extends to the following constructions \cite{Piroli2020}.

Here we present similar constructions for hierarchical dual-unitary gates. These subsume several existing constructions based on spacetime lattice structures and contain non-Clifford dynamics -- a prerequisite for quantum chaotic dynamics and generation of operator entanglement. We expect these gates to be ergodic for `generic' choices of complex Hadamard matrices. 
From a practical point of view, these circuits are naturally constructed out of one-site unitary gates (``kicks") and two-site diagonal gates (``controlled phases"), making them straightforward to implement on quantum simulation platforms. As an important subclass, these contain kicked Ising models, where the kicks correspond to pulses of longitudinal and transverse fields, and the controlled phases correspond to Ising interactions.

\subsubsection{Complex Hadamard matrices}

A complex Hadamard matrix $H$ of order $q$ is a $q \times q$ matrix consisting of complex entries having unit modulus with mutually orthogonal rows (columns), such that
\begin{align}
    H H^{\dagger} = H ^{\dagger} H = q \mathbbm{1}, \qquad |H_{ab}|=1, \quad \forall a,b=1\dots q\,.
\end{align}
Such matrices exist for all values of $q$, see e.g. Ref.~\cite{Tadej_Karol_2006}. An important example of a complex Hadamard matrix of order $q$ is given by the Fourier matrix:
\begin{equation}
H_{ab}=\omega^{ab}, \qquad \omega = \exp(2\pi i/q)\,.
\end{equation}
When $q=2^k$ for some positive integer $k$, the above matrix is (up to a scale factor) the $k$-qubit Fourier matrix or gate, which has important applications in phase estimation and period finding \cite{Nielsen_Chuang_book} as well as in Shor's factoring algorithm \cite{Shor_1997}. 
Complex Hadamard matrices are classified by allowing arbitrary permutation of rows and columns and multiplication by diagonal matrices consisting of arbitrary phases. While the Fourier matrix leads to Clifford dynamics, such a dephasing generically leads to non-Clifford dynamics and the gates constructed here are expected to be ergodic in that case. We refer interested readers to Ref.~\cite{Tadej_Karol_2006} for an online catalog of different families of inequivalent Hadamard matrices.

\subsection{Square lattice}

For completeness, we briefly review the construction of dual-unitary gates out of CHMs.
Dual-unitary matrices $U$ with matrix elements $U_{ab,cd}$ can be constructed out of four Hadamard matrices as
\begin{align}\label{eq:dualUfromH}
   U_{ab,cd}  = \vcenter{\hbox{\includegraphics[height = .2\columnwidth]{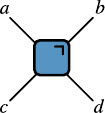}}}  = \frac{1}{q}\vcenter{\hbox{\includegraphics[height = .2\columnwidth]{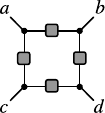}}} = \frac{1}{q}H_{ab}H_{bd}H_{cd}H_{ac},
\end{align}
where we have taken the four Hadamard matrices to be equal, but this is not a requirement. The black circles denote delta tensors, i.e.
\begin{align}
    \vcenter{\hbox{\includegraphics[height = .12\columnwidth]{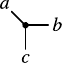}}} \,= \delta_{a,b,c}.
\end{align}
The gray squares denote Hadamard matrices. Each vertical Hadamard matrix acts as a single-site unitary kick operator,
\begin{align}
\frac{1}{\sqrt{q}}\,\vcenter{\hbox{\includegraphics[height = .13\columnwidth]{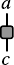}}} \,=\, \frac{1}{\sqrt{q}}H_{ac}\,,
\end{align}
whereas the horizontal Hadamard matrices act as two-site controlled phase gate with
\begin{align}
\vcenter{\hbox{\includegraphics[height = .15\columnwidth]{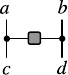}}}\, = \,  H_{ab} \delta_{ac}\delta_{bd}.
\end{align}
The dual-unitarity is apparent from the graphical notation: The gate~\eqref{eq:dualUfromH} is symmetric when seen along either the vertical (time) direction or the horizontal (space) direction. The resulting circuit for the full lattice then follows as, up to specific boundary conditions, 
\begin{align}\label{eq:squarelattice}
    \vcenter{\hbox{\includegraphics[width = .52\columnwidth]{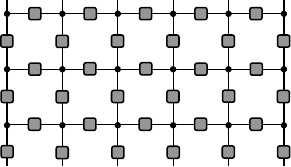}}}\,,
\end{align}
alternating single-site kicks with global nearest-neighbor phases. Here the three-leg delta tensors have been merged into four-leg delta tensors:
\begin{align}
    \vcenter{\hbox{\includegraphics[height = .16\columnwidth]{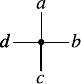}}} = \delta_{a,b,c,d}.
\end{align}

\subsection{Honeycomb lattice}

Motivated by the decomposition of the CNOT gate in terms of the Hadamard matrix, we consider two-site gates of the form
\begin{align}
    \vcenter{\hbox{\includegraphics[height = .2\columnwidth]{figs/fig_Uabcd}}} = \frac{1}{q} \,\,\vcenter{\hbox{\includegraphics[height = .22\columnwidth]{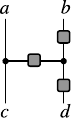}}}\,, \label{eq:CNOT}
\end{align}
where the grey boxes correspond to arbitrary CHM, such that
\begin{align}
    U_{ab,cd}  = \frac{\delta_{ac}}{q} \sum_{f=1}^q H_{af}H_{bf}H_{df}\,. 
\end{align}
For $q=2$ and choosing the complex Hadamard matrix to be the standard Hadamard matrix from quantum information,
\begin{align}\label{eq:Hadamard_qubits}
    H = \begin{pmatrix}
        1 & 1 \\
        1 & -1
    \end{pmatrix}\,,
\end{align}
the resulting gate can be directly checked to return the CNOT gate
\begin{align}
    U = \begin{pmatrix}
        1 & 0 & 0 & 0 \\
        0 & 1 & 0 & 0 \\
        0 & 0 & 0 & 1 \\
        0 & 0 & 1 & 0         
    \end{pmatrix}\,.
\end{align}
For any choice of CHMs the gates constructed in this manner satisfy the \LL{2} condition, where we now take the gray squares to represent $H \otimes H^*$, similar to the folded notation of Eq.~\eqref{fig:folded_gate}. Using the unitarity of the CHMs we find that:
\begin{align}
   \frac{1}{q^4} \, \vcenter{\hbox{\includegraphics[height = .24\columnwidth]{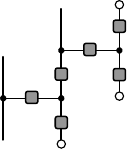}}} \,=\, \frac{1}{q} \, \vcenter{\hbox{\includegraphics[height = .24\columnwidth]{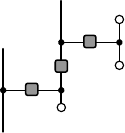}}} \,=\, \frac{1}{q^{3/2}} \, \vcenter{\hbox{\includegraphics[height = .24\columnwidth]{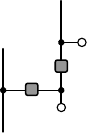}}} \label{eq:HadL2Condition}
\end{align}
In the last expression we can use that these matrix elements are phases, which cancel out exactly due to the contractions with delta tensors in the rightmost gate, to obtain
\begin{align}
    \frac{1}{q^4} \, \vcenter{\hbox{\includegraphics[height = .24\columnwidth]{figs/fig_Had_L2_d1}}} \,=\, \frac{1}{\sqrt{q}} \, \vcenter{\hbox{\includegraphics[height = .24\columnwidth]{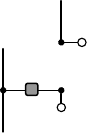}}} \,=\, \frac{1}{\sqrt{q}} \, \vcenter{\hbox{\includegraphics[height = .14\columnwidth]{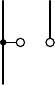}}} \label{eq:HadL2Condition}
\end{align}
This returns the \LL{2} condition. As such, these constructions define a continuous family of \LL{2} gates in arbitrary local Hilbert space dimension with Schmidt rank $q$ that comprises both Clifford and non-Clifford gates.

Constructing a brickwork circuit of the form \eqref{eq:def_brickwork} out of the gates \eqref{eq:CNOT} and combining delta tensors, this yields an evolution operator of the form
\begin{align}
    \vcenter{\hbox{\includegraphics[height = .38\columnwidth]{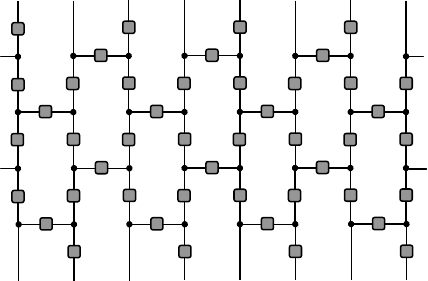}}}\,, \label{eq:L2_brickwork}
\end{align}
which can be represented as a honeycomb lattice by a coordinate transformation 
\begin{align}
    \vcenter{\hbox{\includegraphics[height = .38\columnwidth]{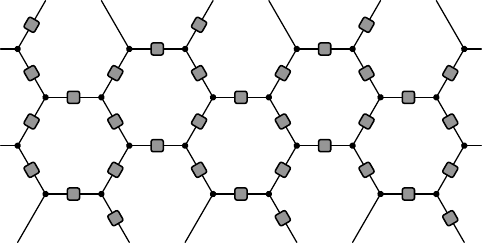}}}\,. \label{eq:L2_hexagon}
\end{align}
For $q=2$ this construction reduces to the construction of Ref.~\cite{Liu2023} and we refer the reader to Ref.~\cite{Liu2023} for a detailed discussion of the resulting entanglement dynamics.

Expressed in the above way, the six-fold rotation symmetry of the honeycomb lattice makes the existence of three unitary axes of time manifest. Notice that the \LL{2} property is preserved if the CHMs are different on each link of the lattice. This enables the construction of circuits with larger unit cells, and disorder in space or time.

\subsection{Triangular lattice}

As the solvability of the honeycomb construction stems from its sixfold rotation symmetry leading to three arrows of time, it is natural to turn to other lattices with the same property. In order to do so we now consider two-site gates constructed out of three complex Hadamard matrices by arranging them as
\begin{align}
     \vcenter{\hbox{\includegraphics[height = .2\columnwidth]{figs/fig_Uabcd}}} =  \frac{1}{\sqrt{q}} \,\,\vcenter{\hbox{\includegraphics[height = .2\columnwidth]{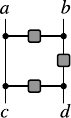}}}\,, \label{eq:triangle}
\end{align}
such that
\begin{align}
    U_{ab,cd} = \frac{\delta_{ac}}{\sqrt{q}} H_{ab}H_{ad}H_{bd}\,.
\end{align}
Choosing the Hadamard matrix as Eq.~\eqref{eq:Hadamard_qubits}, we obtain a unitary gate
\begin{align}
    U = \frac{1}{\sqrt{2}}
    \begin{pmatrix}
        1 & 1 & 0 & 0 \\
        1 & -1 & 0 & 0 \\
        0 & 0 & 1 & -1 \\
        0 & 0 & -1 & -1
    \end{pmatrix}\,.
\end{align}
More complicated gates can again be constructed from different choices of Hadamard matrix.

This class of gates satisfies the \LL{2} property in both light-cone directions. Consider first contractions from the left:
\begin{align}
    \frac{1}{q^2} \,\vcenter{\hbox{\includegraphics[height = .24\columnwidth]{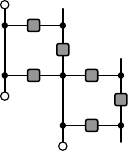}}}\, =\, \frac{1}{q^{5/2}} \,\vcenter{\hbox{\includegraphics[height = .24\columnwidth]{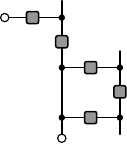}}} \,= \,\frac{1}{q^{5/2}} \,\vcenter{\hbox{\includegraphics[height = .24\columnwidth]{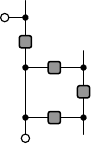}}}\,,
\end{align}
making use of cancelling phases in the first equation and unitarity in the second. Opposing phases again cancel in the final expression, such that we find
\begin{align}
    \frac{1}{q^2} \,\vcenter{\hbox{\includegraphics[height = .24\columnwidth]{figs/fig_Had_L2_d6}}} \,=\,  \frac{1}{q^{3/2}} \,\vcenter{\hbox{\includegraphics[height = .2\columnwidth]{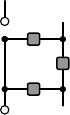}}}\,,
\end{align}
establishing \LL{2} in one direction. For contractions starting from the right, we rather have that
\begin{align}
    \frac{1}{q^2} \, \vcenter{\hbox{\includegraphics[height = .24\columnwidth]{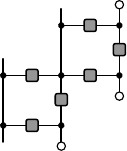}}}\,
      &=\,\frac{1}{q} \, \vcenter{\hbox{\includegraphics[height = .24\columnwidth]{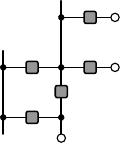}}}\,
       =\,\frac{1}{q^{3/2}} \, \vcenter{\hbox{\includegraphics[height = .15\columnwidth]{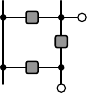}}}\, \nonumber\\
       &=\,\frac{1}{\sqrt{q}} \,\vcenter{\hbox{\includegraphics[height = .15\columnwidth]{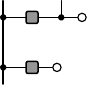}}}\, 
        = \,\frac{1}{\sqrt{q}} \,\vcenter{\hbox{\includegraphics[height = .14\columnwidth]{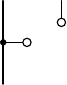}}}\, 
\end{align}
by repeatedly using either unitarity or cancelling phases, establishing \LL{2} in the other direction.

A brickwork circuit constructed out of these gates now returns an evolution operator with Hadamard matrices on the links of a triangular lattice,
\begin{align}
    \vcenter{\hbox{\includegraphics[height = .38\columnwidth]{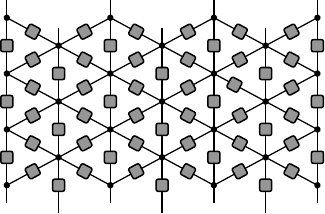}}}\,. \label{eq:L2_triangular}
\end{align}
The presence of multiple unitary directions is again clear from the symmetry of the lattice.

\subsection{Sheared square lattice}
\begin{figure}[t]
\centering
\includegraphics[width=0.45\textwidth]{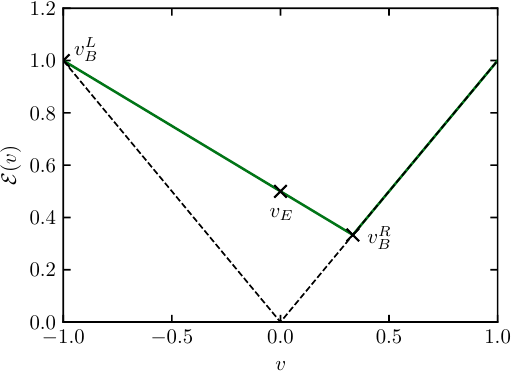}
\caption{Entanglement line tension for the circuit defined from CHM on the sheared square lattice. The gate satisfies \LL{2} in the left and \LL{3} in the right light cone direction. The square lattice structure enables the exact evaluation of the ELT even in the otherwise inaccessible region.}
\label{fig:elt_L2L3}
\end{figure}

We have seen in the previous section that, by passing from constructions based on four Hadamard matrices per gate to constructions based on three matrices, the gate goes from \LL{1} to \LL{2}. It seems therefore natural to simplify the circuit further by constructing a two-site gate from two Hadamard matrices. We arrange one of them on a horizontal and the other on a vertical link, as follows
\begin{align}
     \vcenter{\hbox{\includegraphics[height = .2\columnwidth]{figs/fig_Uabcd}}} =  \frac{1}{\sqrt{q}} \,\,\vcenter{\hbox{\includegraphics[height = .2\columnwidth]{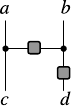}}}\,, \label{eq:L2L3}
\end{align}
such that
\begin{align}
    U_{ab,cd} = \frac{\delta_{ac}}{\sqrt{q}} H_{ab}H_{bd}\,.
\end{align}
E.g. for the Hadamard matrix from Eq.~\eqref{eq:Hadamard_qubits} we find that 
\begin{align}
    U = \frac{1}{\sqrt{2}}
    \begin{pmatrix}
        1 & 1 & 0 & 0 \\
        1 & -1 & 0 & 0 \\
        0 & 0 & 1 & 1 \\
        0 & 0 & -1 & 1
    \end{pmatrix}\,.
\end{align}
The resulting gates still satisfy the \LL{2} condition in the left light-cone direction, where the derivation is similar as for the previous construction, but in the right light-cone direction these now fulfill the \LL{3} condition: 
\begin{align}
    \frac{1}{q^3} \,\vcenter{\hbox{\includegraphics[height = .24\columnwidth]{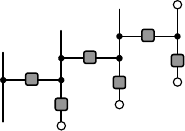}}} &\,=\, \frac{1}{\sqrt{q}} \, \vcenter{\hbox{\includegraphics[height = .12\columnwidth]{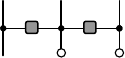}}} \nonumber\\
    &\,=\, \frac{1}{\sqrt{q}} \,\vcenter{\hbox{\includegraphics[height = .12\columnwidth]{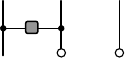}}}. 
    \label{eq:HadL3Condition}
\end{align}
Arranging these gates in a brickwork pattern leads to a geometry which locally has the connectivity of a square lattice, but due to its boundary conditions it globally differs from a rectangle. The brickwork circuits leads to a lattice of the form
\begin{align}
    \vcenter{\hbox{\includegraphics[height = .35\columnwidth]{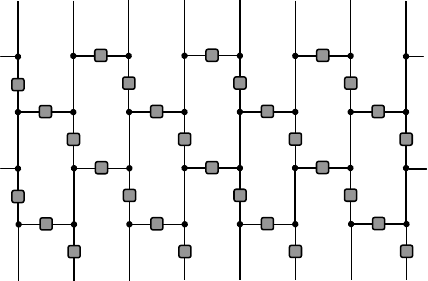}}} \,,\label{eq:L2L3_brickwork}
\end{align}
where delta tensors can again be merged to return an evolution operator 
\begin{align}
    \vcenter{\hbox{\includegraphics[height = .5\columnwidth]{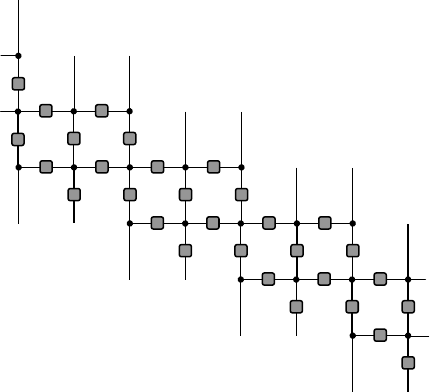}}}\,. \label{eq:L2L3_square}
\end{align}
This lattice is locally equivalent to the dual-unitary lattice from Eq.~\eqref{eq:squarelattice}, such that various results from dual-unitarity directly extend to this lattice.
Purely using the constraints from generalized dual-unitarity, the spacetime region enclosed by the rays $v=0$ and $v=1/3$ would be inaccessible to the methods presented so far. However, the square lattice structure -- a ``hidden'' dual unitarity -- endows the circuit with a stronger notion of solvability than the other examples discussed here.
As one example, due to the hidden dual-unitary structure the correlation functions are only supported on the rays $v=-1$ and $v=1/3$.

A Clifford representative of this class of gates was constructed in Ref.~\cite{Sommers2024}, where the ELT [see Fig.~\ref{fig:elt_L2L3}] was computed based on the symmetry directions of the lattice and it was also commented that such circuits correspond to a sheared dual-unitary circuits. Using the Hadamard representation, this result can be directly reproduced and extended to non-Clifford evolution. The resulting ELT is illustrated in Fig.~\ref{fig:elt_L2L3}. Notice that when imposing the \LL{2} and \LL{3} conditions for the left and right light cone, this form of the ELT is the only one allowed by convexity. For a further discussion of this form of the ELT and operator dynamics in the Clifford version of this circuit, we refer the reader to Ref.~\cite{Sommers2024}.

\section{Existence and construction of generalized dual-unitary gates}
\label{sec:construction}

In this section we detail some of the properties of the Schmidt spectrum \eqref{eq:Schmidt} of \LL{2} gates and use these as a guiding principle to obtain constructions of \LL{2} gates beyond the previously discussed Hadamard constructions. In the two-qubit case an exhaustive parametrization exists of both dual-unitary gates and \LL{2} gates, but going to larger Hilbert spaces can allow for a phenomenology inaccessible in qubits (see e.g. Ref.~\cite{Aravinda2021} for dual-unitary gates). Here we discuss new constructions of \LL{2} for higher local Hilbert space dimensions by taking insights from several existing constructions of dual-unitary gates, which will allow us to e.g. generalize the CNOT construction for qubits to larger Hilbert spaces in different ways. These results are meant to be an initial exploration, showing how different Schmidt ranks and hence different entanglement velocities can be obtained, and will be expanded on in later works.

The \LL{2} conditions strongly restrict the Schmidt spectrum: all such gates have a non-maximal Schmidt rank, i.e. $\R < q^2$, and the Schmidt spectrum is flat, i.e. all nonzero Schmidt values are equal. 
The first property can be directly shown by noting that a maximal Schmidt rank implies that the inverse of $\tilde{U}$ exists, where $\tilde{U}$ is the space-time dual of $U$ defined as $\tilde{U}_{ab,cd} = U_{ac,bd}$. Representing this (folded) inverse as a green tensor, we can contract the hierarchical condition~\eqref{eq:L2DIag} with this inverse acting on the two lowest Hilbert spaces in the diagram to obtain
\begin{align}
    \vcenter{\hbox{\includegraphics[height = .16\columnwidth]{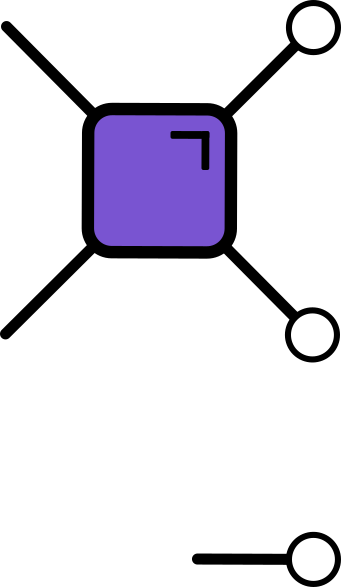}}}\, = \,\vcenter{\hbox{\includegraphics[height = .16\columnwidth]{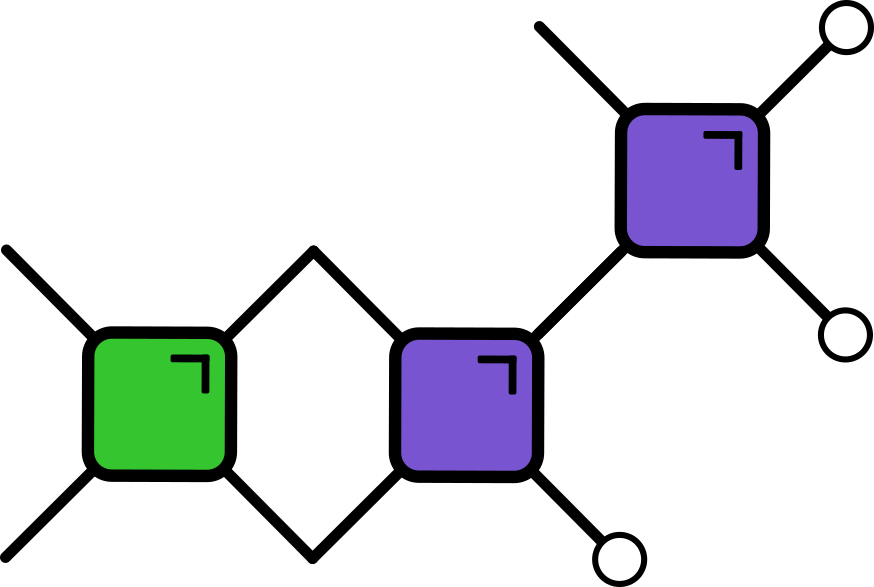}}}\, = \,\vcenter{\hbox{\includegraphics[height = .16\columnwidth]{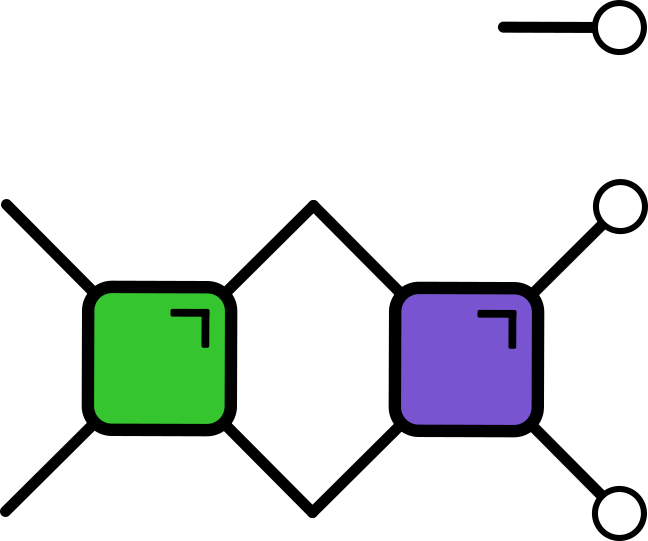}}}\, = \,\vcenter{\hbox{\includegraphics[height = .16\columnwidth]{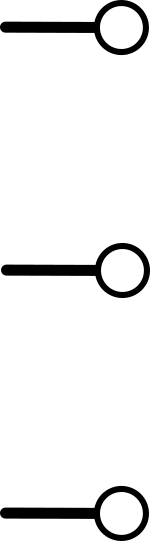}}}.
\end{align}
The above equation factorizes and is satisfied if and only if $U$ is dual unitary. Therefore, any unitary gate $U$ with full Schmidt rank satisfies the $\mathcal{L}_2$ conditions if and only if it is a dual unitary. As such, all \LL{2} gates for local Hilbert space dimension $q$  have Schmidt rank less than $q^2$. 

The flatness of the Schmidt spectrum was proven in Ref.~\cite{Foligno2023}. A defining property of dual-unitary gates is that these have a flat Schmidt spectrum with maximal rank i.e., $\lambda_{i}=1$  for all $i=1 \dots q^2$. For hierarchical \LL{2} gates we conversely find that these gates satisfy $\lambda_{i}=q/\sqrt{\R}$ for all $i=1,\dots \R<q^2$, with a nonzero amount of vanishing Schmidt values. The maximal rank $\R = q^2$ corresponds to dual-unitaries and the opposite limit of $\R=1$ corresponds to unentangled products of one-site unitaries. 
While this is a necessary condition for \LL{2}, it turns out to not be a sufficient condition. This is a consequence of the \LL{2} conditions breaking local invariance, as locally equivalent gates share the same Schmidt spectrum. In the remainder of this section we will present systematic constructions of gates with a flat Schmidt spectrum with non-maximal rank and identify the gates that satisfy the \LL{2} condition. 

\subsection{\LL{2} gates with Schmidt rank $q$}

Consider a controlled unitary gate on $\mathbb{C}^q \otimes \mathbb{C}^q$ given by
\begin{equation}
    U=\sum_{i=1}^{q} \ket{i}\bra{i} \otimes u_i,
    \label{eq:GenContU}
\end{equation}
where $u_i \in \mathbb{U}(q)$. By choosing the unitary gates to satisfy $\tr(u_i^{\dagger}u_j) = \delta_{ij}$ it can be directly checked that these gates have a flat Schmidt spectrum with rank $\R$. The CNOT gate can be extended to arbitrary $q$ by choosing $u_i=X^i$, where $X \ket{k}=\ket{k \oplus 1}$ is the shift operator ($\oplus$ denotes addition $\bmod\, q$). The resulting generalized CNOT gates are well known and given by
\begin{align}
    U_{\text{CNOT}}=\sum_{i=1}^{q} \, \ket{i}\bra{i} \otimes X^{i-1}.
\end{align}
These have Schmidt rank $\R = q$ and satisfy the \LL{2} properties for arbitrary $q$, as also observed in Ref.~\cite{Yu2024}. This result can be checked by direct calculation. Furthermore, for all such controlled unitary gates the \LL{2} property is invariant under local multiplication with enphased permutation gates. For a possible permutation $P$ of the $i$ basis elements and a set of phases $\phi_i,i=1\dots 1$ we can define an enphased permutation gate $u_{P}$ as
\begin{align}
    \bra{a}u_{P}\ket{b} =  e^{i\phi_a} \delta_{b,P(a)}.
\end{align}
A set of \LL{2} gates can then be obtained as
\begin{align}
    (u_{P_1} \otimes u_{P_2}) U_{\text{CNOT}},
\end{align}
where both the two permutations $P_1$ and $P_2$ and the two corresponding sets of phases can be chosen independently. That this parametrization returns an \LL{2} gate can be understood by explicitly rewriting the hierarchical conditions \eqref{eq:L2DIag} for a controlled gate. Controlled gates of the form \eqref{eq:GenContU} can be graphically represented as
\begin{align}
    \vcenter{\hbox{\includegraphics[width = 0.3\columnwidth]{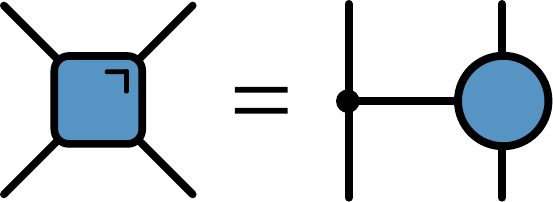}}}\,.
\end{align}
The hierarchical condition \eqref{eq:L2DIag} reads
\begin{align}
    \vcenter{\hbox{\includegraphics[width = 0.45\columnwidth]{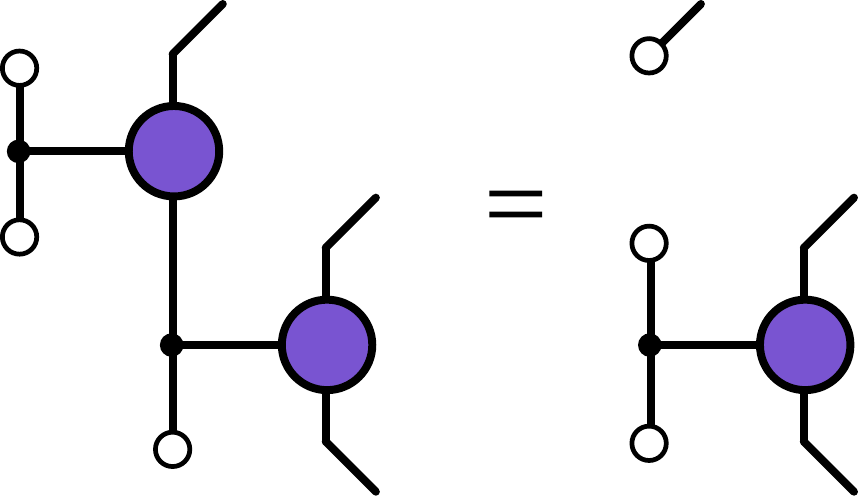}}}\,\,,
\end{align}
where the folded gates now correspond to $U \otimes U^*$ (and hence $\alpha=1$). It can be easily checked that a sufficient condition for this equation to hold is that
\begin{align}
    \vcenter{\hbox{\includegraphics[width = 0.22\columnwidth]{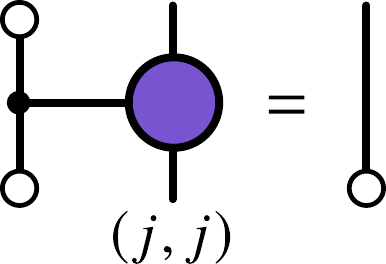}}}\,, \qquad \forall j =1\dots q\,,
\end{align}
where $(j,j)$ denotes the folded version of $\ket{j}\bra{j}$. Expressing the folded diagram algebraically, this equation corresponds to
\begin{align}
    \sum_{i=1}^q u_i \ket{j} \bra{j} u_i^{\dagger} = \mathbbm{1}, \qquad \forall j=1\dots q.
\end{align}
This equation is satisfied whenever $\{u_i\ket{j},i=1\dots q\}$ forms a complete basis for each choice of $j$, as can be directly checked to be the case for the above construction.

The CNOT example for $q=2$ can be extended to higher $q$ in a different way for even local Hilbert space dimension by consider the controlled $X$ gate as given by
\begin{equation}
    CX=\sum_{i=1}^{q} \ket{i}\bra{i} \otimes W_i,
    \label{eq:Rank2UniX}
\end{equation}
where $W_i=\mathbbm{1}_q$ for $i\bmod 2 =0$ and $W_i=X$ otherwise. This permutation gate now has Schmidt rank $\R=2$ with $\lambda_0=\lambda_1=q/
\sqrt{2}$ and again satisfies the \LL{2} constraints. This construction can again be generalized by considering two permutations $P$ and $Q$, where $P$ preserves the parity, $P(i) = i  \bmod 2$ and $Q$ switches the parity, $Q(i) = i+1  \bmod 2$. The resulting gates can again be enphased, where we define $U_{a,b} = e^{i\phi_a} \delta_{b,P(a)}$ and $V_{a,b} = e^{i\psi_a} \delta_{b,Q(a)}$, resulting in a \LL{2} gate
\begin{equation}
    U=\sum_{i=1}^{q} \ket{i}\bra{i}  \otimes W,
    \label{eq:Rank2UniGen}
\end{equation}
where $W_i=U$ for $i\bmod 2 =0$ and $W_i=V$ otherwise.

We note that all \LL{2} gates in local dimension $q=2$ are T-dual and satisfy Eq.~\eqref{eq:Tdual}. For local dimensions $q \geq 4$, there exist \LL{2} gates that are not T-dual. For example, in local dimension $q=4$ consider the following permutation gate satisfies the \LL{2} condition:
\begin{equation}
    \mathcal{P}=(CX) S (CX) S, \label{eq:schmidtrankq_gate}
\end{equation}
where $CX$ is given by Eq.~\eqref{eq:Rank2UniX} and $S$ is the SWAP gate. The resulting permutation gate $\mathcal{P}$ has Schmidt rank $\R =4$ with $\lambda_0=\cdots=\lambda_3=2$, but direct calculation shows that it is not T-dual.

\subsection{\LL{2} gates with Schmidt rank two}

In the previous subsection, we explicitly showed how \LL{2} gates with Schmidt rank two can be systematically constructed for an even local Hilbert space dimension. It is a natural question to ask if this construction can be extended to odd local Hilbert space dimension. Here we will show that unitary gates with flat Schmidt spectrum and Schmidt rank two do not exist for odd local Hilbert space dimensions. Consequently, \LL{2} gates with Schmidt rank two do not exist in this case. 

The proof of the above result relies on the fact that every bipartite unitary gate with $\R = 2$ is locally equivalent to a controlled unitary \cite{Cohen2013}. Local equivalence does not change the Schmidt spectrum, so we can consider the Schmidt spectrum of
\begin{equation}
    U=\sum_{i=1}^{q} \ket{i}\bra{i} \otimes u_i,
\end{equation}
where $u_i \in \mathbb{U}(q)$. For Schmidt rank two there are only two linearly independent $u_i$'s and, without loss of generality, we assume there are only two distinct blocks. Among these two distinct blocks one of the blocks can always be chosen to be the identity matrix by a local change of basis. As we are interested in unitary gates with a flat Schmidt spectrum this implies that the other (orthonormal) unitary must be traceless. Therefore, we can express the above controlled unitary as
\begin{equation}
    U= \left(\sum_{j=1}^{q_1} \ket{j}\bra{j} \right) \otimes \mathbbm{1}_q + \left( \mathbbm{1}_q-\sum_{j=1}^{q_1} \ket{j}\bra{j}\right) \otimes u, \quad 1<q_1<q,
\end{equation}
where we have separated the terms based on whether the action on second qudit is trivial or not. From the above equation we write the Schmidt decomposition as follows,
\begin{equation}
    U=\sqrt{q\, q_1}\, \left(M \otimes \frac{\mathbbm{1}_q}{\sqrt{q}}\right)+\sqrt{q\,(q-q_1)}\,\left(M^{\perp} \otimes \frac{u}{\sqrt{q}}\right),
\end{equation}
where 
\begin{align}
  M=\frac{1}{\sqrt{q_1}}\sum_{j=1}^{q_1} \ket{j}\bra{j},\qquad M^{\perp}=\frac{\mathbbm{1}_q-\sqrt{q_1}\,M}{\sqrt{q-q_1}},
\end{align}
are properly normalized. The Schmidt coefficients can be read off as $\lambda_1=\sqrt{q\,(q-q_1)}$ and $\lambda_2=\sqrt{q\, q_1}$. The condition for flat Schmidt spectrum $\lambda_1=\lambda_2$ then leads to
\begin{align}
\sqrt{q\,(q-q_1)} & = \sqrt{q\, q_1} \quad
\implies \quad  q_1=\frac{q}{2}.
\end{align}
Therefore, $\lambda_1=\lambda_2$ is possible only when $q$ is divisible by 2 since $q_1$ is an integer dimension.

\subsection{\LL{2} gates with larger Schmidt ranks in non-prime dimensions}
\subsubsection{Tensor product constructions}
The tensor product of two dual-unitary gates results in a dual-unitary gate acting on the composite Hilbert space~\cite{Borsi2022}, and the same result applies for hierachical dual-unitary gates.

\begin{figure}
\centering
\includegraphics[width = 0.45\textwidth]{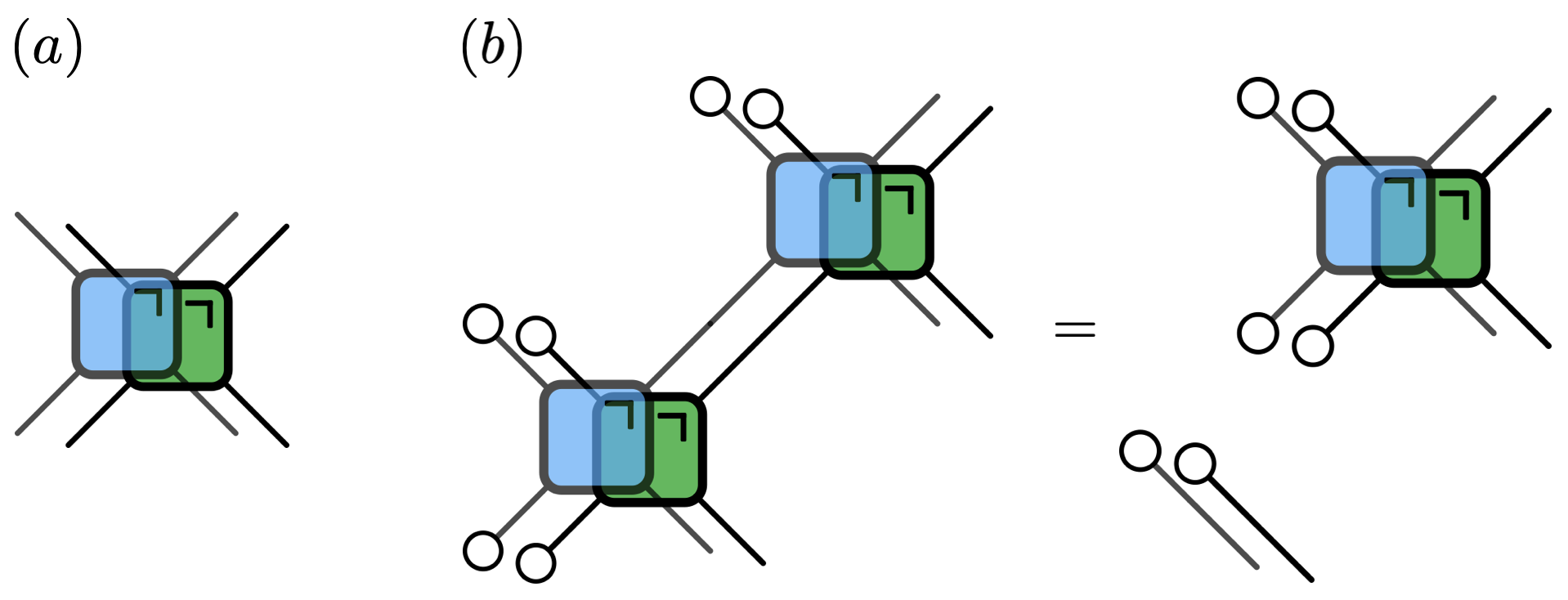}
\caption{Tensor product construction of $\mathcal{L}_2$ gates in composite dimension $q=q_1q_2$ using $\mathcal{L}_2$ gates in  smaller local dimensions $q_1$ and $q_2$. The colors (blue and green) are used to distinguish $\mathcal{L}_2$ gates in local dimension $q_1$ and $q_2$ respectively.}
\label{fig:L_2_ten_prod}
\end{figure}

Let $V_1$ and $V_2$ be $\mathcal{L}_2$ gates for local Hilbert space dimension $q_1$ and $q_2$ respectively. Since the Schmidt spectrum of both $V_1$ and $V_2$ is flat, the same immediately holds for their tensor product. 
Furthermore, the resulting gate is an $\mathcal{L}_2$ gate by construction, as shown in Fig.~(\ref{fig:L_2_ten_prod}).
Denoting the Schmidt values of $V_1$ as $\lambda$ and those of $V_2 = \mu$, the Schmidt values of $U$ all equal to $\lambda \mu $. Consequently, the Schmidt rank of $U$ is the product of the Schmidt ranks of $V_1$ and $V_2$; $\mathcal{R} = \mathcal{R}_1\mathcal{R}_2$.

In this way hierarchical gates can be constructed from gates acting on smaller Hilbert spaces whenever $q$ is a non-prime local dimension; $q=q_1q_2;q_1\geq q_2>1$. By taking an appropriate tensor product of the $\mathcal{L}_2$ gates in smaller local dimensions $q_1$ and $q_2$, we obtain $\mathcal{L}_2$  gates in a larger (non-prime) local dimension $q$. 
Furthermore, it is possible to take tensor products of gates in different levels of the hierarchy, where it similarly follows that the resulting gate will satisfy the equation for the highest level of the hierarcy, e.g. taking the tensor product of a dual-unitary gate and an \LL{2} gate returns an \LL{2} gate.
Taking the tensor product of gates belonging to different hierarchies we can obtain $\mathcal{L}_2$ gates of different Schmidt ranks and hence entanglement velocities in non-prime dimensions. This is shown in Table 1 and Table 2 for local dimensions $q=4$ and $6$, respectively. These results can also be understood by noting that the dynamics in these different Hilbert spaces effectively decouples.

\begin{table}
\caption{$\mathcal{L}_2$ gates in local dimension $q=4$ can be obtained from two-qubit $\mathcal{L}_2$ using the tensor product construction discussed in the main text.}
$$
\begin{array}{c|c|c|c|c}
\hline
V_1; q_1=2 & V_2; q_2=2   & \mathcal{R}_1 & \mathcal{R}_2 & \mathcal{R}_1\mathcal{R}_2 \\
\hline 
\hline
\mathbb{I} & \text{CNOT} & 1 & 2 & 2 \\
\text{CNOT} & \text{CNOT} & 2 & 2 & 4 \\
\text{CNOT} & \text{Dual unitary} & 2 & 4 & 8 \\
\text{Dual unitary} & \text{Dual unitary} & 4 & 4 & 16 \\
\hline
\end{array}
$$
\end{table}
\begin{table}
\caption{$\mathcal{L}_2$ gates in local dimension $q=6$ obtained from the tensor product of two-qubit ($q_1^2=4$) and two-qutrit ($q_2^2=9$) $\mathcal{L}_2$ gates.}

$$\begin{array}{c|c|c|c|c}
\hline
V_1; q_1=2 & V_2; q_2=3   & \mathcal{R}_1 & \mathcal{R}_2 & \mathcal{R}_1\mathcal{R}_2 \\
\hline 
\hline
 \text{CNOT} & \mathbb{I} & 2 & 1 & 2 \\
\mathbb{I} & \text{CNOT} & 1 & 3 & 3 \\
\text{Dual unitary}  & \mathbb{I} & 4 & 1 & 4 \\
\text{CNOT} & \text{CNOT} & 2 & 3 & 6 \\
\mathbb{I} & \text{Dual unitary} & 1 & 9 & 9 \\
\text{Dual unitary} & \text{CNOT} & 4 & 3 & 12 \\
\text{CNOT} & \text{Dual unitary} & 2 & 9 & 18 \\
\text{Dual unitary} & \text{Dual unitary} & 4 & 9 & 36 \\
\hline
\end{array}
$$
\end{table}
\subsubsection{Block-diagonal or controlled \LL{2} gates}
Here we discuss the construction of \LL{2} gates using block-diagonal or controlled unitaries for non-prime local dimension; $q=p_1p_2$, such that $\mathbb{C}^q \otimes \mathbb{C}^q \sim \mathbb{C}^{p_1} \otimes \mathbb{C}^{q p_2} \sim \mathbb{C}^{p_2} \otimes \mathbb{C}^{q p_1}$. Consider a block-diagonal unitary on $\mathbb{C}^{p_1} \otimes \mathbb{C}^{qp_2}$ constructed out of a set of $p_1$ number of one-site unitary matrices $V_k\in \mathbb{U}(qp_2)$ given by
\begin{align}
    U & = \sum_{k=1}^{p_1} \ket{k}\bra{k} \otimes V_k =\left(
     \begin{array}{c|c|c|c}
          V_1 & \cdots & \cdots & \cdots    \\
          \hline
          \cdots & V_2 & \cdots & \cdots    \\
          \hline
          \cdots & \cdots & \cdots & \cdots    \\
          \hline
          \cdots & \cdots & \cdots & V_{p_1}    \\
     \end{array}
     \right).
     \label{eq:blockdiagU}
\end{align}
As we are interested in unitary matrices for which the Schmidt spectrum is flat, this necessarily requires that the unitary matrices $V_k$ acting on $\mathbb{C}^{p_2} \otimes \mathbb{C}^{q}$ are maximally entangled i.e., the unitary matrices $V_k$ themselves have flat Schmidt decomposition. Therefore, the construction of block-diagonal unitaries of the form given in Eq.~\eqref{eq:blockdiagU} having a flat Schmidt spectrum reduces to finding maximally entangled unitary gates in $\mathbb{C}^{p_2} \otimes \mathbb{C}^{q}$. Similarly, the construction of block-diagonal unitary gates on $\mathbb{C}^{q_2} \otimes \mathbb{C}^{p_1q}$ with flat Schmidt spectrum reduces to finding maximally entangled unitary gates in $\mathbb{C}^{p_1} \otimes \mathbb{C}^{q}$. 

A well-known example of a maximally entangled operator on $\mathbb{C}^{q_1} \otimes \mathbb{C}^{q_2}$ is the quantum Fourier transform or Fourier gate \cite{Tyson2003}. The Fourier gate $F_{q_1 \times q_2}: \mathbb{C}^{q_1} \otimes \mathbb{C}^{q_2} \mapsto \mathbb{C}^{q_1} \otimes \mathbb{C}^{q_2}$ is defined as
\begin{equation}
  \bra{m,n} F_{q_1 \times q_2} \ket{m',n'}=\frac{1}{\sqrt{q_1q_2}} \omega^{((m-1)q_2+n-1)((m'-1)q_2+n'-1)},
\end{equation}
where $\omega=\exp[{2\pi i}/{(q_1q_2)}]$ and $m,m'$ take values from 1 to $q_1$ and $n,n'$ take values from 1 to $q_2$. If $q_1 \leq q_2$, then $F_{q_1 \times q_2}$ is maximally entangled if $q_1$ divides $q_2$ i.e., $q_2 \mod q_1=0$ \cite{Tyson2003}. Using maximally entangled unitaries such as the Fourier gate defined above, we illustrate the above construction of \LL{2} gates in local dimension $q=4$ in App.~\ref{app:blockdiag}.

\subsection{Entangling properties of $\mathcal{L}_2$ gates}

In this section we study the entangling properties of the $\mathcal{L}_2$ gates discussed in the previous section. A useful quantity to describe entangling property of a bipartite unitary operator is the entangling power \cite{Zanardi2001}. The entangling power of a bipartite unitary $U$ is defined as the average entanglement it creates when applied to product states sampled from a uniform distribution. Taking the linear entropy as a measure of entanglement, the entangling power of $U$, denoted as $EP(U)$, simplifies to the following expression:
\begin{widetext}
\begin{align}
EP(U) & =\frac{q^2}{q^2-1}\left[\left(1+\frac{1}{q^2}\right)-\frac{1}{q^4}\left[\Tr \left(\tilde{U}\tilde{U}^{\dagger} \tilde{U}\tilde{U}^{\dagger}\right)
+\Tr \left(U^{\Gamma}U^{\Gamma \dagger} U^{\Gamma}U^{\Gamma \dagger}\right) \right]\right] = \frac{q^2}{q^2-1}\left[\left(1+\frac{1}{q^2}\right)-\frac{1}{q^4}\sum_{i} \left( \lambda_i^4+\gamma_i^4 \right)\right],
\end{align}
\end{widetext}
where the $\lambda_i$ are singular values of $\tilde{U}$ and $\gamma_i$ are the singular values of $U^{\Gamma}$. As such, $\lambda_i$ are the Schmidt values of $U$ and it can be similarly shown that $\gamma_i$ are the Schmidt values of $US$ \cite{Bhargavi2017}, where $S$ is the swap gate. In diagrammatic notation,
\begin{equation}
EP(U)=\frac{q^2}{q^2-1}\left[\left(1+\frac{1}{q^2}\right)-\left( \vcenter{\hbox{\includegraphics[width = .1\columnwidth]{figs/bubble.png}}}\, + \vcenter{\hbox{\includegraphics[width = .16\columnwidth]{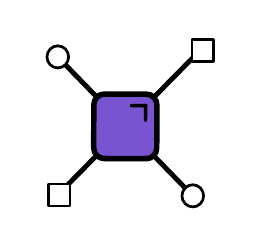}}}\, \right)\right].
\end{equation}
As $EP(U)$ involves four copies of $U$ the folded gate here represents two copies of $U \otimes U^*$, again with implicit $\alpha=2$. In our convention $0 \leq EP(U) \leq 1$, with the lower bound satisfied by local gates and the SWAP gate as these do not create any entanglement when applied to product states. Although local gates and the swap gate both have entangling power equal to zero, these gates can clearly give rise to highly different dynamics. In order to distinguish such gates we can consider the so-called gate-typicality~\cite{Bhargavi2017}:
\begin{equation}
GT(U)=\frac{q^2}{2(q^2-1)}\left[\left(1-\frac{1}{q^2}\right)-\frac{1}{q^4}\sum_{i} \left( \lambda_i^4-\gamma_i^4 \right)\right],
\end{equation}
In diagrammatic notation,
\begin{equation}
GT(U)=\frac{q^2}{2(q^2-1)}\left[\left(1-\frac{1}{q^2}\right)-\left( \vcenter{\hbox{\includegraphics[width = .1\columnwidth]{figs/bubble.png}}}\, - \vcenter{\hbox{\includegraphics[width = .16\columnwidth]{figs/bubble_2.png}}}\, \right)\right].
\end{equation}

For dual unitary gates we have that
\begin{align}\vcenter{\hbox{\includegraphics[width = .1\columnwidth]{figs/bubble.png}}}=1/q^2.
\end{align}
Therefore the entangling power and gate-typicality simplify to the following expressions
\begin{align}
EP(U) & =\frac{q^2}{q^2-1}\left[1-\vcenter{\hbox{\includegraphics[width = .16\columnwidth]{figs/bubble_2.png}}}\,\right], \\
GT(U) & =\frac{q^2}{2(q^2-1)}\left[\left(1-\frac{2}{q^2}\right)+\vcenter{\hbox{\includegraphics[width = .16\columnwidth]{figs/bubble_2.png}}}\, \right].
\end{align}
From the above expressions it follows that for dual unitary gates $EP(U)$ and $GT(U)$ satisfy a simple relation, 
$
GT(U)=1-EP(U)/2$.
It can similarly be shown that for T-dual gates
$
GT(U)=EP(U)/2
$. The upper bound for entangling power; $EP(U)=1$, is attained if and only if $U$ satisfies both dual unitarity and T-dual unitarity. Such unitary gates exist for all local dimensions $q>2$ \cite{Rather2022,Rather2020} , and are called perfect tensors or 2-unitary gates~\cite{Goyeneche2015}.

\begin{figure}
\centering
\includegraphics[scale=0.9]{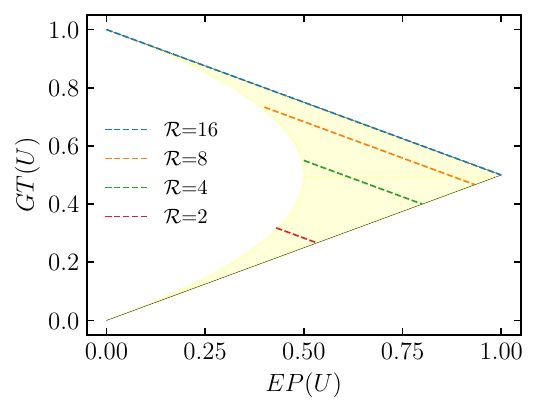}
\caption{$EP(U)$ and $GT(U)$ values of $\mathcal{L}_2$ gates lie on straight lines with slope equal to 1/2 and intercept determined by the Schmidt rank $\mathcal{R}$. This is shown here in local dimension $q=4$ for $\mathcal{R}=2,4,8$, and $16$. The $EP(U)$ and $GT(U)$ values for general bipartite unitary gates can lie on and inside the light-shaded yellow region.}
\label{fig:ep_gt_L_2}
\end{figure}

As $\mathcal{L}_2$ gates have a flat Schmidt spectrum, the expressions for entangling power and gate-typicality simplify as in the dual unitary case. Let $\mathcal{R}$ be the Schmidt rank of a given $\mathcal{L}_2$ gate. Then due to the flatness of the Schmidt spectrum it follows that
\begin{align}
\vcenter{\hbox{\includegraphics[width = .1\columnwidth]{figs/bubble.png}}}=\frac{1}{\mathcal{R}}.
\end{align}
This expression leads to the following linear relation between entangling power and gate-typicality for $\mathcal{L}_2$ gates
\begin{equation}
    GT(U)+\frac{1}{2}EP(U)=\frac{1-\frac{1}{\mathcal{R}}}{1-\frac{1}{q^2}}.
\end{equation}
The $EP(U)$ and $GT(U)$ values of an $\mathcal{L}_2$ gate always lie on a straight line with slope equal to 1/2, whose intercept is determined by its Schmidt rank. This result is illustrated in Fig.~\ref{fig:ep_gt_L_2} for local dimension $q=4$.

For a given local dimension $q$ and Schmidt rank $\mathcal{R}$, the maximum entangling power for $\mathcal{L}_2$ gates is achieved by gates that also satisfy T-duality. For these special $\mathcal{L}_2$ gates, 
\begin{align}
\vcenter{\hbox{\includegraphics[width = .16\columnwidth]{figs/bubble_2.png}}}=\frac{1}{q^2}\,.
\end{align}
The maximum entangling power for these $\mathcal{L}_2$ gates, denoted as $EP^{\text{max}}(\mathcal{R})$, is given by
$$ EP^{\text{max}}(\mathcal{R})=\frac{q^2}{q^2-1}\left(1-\frac{1}{\mathcal{R}} \right).$$
From the above equation it is easy to see that $EP^{\text{max}}(\mathcal{R})=1$ (the maximum value in our convention) is achieved when $\mathcal{R}=q^2$, corresponding to perfect tensors.

\begin{figure}
\centering
\includegraphics[scale=0.9]{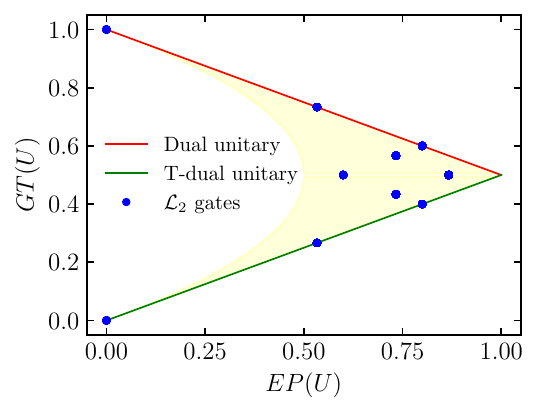}
\caption{$EP(U)$ and $GT(U)$ values of $\mathcal{L}_2$ gates: The discrete set of points shown in blue correspond to $\mathcal{L}_2$ gates obtained by taking the tensor product of two-qubit permutations. The $EP(U)$ and $GT(U)$ values of dual unitary and T-dual gates lie on straight lines shown in red and green, respectively.}
\label{fig:L_2_Perm_q_4}
\end{figure}

\begin{figure}
\centering
\includegraphics[scale=0.9]{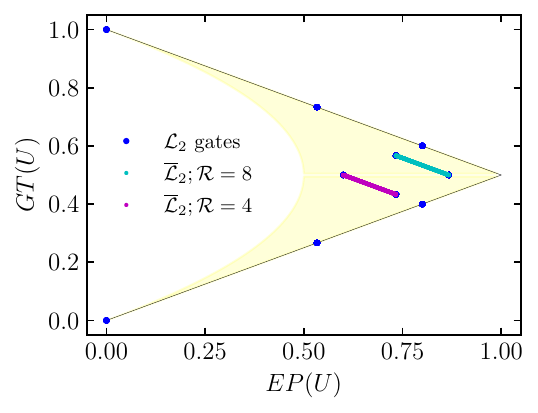}
\caption{$EP(U)$ and $GT(U)$ values of continuous families of \LL{2} gates. \LL{2} gates with $\mathcal{R}=4$ (shown in cyan) are obtained by taking tensor product of two-qubit dual unitaries with the identity matrix. Similarly, \LL{2} gates with $\mathcal{R}=8$ (shown in purple)  are obtained by taking tensor product of two-qubit dual unitaries with the two-qubit CNOT gate. The number of realizations of random two-qubit dual unitaries used in both cases is $10^2$.}
\label{fig:L_2_continuous_q_4}
\end{figure}

Using the tensor product construction discussed in the previous section, we obtain $\mathcal{L}_2$ permutation gates in $q=4$ from two-qubit permutation gates. There are $4!=24$ total number of two-qubit permutation gates and the number of $\mathcal{L}_2$ permutation gates in $q=4$ obtained is $24^2=576$. The entangling power and gate typicality values of these permutation gates of size 16 is shown in Fig.~\ref{fig:L_2_Perm_q_4}. Note that there are only $10$ distinct points on the $EP(U)$-$GT(U)$ plane, indicating that several $\mathcal{L}_2$ permutation gates have the same $EP(U)$ and $GT(U)$ values. These $\mathcal{L}_2$ permutation gates also include dual-unitary permutation gates, corresponding to the three points lying on the red line in Fig.~\ref{fig:L_2_Perm_q_4}. All other points correspond to \LL{2} permutation gates. Interestingly, there are $\mathcal{L}_2$ permutation gates that are neither dual nor T-dual.

Using the tensor product construction we construct two families of \LL{2} gates whose entangling power and gate-typicality take continuous values. One of the families is obtained by taking the tensor product of two-qubit dual unitaries and the identity matrix. The two-qubit dual unitaries we consider are of the form $S\mathcal{D}$, where $S$ is the two-qubit SWAP gate and $\mathcal{D}$ is a diagonal unitary. The resulting \LL{2} gates have Schmidt rank 4 and the corresponding $EP(U)$, $GT(U)$ values are shown in Fig.~\ref{fig:L_2_continuous_q_4}. Similarly, the other family is obtained by taking the tensor product of two-qubit dual unitaries and the CNOT gate. The resulting \LL{2} gates have Schmidt rank 8 and the corresponding $EP(U)$, $GT(U)$ values are shown in Fig.~\ref{fig:L_2_continuous_q_4}. In both cases we take $10^2$ realizations of random two-qubit dual unitaries.

\subsection{Possible Schmidt ranks for \LL{2} gates} 

The constructions of \LL{2} gates discussed above always lead to \LL{2} gates whose Schmidt rank $\mathcal{R}$ divides $q^2$, i.e., $q^2 \mod \mathcal{R}=0$ for a given local dimension $q$. Based on these results we conjecture the following: 
{\em For a given local dimension $q$ the only allowed Schmidt ranks for entangling \LL{2} gates are the non-trivial factors of $q^2$.}

For prime-dimensional qudits having local dimension $q$, the above conjecture states that that the only allowed Schmidt rank for entangling \LL{2} gates is $q$. If this conjecture holds this implies that the only entanglement velocity for \LL{2} circuits composed of prime-dimensional qudits is given by
\begin{align}
    v_E=\frac{\log q}{2 \log q}=\frac{1}{2}\,.
\end{align}
Below we discuss some results that are in support of the above conjecture.

For local dimension $q=2$ (qubit case) all \LL{2} gates are known~\cite{Yu2024}. All entangling \LL{2} are equivalent (up to multiplication by single-qubit unitaries) to the two-qubit CNOT gate, for which $\mathcal{R}=2$. As such, the above conjecture holds trivially in the qubit case. 
For odd local dimensions we have already shown above that Schmidt-rank two \LL{2} gates do not exist, consistent with the conjecture. 
A subset of unitary gates is given by permutation matrices, for which the dynamics reduces to a purely classical dynamics. A proof for our conjecture is possible for arbitrary values of $q$ when the \LL{2} gates are restricted to be permutations, as shown in Appendix~\ref{app:Schmidt}.

\subsection{Schmidt decomposition of $\mathcal{L}_2$ gates}
To conclude this section, we note an additional restriction on the Schmidt decomposition of general $\mathcal{L}_2$ gates implicit in the results of Ref.~\cite{Foligno2023}. While we have already discussed the flatness of the Schmidt spectrum, the results of Ref.~\cite{Foligno2023} additionally imply that all matrices $X_i$ and $Y_i$ can be chosen to be (proportional to) unitary matrices, such that we can write
\begin{align}
U = \frac{q}{\sqrt{\R}} \sum_{i=1}^{\R} X_i \otimes Y_i,
\end{align} 
with both $X_i$ and $Y_i \in \mathbb{U}(q)$ (up to an overall prefactor $\sqrt{q}$). Note however that, due to the degenerate Schmidt spectrum, this decomposition is not unique. 
From the Schmidt decomposition of $U$, it follows that
\begin{equation}
    \tilde{U}\tilde{U}^{\dagger}=\frac{q^2}{\R}\sum_{i=1}^{\R} \ket{X_i}\bra{X_i^*},
\end{equation}
where $\ket{X_i} \in \mathbb{C}^q \otimes \mathbb{C}^q$ is vectorization of the operator basis; $\ket{X_i}=\sum_{k,l=1}^q \bra{k}X_i\ket{l} \;\ket{l}\otimes \ket{k}$, and `*' denotes the complex conjugation in the computational basis. As shown in Ref.~\cite{Foligno2023}, the corresponding orthogonal projectors satisfy
\begin{equation}
    \tr_2(\ket{X_i}\bra{X_i^*})=\tr_1(\ket{X_i}\bra{X_i^*})=\frac{1}{q}\mathbbm{1}_q,
\end{equation}
where $\tr_i$ denotes the partial trace with respect to the $i$-th qudit. From the above equation we infer that bipartite pure states $\ket{X_i}$ have maximally mixed single-qudit marginals or, equivalently, $\ket{X_i}$ are maximally entangled states. As maximally entangled states are isomorphic to unitary matrices \cite{Zanardi2001} this means that the basis operators $X_i$ appearing in the Schmidt decompositon of  $\mathcal{L}_2$ gates can always chosen to be unitary (up to an overall constant). This argument directly extends to the set of $Y_i$ matrices. A similar result for a particular class of dual unitaries obtained from diagonal unitaries was shown in Ref.~\cite{Brahmachari2022}, for which both operator bases can be chosen to be unitary.

\section{Conclusion and Outlook}
\label{sec:conc}

We have investigated entanglement and operator growth in a class of solvable models of chaotic dynamics by two complementary means. First, by recourse to an effective coarse-grained description, EMT, for whose central quantity, the ELT, we give an exact expression. Second, by direct investigation of the characteristic dynamical probes. Our results reveal the rich physics of hierarchically generalized dual-unitary circuits, displaying behavior expected from generic systems, such as non-maximal entanglement growth and information scrambling, while retaining some of the pathologies of dual-unitary circuits, such as maximum-velocity information transport. \LL{2} circuits also saturate general bounds on entanglement growth and have a quantized entanglement velocity, further indicating their special place in many-body dynamics. We link this saturation to the observation that in these models the local entanglement production is directly related to the information transport. Furthermore, the higher levels of the hierarchy behave in new and unexpected ways, showing coexistence of lines and areas of correlations in spacetime, as well as a transition in temporal entanglement scaling.

A special subclass of hierarchical gates can be constructed out of complex Hadamard matrices, presenting both a general way of constructing hierarchical gates for arbitrary local Hilbert space dimension and a set of gates with additional solvability properties. Dual-unitary gates constructed out of complex Hadamard matrices have played an important role in the study of e.g. kicked spin chains, measurement-based quantum computing, and deep thermalization, and we expect that the constructions in this work can play a similar role for hierarchical gates -- especially so since the scarcity of constructions of appropriate gates presented a major obstruction towards the (numerical) investigation of hierarchical models so far.

The exact result for the ELT enabled us to perform non-trivial checks on the validity of EMT in a class of microscopic Floquet lattice models, confirming the predictions of the effective theory. It would be desirable to shed further light on the connection between the ELT and microscopics, as well as to have efficient means of extracting the information from numerics or experiment -- where the considered probes of entanglement and operator dynamics are directly accessible in current quantum computing setups. The connection between hierarchical dual-unitary gates and kinetically constrained models, e.g. the quantum East model, would also be interesting to explore further.

Many open questions remain concerning the physics of higher levels of the hierarchy of generalized dual-unitary circuits. We have shown that these retain exact solvability above a threshold velocity, but the dynamics below this threshold calls for the application of different techniques in general. It seems conceivable that the higher the level of the hierarchy the more generic the possible behavior, however it is not yet clear which restrictions the hierarchical conditions place on the dynamics in the inaccessible region.  A possible remedy, utilized in the present work, is to relax the hierarchical condition to a single light-cone direction, reducing the area of solvability but increasing the space of gates, or impose additional constraints on the gates through the construction out of complex Hadamard matrices. Nevertheless, examples of \LL{k\geq3} gates with non-pathological properties would be desirable.

In the final stages of this work, Ref.~\cite{Foligno2023} appeared online, which similarly calculates the entanglement line tension and discusses entanglement growth and operator spreading in hierarchical dual-unitary circuits. Where our works overlap they agree. Additionally, Ref.~\cite{Foligno2023} presents a proof of the flatness of the Schmidt spectrum of \LL{2} gates, a result we independently conjectured based on numerical observations and use throughout this work. 

\begin{acknowledgements}
    We acknowledge useful discussions with Felix Fritzsch, Marko Ljubotina, and Xhek Turkeshi. We are grateful to the authors of Ref.~\cite{Foligno2023} for discussing their results with us before publication. The numerical simulations presented in Sec.~\ref{sec:ent_growth} were performed using the \textsc{ITensor} library~\cite{itensor}.
\end{acknowledgements}

\appendix

\section{Parametrization of two-qubit gates}
\label{app:parametrization}

In this Appendix we collect the parametrizations of \LL{2} and \LL{3} gates in local dimension $q=2$. These parametrizations are derived in Ref.~\cite{Yu2024}.
Non-trivial qubit \LL{2} gates are of the form
\begin{align}
    U = (u_1\otimes u_2) e^{i\frac{\pi}{4}ZZ},
\end{align}
where the local gates $u_i$ are parametrized as
\begin{equation}
    u_i = \exp\left(i r_i(\sin\theta_i\cos\phi_i \sigma_x + \sin\theta_i\sin\phi_i \sigma_y + \cos\theta_i\sigma_z)\right),
\end{equation}
and the parameters satisfy the relation
\begin{equation}
    \sqrt{2}\sin r_i \sin\theta_i = \pm 1.
\end{equation}
Qubit \LL{3} gates are of the form
\begin{align}
    U &= (v_1\otimes v_2)e^{-iJ_z ZZ},  &v_i &= \cos\phi_i X +\sin\phi_i Y, \label{eq:l3_qubit}\\
    J_z&\in \left[0,\frac{\pi}{4}\right],  &\phi_i&\in [0,2\pi]. \nonumber
\end{align}

\section{Overlaps of leading eigenvectors}
\label{app:overlaps}

In the following we present the computation of the overlaps of the leading eigenvectors of the LCTM for \LL{2} circuits. We prove that the overlap matrix is a Hankel matrix. The arguments can be generalized to \LL{k} circuits in a straightforward manner.

Unitarity enables the computation of the overlap between a staircase vector and the trivial leading eigenvector
\begin{equation}
    \rbraket{\circ_n}{s_n} = \rbraket{\Tilde{s}_n}{\Box_n} = \vcenter{\hbox{\includegraphics[width = .3\columnwidth]{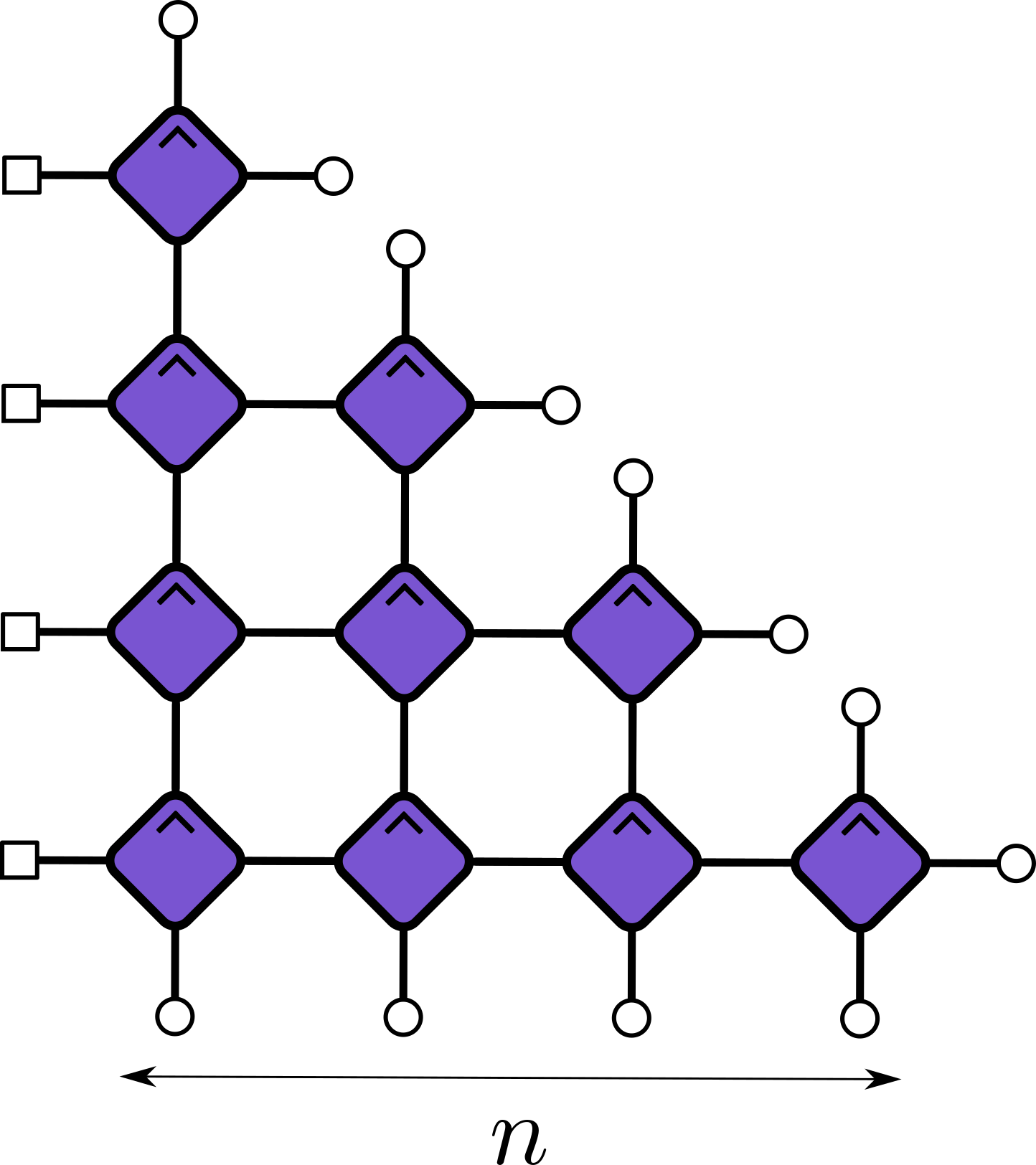}}} = \frac{1}{q^n}.
\end{equation}
Consider now the general overlap $\rbraket{\ell_i}{r_j}=\rbraket{\Tilde{s}_i\circ_{n-i}}{\Box_{n-j}s_j}$ and take $i\leq j$ without loss of generality. For $i+j\leq n$ the overlaps factorize
\begin{equation}
    \rbraket{\ell_i}{r_j} = \rbraket{\Tilde{s}_i}{\Box_i} \rbraket{\circ_{n-i-j}}{\Box_{n-i-j}}  \rbraket{\circ_j}{s_j} = \frac{1}{q^n}.
\end{equation}

To treat the case $i+j>n$ where the staircases overlap, we proceed in two steps. First, the entries of the last column of the overlap matrix are computed, before the Hankel property, that fixes all remaining entries, is proved. The entries of the last column are
\begin{equation}
    \rbraket{\ell_k}{r_n} =  \vcenter{\hbox{\includegraphics[width = .3\columnwidth]{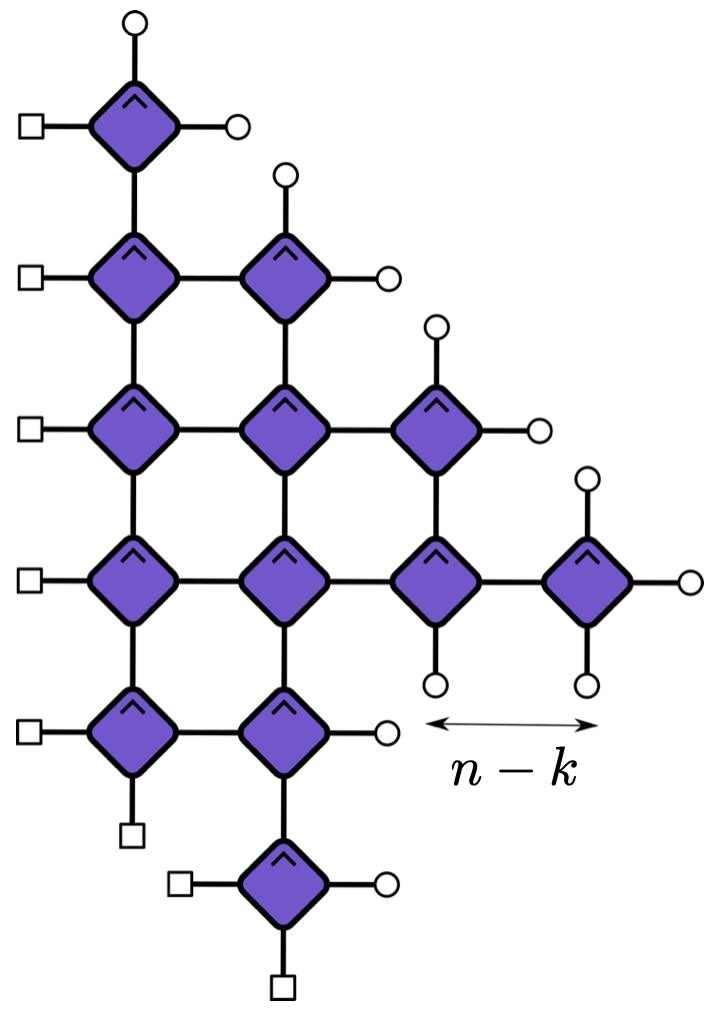}}} = \frac{b_1^k}{q^n}.
\end{equation}
The Hankel property is expressed as
\begin{equation}
    \rbraket{\ell_i}{r_j} = \rbraket{\ell_{i+r}}{r_{j-r}}, \quad k\leq j,\quad r\leq j-k.
\end{equation}
Consider any $\rbraket{\ell_i}{r_j}$ where $i+j>n$. Unitarity enables the reduction to a diagram of the following form
\begin{align}
    \rbraket{\ell_i}{r_j} = \frac{1}{q^n}\,\vcenter{\hbox{\includegraphics[width = .3\columnwidth]{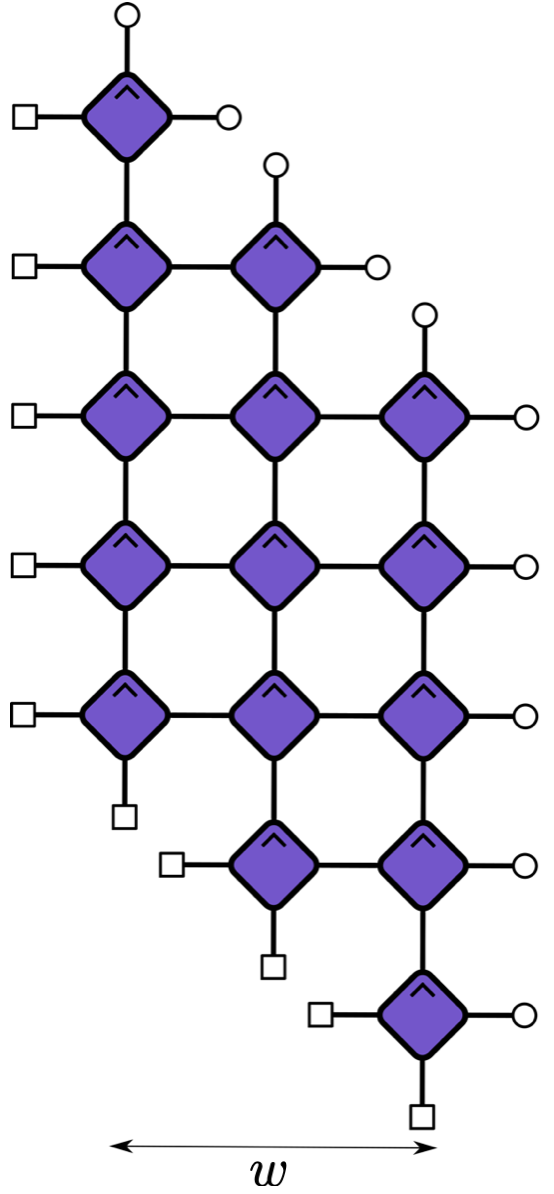}}} = \frac{b_1^w}{q^n}
\end{align}
Due to the \LL{2} property, its value depends only on the width $w$ of the diagram. Inspection of the diagram reveals $w=i+j-n$. Hence, the width is invariant under the transformation $i\rightarrow i+r,j\rightarrow j-r$. This proves the Hankel property.

\section{Tripartite information}
\label{app:tripartite}

The tripartite information $I^{(3)}$ is a quantity derived from the time-evolution operator to quantify information scrambling in quantum many-body systems. Hereby, $-I^{(3)}$ corresponds to the amount of information that cannot be recovered from local measurements on the output subsystem~\cite{Hosur2016}. The R\'{e}nyi-2 tripartite information is represented diagrammatically through Eq.~\eqref{eq:z_alpha} and through~\cite{Bertini2020a}
\begin{equation}
   \Tilde{Z}_2(m,n) = \vcenter{\hbox{\includegraphics[height = .4\columnwidth]{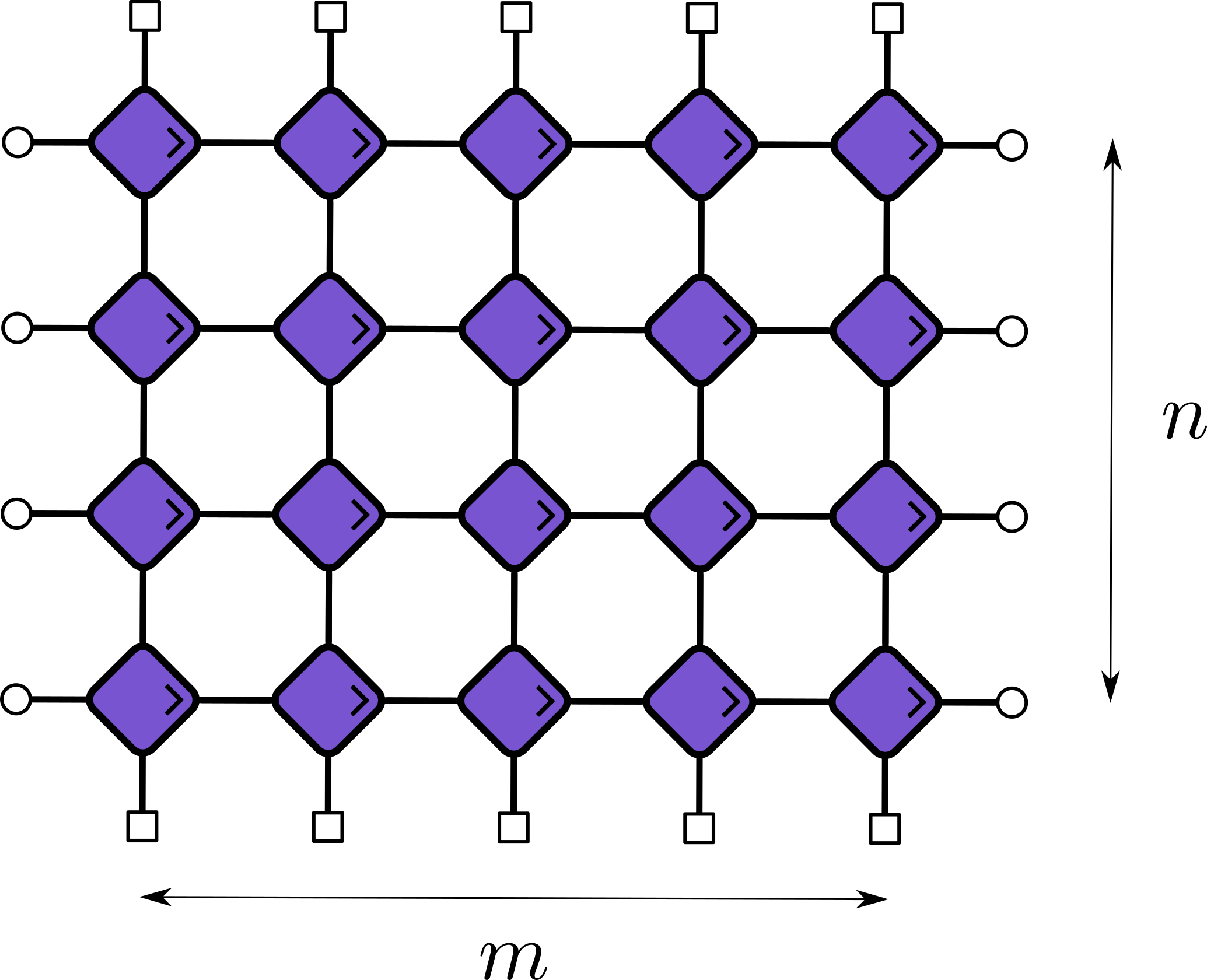}}}, \label{eq:ti_diag}
\end{equation}
as
\begin{equation}
    I^{(3)}(x,t) = \log\left(q^{m+n}Z_2(m,n)\right) + \log\left(\Tilde{Z}_2(m,n)\right).
\end{equation}
For chaotic \LL{2} circuits we assume that the transfer matrix generating Eq.~\eqref{eq:ti_diag} possesses no leading eigenvectors beyond the trivial one fixed by unitarity. Focusing on the right light-cone edge this implies
\begin{equation}
    \lim_{m\rightarrow\infty}\Tilde{Z}_2(m,n) = q^{-2n}. 
\end{equation}
This result, together with Eq.~\eqref{eq:l2_diagram_te_opent}, yields Eq.~\eqref{eq:ti_l2}.

\section{Finite butterfly velocity with vanishing subleading eigenvalues}
\label{app:butterfly}

In the following we argue that it is possible to have a circuit with non-zero finite butterfly velocity $v_B>0$ where all non-trivial eigenvalues of the LCTM are zero. The necessary condition is that the size of the largest Jordan block grows sufficiently fast with $n$. Assume the largest Jordan block grows linearly with the width $n$ of the LCTM, $m\sim\alpha n$. Asymptotically we have
\begin{equation}
    v_B = \lim_{n\rightarrow\infty}\frac{m-n}{m+n} = \frac{\alpha-1}{\alpha+1}.
\end{equation}
Hence, if $\alpha>1$ the butterfly velocity is non-zero.

For the class of qubit \LL{3} gates defined by Eq.~\eqref{eq:l3_qubit} we numerically observe $m=n+1$ leading to $v_B=0$. Upon dressing one of the legs with a generic unitary, we observe $m=2n$ consistent with $v_B=1/3$.

\section{Illustration of the block-diagonal construction in local dimension $q=4$}
\label{app:blockdiag}

\begin{figure}[t]
    \centering
    \includegraphics[width = .8\columnwidth]{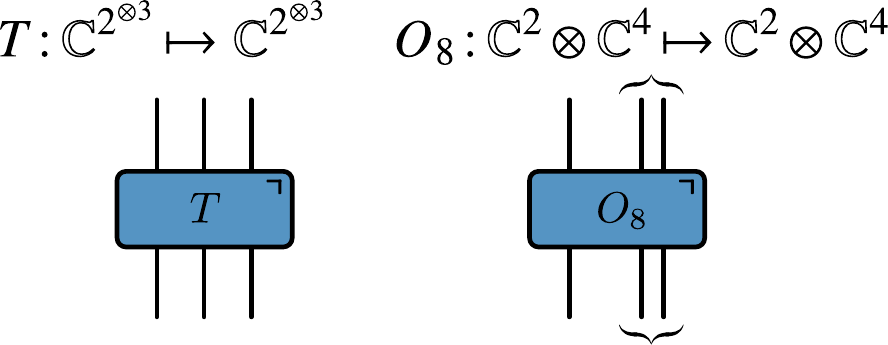}
    \caption{A six-qubit perfect tensor $T$ with three input and three output legs is reshaped into a tensor $O_8$ which has two input and two output legs. The two legs of $T$ can be combined to return a composite Hilbert space with dimension $4$.}
    \label{fig:O8mat}
\end{figure}

For local dimension $q=4=2\times2$, consider the following unitary gate
\begin{align}
    U & = \sum_{k=1}^{2} \ket{k}\bra{k} \otimes F_{2 \times 4}  =\left(
     \begin{array}{c|c}
          F_{2 \times 4} & {\bf 0} \\
          \hline
          {\bf 0} &   F_{2 \times 4}
     \end{array}
     \right).
\end{align}
The matrix form of $F_{2\times 4}$ is given by $$F_{2\times 4}=\frac{1}{\sqrt{8}} \left(\begin{array}{cccc|cccc}
1  &  1  &  1  &  1  &  1  &  1  &  1  &  1 \\
        1  &  \omega &  \omega^2 &  \omega^3  &  \omega^4  &  \omega^5  &  \omega^6 &  \omega^7 \\
        1  &  \omega^2 &  \omega^4  &  \omega^6 &  1  &  \omega^2 &  \omega^4  &  \omega^6 \\
        1  &  \omega^3  &  \omega^6 &  \omega &  \omega^4  &  \omega^7 &  \omega^2 &  \omega^5 \\
\hline
        1  &  \omega^4  &  1  &  \omega^4  &  1  &  \omega^4  &  1  &  \omega^4 \\
        1  &  \omega^5  &  \omega^2 &  \omega^7 &  \omega^4  &  \omega &  \omega^6 &  \omega^3 \\
        1  &  \omega^6 &  \omega^4  &  \omega^2 &  1  &  \omega^6 &  \omega^4  &   \omega^2 \\
        1  &  \omega^7 &  \omega^6 &  \omega^5  &  \omega^4  &  \omega^3  &  \omega^2 &  \omega
\end{array}
\right),
$$
where $\omega=\exp(2 \pi i/8)$. It can be easily checked that $F_{2\times 4}$ has four orthogonal $4 \times 4$ blocks each having norm equal to $\sqrt{2}$, and thus has all four Schmidt values equal to $\sqrt{2}$. Therefore the above unitary of size 16 constructed from $F_{2\times 4}$ has Schmidt rank $\R=4$ and has a flat Schmidt spectrum by construction. It can again be checked by direct calculation that this presents an \LL{2} gate. One can obtain an \LL{2} gate with Schmidt rank eight by permuting the rows and columns of one of the $F_{2 \times 3}$ blocks appearing in above equation in such a way that the resultant $8\times 8$ unitary matrix has orthonormal blocks with that of $F_{2 \times 4}$. One of the examples is the following gate
\begin{equation}
    U'= \left(
     \begin{array}{c|c}
          F_{2 \times 4} & {\bf 0} \\
          \hline
          {\bf 0} &  F_{2 \times 4}'
     \end{array}
     \right),
\end{equation}
where $F_{2 \times 4}'=X^2 F_{2 \times 4} X^2$ and $X$ is the shift operator on $\mathbb{C}^8$; $X \ket{k}=\ket{(k+1)\bmod 8}$. The unitary $U'$ has Schmidt rank $\R=8$ with flat Schmidt spectrum, $\lambda_{1}=\lambda_{2}=\cdots=\lambda_{8}=\sqrt{2}$, and also satisfies the \LL{2} conditions.

Another way to obtain maximally entangled unitaries on $\mathbb{C}^2 \otimes \mathbb{C}^4$ is using the six-qubit perfect tensor. By definition, a perfect tensor \cite{Pastawski2015} defines an isometry under arbitrary partitioning of its indices into two disjoint sets. 
The six-qubit perfect tensor defines a special pure state of six qubits that has maximal entanglement with respect to all bipartitions \cite{Goyeneche2015,Pastawski2015}. Such states, also known as absolutely maximally entangled (AME) states, have useful applications in quantum error correction, quantum teleportation, and quantum secret sharing \cite{Helwig2012}. A six-qubit AME state can be obtained by the superposition of the logical states of the well-known five-qubit error correcting code \cite{Laflamme1996}. A convenient representation of the six-qubit perfect tensor exists in terms of a real Hadamard matrix of size 8 given by \cite{Goyeneche2015}
\begin{equation}
    O_8=\frac{1}{\sqrt{8}}
    \left(\begin{array}{cccc|cccc}
    - & - & - &+ & - &+ &+ &+ \\
 - & - & - &+ &+ & - & - & - \\
 - & - &+ & - & - &+ & - & - \\
+ &+ & - &+ & - &+ & - & - \\
 \hline
 - &+ & - & - & - & - &+ & - \\
+ & - &+ &+ & - & - &+ & - \\
+ & - & - & - &+ &+ &+ & - \\
+ & - & - & - & - & - & - &+
    \end{array}
    \right),
\end{equation}
where `$+$' denotes $+1$ and `$-$' denotes $-1$. The above representation is given in the three-qubit computational basis; $\left\lbrace \ket{000},\ket{001},\cdots,\ket{110},\ket{111}\right\rbrace$. For example, $\bra{000}O_8\ket{000}=-1/\sqrt{8},\bra{000}O_8\ket{001}=-1/\sqrt{8},\cdots,\bra{111}O_8\ket{111}=1/\sqrt{8}$. 

We now consider the mapping $O_8: \mathbb{C}^2 \otimes \mathbb{C}^4 \mapsto \mathbb{C}^2 \otimes \mathbb{C}^4$, as indicated by partitioning the $O_8$ matrix into $4 \times 4$ blocks. This is also illustrated in Fig.~\ref{fig:O8mat} in which we group together a pair of both input and output legs of a six-qubit perfect tensor resulting into a bipartite unitary on $\mathbb{C}^2 \otimes \mathbb{C}^4$. One can easily check that the four $4 \times 4$ blocks in $O_8$ are orthogonal and its Schmidt spectrum is flat; $\lambda_1=\lambda_3=\lambda_3=\lambda_4=2$.

In order to obtain \LL{2} gates using the $O_8$ matrix, we consider the block-diagonal unitary $U \in \mathbb{U}(16)$ given by
\begin{widetext}
\begin{align}
    U & =O_8\oplus O_8 =\frac{1}{\sqrt{8}}\left(\begin{array}{cccc|cccc|cccc|cccc}
    -& -& -& + & -& + & + & + & . & . & . & . & . & . & . & . \\
 -& -& -& + & + & -& -& -& . & . & . & . & . & . & . & .\\
 -& -& + & -& -& + & -& -& . & . & . & . & . & . & . & . \\
 + & + & -& + & -& + & -& -& . & . & . & . & . & . & . & .\\
 \hline
 -& + & -& -& -& -& + & -& . & . & . & . & . & . & . & . \\
 + & -& + & + & -& -& + & -& . & . & . & . & . & . & . & .\\
 + & -& -& -& + & + & + & -& . & . & . & . & . & . & . & . \\
 + & -& -& -& -& -& -& + & . & . & . & . & . & . & . & . \\
 \hline
 . & . & . & . & . & . & . & . & -& -& -& + & -& + & + & + \\
 . & . & . & . & . & . & . & . & -& -& -& + & + & -& -& -\\
 . & . & . & . & . & . & . & . & -& -& + & -& -& + & -& -\\
  . & . & . & . & . & . & . & . &+ & + & -& + & -& + & -& -\\
 \hline
  . & . & . & . & . & . & . & . &-& + & -& -& -& -& + & -\\
  . & . & . & . & . & . & . & . &+ & -& + & + & -& -& + & -\\
 . & . & . & . & . & . & . & . & + & -& -& -& + & + & + & -\\
  . & . & . & . & . & . & . & . &+ & -& -& -& -& -& -& + 
 \end{array}
    \right).
\end{align}
\end{widetext}
This unitary gate clearly has Schmidt rank four and has flat Schmidt spectrum; $\lambda_1=\cdots=\lambda_4=2$. The maximal Schmidt rank possible for a block-diagonal unitary matrix consisting of two $8 \times 8$ unitary matrices is eight. In order to obtain Schmidt rank eight, we permute the rows of $O_8$ in such a way that the four $4 \times 4$ blocks of the resultant orthogonal matrix are orthonormal to those of $O_8$. One of the examples of such Schmidt rank eight is given by
\begin{align}
   U' & =\left(\begin{array}{c|c}
    O_8 & {\bf 0} \\
    \hline
     {\bf 0}   & O_8'
   \end{array}
    \right), 
    \label{eq:schmidtrank2q_gate}
\end{align}
where $O_8'=X^2O_8$ and $X$ is the shift operator on $\mathbb{C}^8$; $X \ket{k}=\ket{(k+1)\mod 8}$. This unitary gate has eight orthogonal $4 \times 4$ blocks and consequently has Schmidt rank eight with $\lambda_1=\lambda_2=\cdots=\lambda_8=\sqrt{2}$. Both unitary gates obtained from $O_8$ above belong to \LL{2}.

\section{Possible Schmidt ranks for \LL{2} permutation gates}
\label{app:Schmidt}
Permutation gates are the simplest unitary matrices that have only one non-zero element, equal to 1, in any row and column. Here we focus on bipartite permutation gates of size $q^2$ acting on $\mathbb{C}^q \otimes \mathbb{C}^q$. Consider a permutation gate $P$ whose action on the computational basis states is given by
\[
P \ket{i} \otimes \ket{j}=\ket{a_{ij}} \otimes \ket{b_{ij}};i,j=1,2,\cdots,q,
\]
where $a_{ij}$ and $b_{ij}$ are functions of $i$ and $j$. Another useful representation of bipartite permutation gates is so-called block form given by,
\begin{align}
P & =\sum_{i,j=1}^{q}\ket{i}\bra{j} \otimes p_{ij}, \nonumber \\
 &= \left(\begin{array}{c|c|c|c}
 p_{11} & p_{12} & \cdots & p_{1q} \\
 \hline
 p_{21} & p_{22} & \cdots & p_{2q} \\
 \hline
 \vdots & \vdots & \vdots & \vdots \\
 \hline
  p_{q1} & p_{q2} & \cdots & p_{qq}
  \end{array}
  \right),
  \label{eq:P_block_form}
  \end{align}
where the $p_{ij}$ are $q \times q$ matrices. The block-form of a bipartite unitary gate is useful for determining its Schmidt rank; the Schmidt rank is equal to the number of linearly independent blocks. From the unitarity of $P$ it follows that:
\begin{itemize}
\item
$\sum_{i,j=1}^q ||p_{ij}||^2=q^2$.
\item
$||p_{ij}||^2 \in \left\lbrace 0,1,\cdots,q \right\rbrace$.
 Note that $||p_{ij}||^2$ is the number of non-zero entries in a given block $p_{ij}$ and can be at most $q$.
\end{itemize} 
For general operators $w_1$ and $w_2$ acting on $\mathbb{C}^q$, the following identity holds
\begin{align}
\Tilde{w_1 \otimes w_2}=\ket{w_1}\bra{w_2^*},
\end{align}
where $\ket{w_1}$ ($\ket{w_2}$) is the vectorization of $w_1 \,(w_2)$ and `*' refers to complex conjugation in the computational basis. Using the above identity in Eq.~(\ref{eq:P_block_form}), we obtain
\begin{equation}
\tilde{P}=\sum_{i,j} \ket{i}\ket{j^*} \bra{p_{ij}^*}=\sum_{i,j} \ket{i}\ket{j} \bra{p_{ij}},
\end{equation}
where in the last step we have used $\ket{j^*}=\ket{j}$ and $\ket{p_{ij}^*}=\ket{p_{ij}}$ as $p_{ij}$ are real.
The Schmidt values and Schmidt rank of $P$ are obtained from the singular value decomposition of $\tilde{P}$. We now examine the block-form of hierarchical permutation gates.

\emph{Dual-unitary permutation gates}. 
For a dual-unitary permutation gate $P$ we have that $\tilde{P}$ is also unitary, such that the corresponding blocks satisfy the following condition
\begin{align}
\tr\left(p_{ij}^T p_{i'j'}\right)=\delta_{ii'}\delta_{jj'},
\end{align}
i.e., all $p_{ij}$ are orthogonal to each other and each $p_{ij}$ contain exactly one non-zero entry equal to 1. Due to this property the block-form given in Eq.~\eqref{eq:P_block_form} directly provides the Schmidt decomposition.

\emph{\LL{2} permutation gates}.  
A common property of \LL{2} gates and DU gates is that both have a flat Schmidt spectrum. However, for \LL{2} gates the Schmidt rank cannot be maximal, unlike for DU gates. This implies that the non-zero blocks in \LL{2} permutation gates may or may not be linearly independent. If all the non-zero blocks are linearly independent then the block form provides a Schmidt decomposition. However, if all the blocks  are not linearly independent, i.e., some non-zero blocks repeat, then the block-form given in Eq.~\eqref{eq:P_block_form} itself does not provide Schmidt decomposition for  \LL{2} permutation gates. 
However, Schmidt decomposition for such \LL{2} permutation gates can be obtained by summing over the computational basis states that are associated with a given repeated block. Below we consider these two cases separately.

\emph{All non-zero blocks are linearly independent}. 
Let $\mathcal{R}$ be the Schmidt rank of  given \LL{2} permutation gate. Therefore, the number of non-zero blocks appearing in the block form is equal to $\mathcal{R}$. Due to unitarity the number of non-zero blocks cannot be less than $q$ and the Schmidt rank of these permutation gates satisfies $\mathcal{R}\geq q$. In this case the Schmidt decomposition of \LL{2} permutation gate is given by
\[
P=\sum_{i,j} ||p_{ij}|| \, \ket{i}\bra{j} \otimes \frac{1}{||p_{ij}||} p_{ij},
\]
with Schmidt values given by the norm of blocks; $||p_k||$. As \LL{2} gates have flat Schmidt spectrum, we necessarily have that  $||p_k||^2=\mu$ for all $k=1,2,,\cdots \mathcal{R}$, This leads to the following equation
\[
P=\sqrt{\mu} \sum_{i,j} \ket{i}\bra{j} \otimes \frac{1}{\sqrt{\mu}} p_{ij}.
\]
From the unitarity of $P$ it follows that $||p_{ij}||^2=\mu \in \left\lbrace 1,\cdots,q \right\rbrace$ and $\mathcal{R}\mu=q^2 \implies \mu =q^2/\mathcal{R}$. As $\mu $ is a positive integer, we find that $\mathcal{R}$ must divide $q^2$; $q^2 \mod \mathcal{R}=0$.

\emph{All non-zero blocks are not linearly independent}. 
Let $\mathcal{R}$ be the Schmidt rank of a given \LL{2} permutation gate. The number of linearly independent blocks appearing in the block form is equal to $\mathcal{R}$. Summing over the computational basis states that are associated with a given repeated block, the Schmidt decomposition of \LL{2} permutation gate is given by
\[
P=\sum_{k=1}^\mathcal{R}  ||A_{k}|| \, ||p_{k}|| \;  \left(\frac{1}{||A_{k}||}A_k \otimes \frac{1}{||p_{k}||} p_{k}\right),
\]
with Schmidt values given by $ ||A_{k}|| \, ||p_{k}|| $. As \LL{2} gates have flat Schmidt spectrum, therefore  $ ||A_{k}||^2=\nu \, , ||p_{k}||^2 =\mu$ for all $k=1,2,,\cdots \mathcal{R}$, This leads to the following equation
\[
P=\sqrt{\mu \nu} \sum_{k=1}^\mathcal{R}  \left(\frac{1}{||A_{k}||}A_k \otimes \frac{1}{||p_{k}||} p_{k}\right).
\]
From the unitarity of $P$ it follows that $\mathcal{R}\mu \nu=q^2 \implies {\mu \nu} \mathcal{R}=q^2/\mathcal{R}$. As $\mu$ and $\nu$ are both positive integers, the possible values for $\mathcal{R}$ are given by $q^2 \mod \mathcal{R}=0$.

\bibliography{entanglement_membrane}% Produces the bibliography via BibTeX.
\end{document}